\documentclass[preprint,10pt]{elsarticle}
\usepackage{geometry}
\usepackage[edges]{forest}
\usepackage{tikz}
\usetikzlibrary{arrows.meta,positioning,calc}
\usepackage{amsmath,amsfonts}
\usepackage{graphicx}
\usepackage{url}
\usepackage{xcolor}
\usepackage{booktabs}
\usepackage[hidelinks]{hyperref}
\usepackage{enumitem}
\usepackage{wrapfig}
\usepackage{tabularx}
\usepackage{multirow}
\usepackage{adjustbox}

\usepackage{longtable}
\usepackage{rotating}
\usepackage{subcaption}
\usepackage{float}
\usepackage{placeins}
\usepackage{comment}
\usepackage{cleveref}

\usepackage{forest}
\usepackage[normalem]{ulem} %
\usepackage{xcolor} %

\title{From Entry Point to Impact: A Survey of In-Vehicle Cybersecurity Threats, Mitigations, and Open Research Directions\tnoteref{t1}}
\tnotetext[t1]{Accepted manuscript. Accepted for publication in \emph{Vehicular Communications} (Elsevier), 2026. \copyright~2026. This manuscript version is made available under the CC-BY-NC-ND 4.0 license \url{https://creativecommons.org/licenses/by-nc-nd/4.0/}.}

\author[tntech]{Micah Jones}
\author[rug]{Fadi Mohsen}
\author[memphis]{Christos Papadopoulos}
\author[mercy]{Usman Rauf}
\author[tntech]{Susmit Shannigrahi}

\address[tntech]{Department of Computer Science, Tennessee Technological University, Cookeville, TN, USA}
\address[rug]{Department of Computing Science, University of Groningen, Groningen, The Netherlands}
\address[memphis]{Department of Computer Science, University of Memphis, Memphis, TN, USA}
\address[mercy]{Department of Cybersecurity, Mercy University, New York, NY, USA}

\begin{document}

\begin{abstract}

The increasing complexity and connectivity of modern vehicles have significantly expanded the attack surface of in-vehicle systems. While in-vehicle networks and sensors enable advanced functionality and automation, their underlying designs were developed with limited cybersecurity considerations. As a result, once access to the vehicle is obtained, attackers can exploit internal communication channels and sensor interfaces to manipulate control behavior, disrupt operation, and compromise safety and privacy. Addressing these risks requires a systematic understanding of how in-vehicle attacks are initiated, executed, and mitigated. We use {four}  representative
attack chains: the Jeep Cherokee remote compromise,
the Tesla Free-Fall exploit, GPS spoofing against
 connected-vehicle safety applications{, and the 2024 Kia connected-vehicle backend/API compromise} as driving examples to
ground the analysis across all four stages of the lifecycle: entry points, attack methodologies, operational impacts, and mitigations.

In this paper, we analyze in-vehicle attacks along two dimensions: the
attack lifecycle, comprising the four stages above, and
technology scope spanning diverse in-vehicle communication and sensor
technologies.
We analyze existing defense mechanisms and evaluate their effectiveness and limitations across attack classes. Finally, we synthesize prior work to identify research gaps and underexplored threats where we discuss emerging trends, open issues, and future directions for resilient and safety-aware in-vehicle security solutions.
The primary contribution is this structured end-to-end taxonomy covering attack
entry points, methodologies, operational impacts, and mitigations within a
single framework spanning heterogeneous in-vehicle technologies.
\end{abstract}


\maketitle

\section{Introduction}\label{sec:1-intro}

The rapid growth in functionality and connectivity of modern vehicles has fundamentally changed their internal security posture. What were once largely isolated mechanical systems have evolved into software-driven platforms composed of interconnected electronic control units (ECUs), in-vehicle networks (IVNs), sensors, and cloud-facing services \cite{rathore2022vehicle,canino2025cybersecurity}. Features such as advanced driver assistance, autonomous operation, remote diagnostics, and over-the-air (OTA) updates rely on continuous data exchange and, increasingly, on AI- and ML-based decision pipelines \cite{george2025cyber,santa2025automotive}. While these capabilities enable significant improvements in safety and efficiency, they also expose in-vehicle systems to cyber threats that were not anticipated during their original design \cite{rathore2022vehicle,canino2025cybersecurity}.

\begin{figure}[!htbp]
    \centering    
    \includegraphics[width=0.7\linewidth]{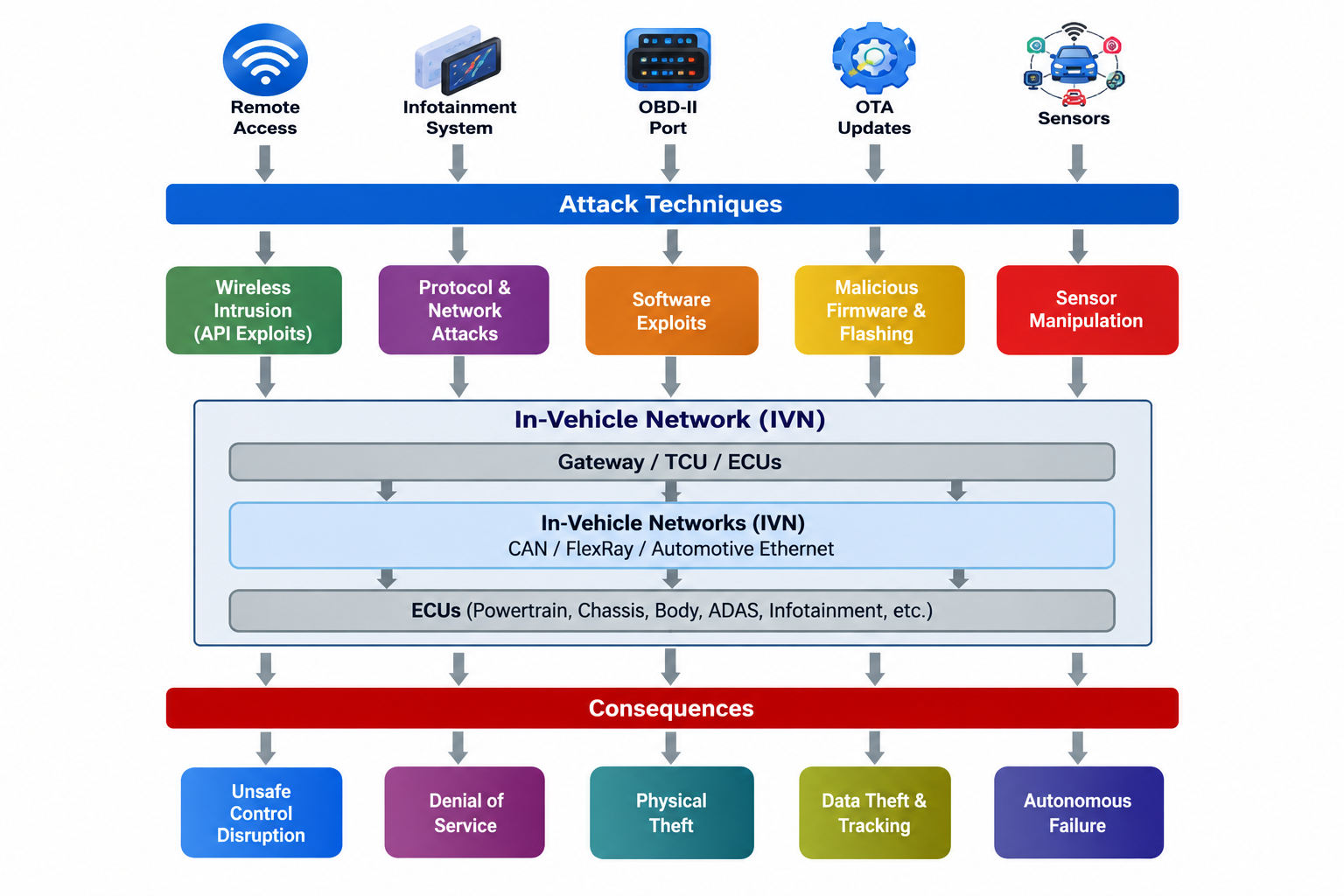}
    \caption{%
Cyber attack pathways in in-vehicle networks}
    \label{fig:pathways}
\end{figure}

Within a modern vehicle, internal and external components are now intertwined as Figure \ref{fig:pathways} shows. %
 IVNs such as CAN, FlexRay, and Automotive Ethernet coordinate time-critical communication among gateways, TCUs, and ECUs responsible for steering, braking, powertrain control, and ADAS \cite{canino2025cybersecurity}. Sensors including cameras, radar, LiDAR, and GPS directly influence perception and control logic, often through learning-based models \cite{george2025cyber}. At the same time, software interfaces and application programming interfaces (APIs) connect the vehicle to mobile applications, backend services, remote diagnostics platforms, and OTA infrastructure via cellular, Wi-Fi, and Bluetooth links \cite{santa2025automotive}. Each of these interfaces introduces potential attack surfaces through which an adversary may gain access to internal vehicle functionality.

The in-vehicle cybersecurity landscape is shaped by a set of architectural and operational assumptions that were reasonable in earlier vehicle designs but are increasingly misaligned with modern, connected, and software-defined platforms. The primary sources of vulnerability considered in this work arise from the following assumptions:

\textbf{Legacy design assumptions:}
\begin{itemize}
  \item \textbf{Assumed physical isolation of IVNs.}
  IVNs were traditionally designed as closed systems, with the implicit assumption that physical access equates to trust. This assumption weakens as remote connectivity and software-mediated access paths increasingly bypass physical boundaries.

  \item \textbf{Lack of built-in cybersecurity mechanisms.}
  Many legacy in-vehicle protocols and components were not designed with authentication, encryption, or access control in mind. These designs assume cooperative participants and provide limited means to distinguish legitimate behavior from misuse once internal access is obtained.
\end{itemize}

\textbf{Architectural trends:}
\begin{itemize}
  \item \textbf{Expanding number of external interfaces.}
  Modern vehicles integrate a growing set of wired and wireless interfaces, including infotainment connectivity, diagnostics, OTA updates, and fleet services. Each additional interface increases the number of entry points that must be secured and consistently mediated.

  \item \textbf{Increasing reliance on third-party software.}
  Vehicle platforms incorporate operating systems, middleware, applications, and libraries developed by external vendors. Deep integration of third-party software into core vehicle functions introduces supply-chain dependencies and trust assumptions that are difficult to validate at runtime.

  \item \textbf{Increasing integration of AI, ML, and data-driven components.}
  Learning-based perception, planning, and control pipelines rely on complex data flows and software dependencies. Their integration into safety-relevant functions expands the internal attack surface and introduces failure modes not addressed by traditional network-centric models.
\end{itemize}

\textbf{Engineering constraints:}
\begin{itemize}
  \item \textbf{Real-time operation and functional safety guarantees.}
  Safety-critical control functions impose strict latency, determinism, and reliability constraints. These requirements limit the deployment of computationally intensive security mechanisms and restrict protocol flexibility, often prioritizing timing correctness over defensive checks.
\end{itemize}

Taken together, these assumptions reflect a fundamental shift in vehicle design from isolated, hardware-centric systems to connected, software-defined platforms. 
Vulnerabilities emerge not from any single component, but from a growing mismatch between hardware- and trust-based architectural assumptions and the increasing volume of functionality introduced through connectivity, software integration, and learning-enabled components.

In-vehicle attacks typically follow a progression from access acquisition to system-level impact. Access may be obtained physically through diagnostic or maintenance interfaces, or remotely through wireless channels, exposed APIs, infotainment systems, or backend services \cite{rathore2022vehicle}. Once access is achieved, attackers can exploit implicit trust within IVNs, inject or manipulate network messages, interfere with sensor data streams, or target AI/ML components by influencing inputs or model behavior \cite{kousar2025deep,christy2025machine}. These actions can degrade system reliability, disable safety functions, or induce unsafe control decisions \cite{george2025cyber,jayaratne2025quantitative}. Beyond immediate safety concerns, compromised vehicles may leak sensitive data, enable persistent surveillance, or be leveraged as part of coordinated attacks across fleets or infrastructure \cite{santa2025automotive}.

\begin{figure}[!htbp]
    \centering
    \adjustbox{width=\linewidth}{\begin{tikzpicture}[
    x=1cm, y=1cm,
    font=\sffamily\scriptsize,
    >={Stealth[length=1.6mm,width=1.3mm]},
    nohyph/.style  ={execute at begin node={\hyphenpenalty=10000\exhyphenpenalty=10000\relax}},
    panel/.style   ={draw=black!60, line width=0.4pt},
    ptitle/.style  ={font=\sffamily\bfseries\normalsize},
    stage/.style   ={draw=black!70, fill=taxblue, font=\sffamily\bfseries\small,
                     minimum height=0.62cm, inner xsep=6pt, align=center, nohyph},
    band/.style    ={draw=black!60, fill=taxblue, minimum height=0.98cm, inner sep=0pt},
    xband/.style   ={draw=black!60, fill=black!10, minimum height=1.22cm, inner sep=0pt},
    hbox/.style    ={draw=black!70, fill=white, align=center, minimum height=0.70cm,
                     inner sep=1.5pt, nohyph},
    item/.style    ={draw=black!45, fill=white, align=center, minimum height=0.62cm,
                     inner sep=2pt, nohyph},
    frame/.style   ={draw=black!60, inner sep=2.2pt},
    xframe/.style  ={draw=black!60, dashed, inner sep=2.2pt},
    flow/.style    ={draw=black!75, line width=0.7pt, ->},
    fan/.style     ={draw=black!65, line width=0.35pt, ->},
    link/.style    ={draw=black!70, line width=0.6pt, dashed, ->},
    xlink/.style   ={draw=black!55, line width=0.4pt, dashed, ->},
  ]
  \definecolor{taxblue}{RGB}{164,197,232}

  \newcommand{\taxgroup}[8]{%
    \node[#7, minimum width=#4cm, anchor=north] (#1-h) at (#2,#3) {};
    \node[hbox, text width=#4cm-0.55cm] at (#1-h.center) {#5};
    \coordinate (#1-i0) at ($(#1-h.south)+(0,-0.03)$);
    \foreach \t [count=\i, remember=\i as \p (initially 0)] in {#6} {%
      \node[item, text width=#4cm-0.6cm, below=0.16cm of #1-i\p, anchor=north] (#1-i\i) {\t};
      \xdef\taxlast{#1-i\i}%
    }%
    \node[#8, fit=(#1-h)(\taxlast)] (#1) {};
  }
  \newcommand{\mitgroup}[5]{%
    \node[stage, text width=3.9cm] (#1) at (16.9,#2) {#3};
    \coordinate (#1-i0) at ($(#1.south)+(0,-0.12)$);
    \foreach \t [count=\i, remember=\i as \p (initially 0)] in {#4} {%
      \node[item, text width=3.6cm, below=#5 of #1-i\p, anchor=north] (#1-i\i) {\t};
      \xdef\taxlast{#1-i\i}%
    }%
    \node[frame, fit=(#1)(\taxlast), inner sep=3pt] (#1-f) {};
  }

  \node[stage] (ep) at (6.65,-1.15)  {Entry Points};
  \node[stage] (am) at (6.65,-8.95)  {Attack Methodologies};
  \node[stage] (im) at (6.65,-14.75) {Impacts};

  \taxgroup{phy}{3.05}{-1.95}{3.50}{Physical Layer}
    {{OBD-II / USB /\\Media Ports},{Perception Sensors\\(LiDAR, Radar, Camera)},
     {GNSS / Tire Pressure\\Monitoring System\\(TPMS) / Ultrasonic /\\Microphone},{EV Charging}}{band}{frame}
  \taxgroup{dl}{6.65}{-1.95}{3.25}{Data-Link Layer}
    {{CAN / CAN FD},{LIN},{FlexRay},{Automotive Ethernet}}{band}{frame}
  \taxgroup{app}{10.10}{-1.95}{3.25}{Application Layer}
    {{Keyless Entry / Bluetooth / Wi-Fi},{Cellular},{V2X},{OTA Updates},
     {Automotive Applications},{Cloud / Backend API}}{band}{frame}

  \taxgroup{man}{2.80}{-9.75}{3.65}{Manipulation}
    {{Signal / Measurement Level},{Message / Frame Level},
     {Firmware / Execution Level},{Control-Logic / Semantic Level}}{band}{frame}
  \taxgroup{mon}{6.65}{-9.75}{3.65}{Monitoring and\\Reconnaissance}
    {{Internal (On-IVN)},{External (Off-Vehicle)}}{band}{frame}
  \taxgroup{dis}{10.50}{-9.75}{3.65}{Disruption}
    {{Internal (On-IVN)},{External (RF / Sensor / Network)}}{band}{frame}

  \taxgroup{mech}{1.30}{-15.55}{3.25}{Mechanical and\\Electrical Anomalies}
    {{Steering / Brake /\\Engine Lockout},{Lane Assist Failure},
     {Auxiliary Actuator\\Disruption}}{band}{frame}
  \taxgroup{false}{4.85}{-15.55}{3.25}{False or Misleading\\Information}
    {{Speedometer and\\Dashboard Manipulation},{Odometer, Tachograph,\\and Governor Tampering},
     {Infotainment and\\Display Deception}}{band}{frame}
  \taxgroup{leak}{8.40}{-15.55}{3.25}{Information Leakage and\\Follow-On Attacks}
    {{Eavesdropping and\\Privacy Leakage},{Door Locks, Alarms, and\\Theft Enablement}}{band}{frame}
  \taxgroup{cross}{11.95}{-15.55}{3.25}{Cross-Cutting\\Impact Properties\\(not a category)}
    {{Availability Loss},{Safety Degradation\\Without Explicit Failure},
     {Cascading\\Cross-Domain Effects},{Driver Trust Erosion},
     {ML / Probabilistic Failures}}{xband}{xframe}

  \mitgroup{mep}{-1.15}{At Entry Points}
    {{Detection at Attack\\Entry Points},{Architectural Isolation and\\Secure Gateways},
     {Wireless Interface Hardening},{Sensor and External\\Signal Validation}}{0.30cm}
  \mitgroup{mam}{-8.95}{Against Methodologies}
    {{Detection of Attack\\Methodologies},{Signal- and Measurement-Level\\Defenses},
     {Message- and Frame-Level\\Protections},{Firmware- and Execution-Level\\Security},
     {Control-Logic and Semantic\\Validation}}{0.16cm}
  \mitgroup{mim}{-14.75}{Impact Containment\\and Recovery}
    {{Impact-Level Detection},{Availability and Disruption\\Mitigation},
     {Fail-Safe Operation and\\Graceful Degradation},{Privacy and Information\\Leakage Mitigation}}{0.30cm}
  \node[stage, fill=black!10, text width=3.9cm] (mdd) at (16.9,-20.25)
        {Data-Driven and Semantic\\Attack Detection};
  \node[item, text width=3.6cm, below=0.12cm of mdd] (mdd-i1) {Applies across all three attack stages};
  \node[xframe, fit=(mdd)(mdd-i1), inner sep=3pt] (mdd-f) {};
  \node[font=\sffamily\scriptsize\itshape, align=center, text width=4.0cm]
        (roles) at (16.9,-21.95) {Defensive roles:\\prevent / detect / contain / recover};

  \draw[panel] (-0.9,0.45) rectangle (13.8,-22.5);
  \draw[panel] (-0.9,0.45) rectangle (13.8,-0.40);
  \node[ptitle] at (6.45,0.02) {Attacks};
  \draw[panel] (14.5,0.45) rectangle (19.3,-22.5);
  \draw[panel] (14.5,0.45) rectangle (19.3,-0.40);
  \node[ptitle] at (16.9,0.02) {Mitigations};

  \foreach \s/\g in {ep/phy,ep/dl,ep/app,am/man,am/mon,am/dis,im/mech,im/false,im/leak}
    \draw[fan] (\s.south) -- (\g-h.north);
  \draw[fan, dashed] (im.south) -- (cross-h.north);
  \draw[flow] (ep.west) -- (-0.5,-1.15) -- (-0.5,-8.38) -- (6.65,-8.38) -- (am.north);
  \draw[flow] (am.west) -- (-0.5,-8.95) -- (-0.5,-14.18) -- (6.65,-14.18) -- (im.north);
  \draw[link] (ep.east) -- (mep.west);
  \draw[link] (am.east) -- (mam.west);
  \draw[link] (im.east) -- (mim.west);
  \draw[xlink] (dis.east) -- (14.30,0 |- dis.east) -- (14.30,0 |- mim-i2.west) -- (mim-i2.west);
  \coordinate (mon-out) at ($(mon.south)+(0.9,0)$);
  \coordinate (mon-low) at ($(dis.south)+(0,-0.35)$);
  \draw[xlink] (mon-out) -- (mon-out |- mon-low) -- (14.00,0 |- mon-low)
              -- (14.00,0 |- mim-i4.west) -- (mim-i4.west);
\end{tikzpicture}}
    \caption{%
 End-to-end taxonomy of in-vehicle attacks: entry points, attack methodologies, impacts, and mitigations.}
    \label{fig:taxonomy}
\end{figure}
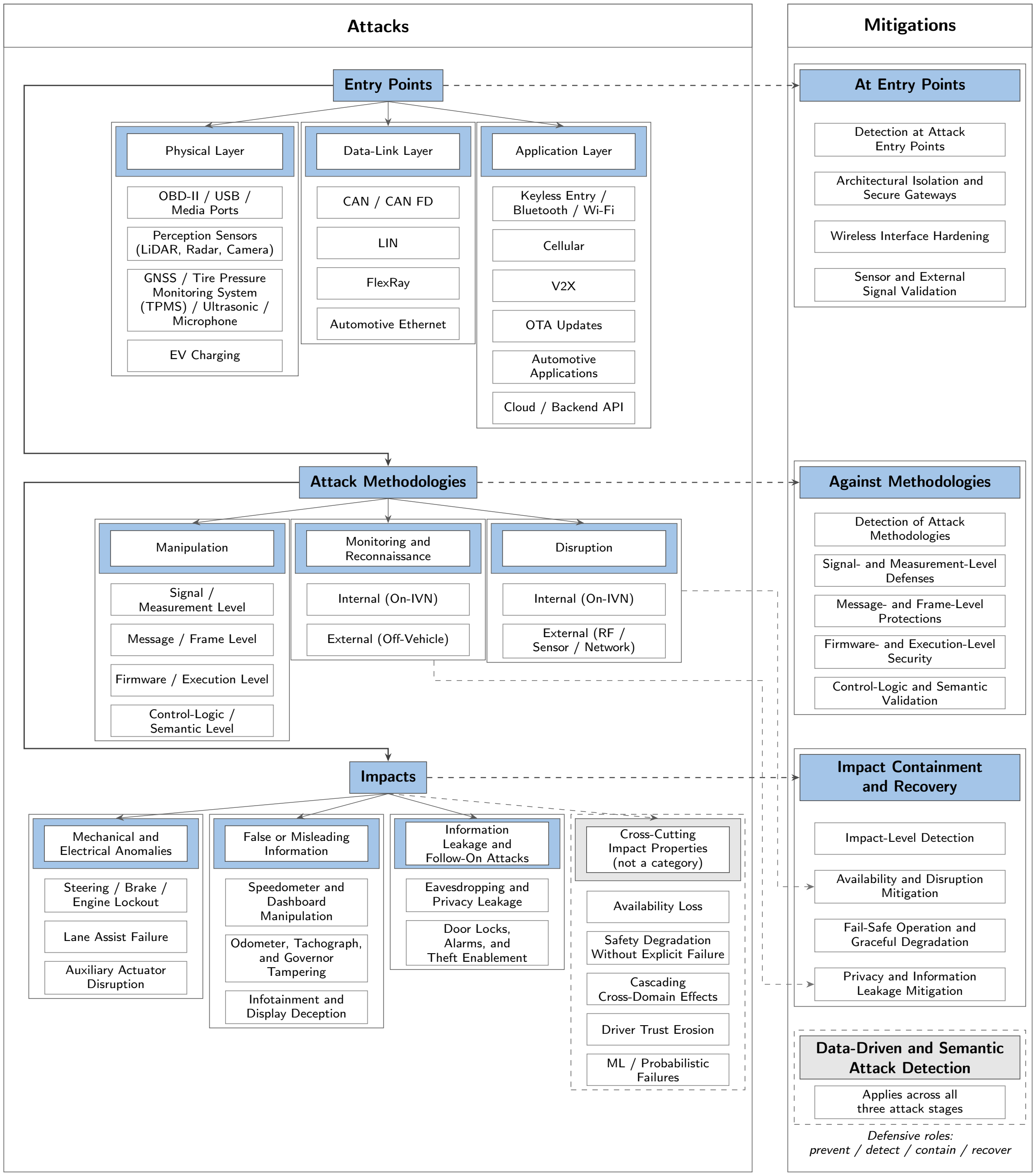

A substantial body of prior work has examined in-vehicle cybersecurity from multiple perspectives, including practical attacks against production vehicles \cite{checkoway2011comprehensive, wright2011hacking}, threat categorization by access vectors \cite{6615528}, entry-point classifications and countermeasures \cite{10.1145/3431233,canino2025cybersecurity}, hierarchical attack taxonomies \cite{info10040148}, protocol-specific vulnerabilities in CAN \cite{DBLP:journals/corr/abs-1802-01725}, and security challenges across broader IVN technologies including FlexRay and Automotive Ethernet \cite{8688625,rathore2022vehicle,7393668}, software-level propagation analyses \cite{LE201817}, subsystem and system-level attack surveys \cite{KIM2021102150}, operational resilience and Vehicle-SOCs \cite{santa2025automotive}, learning-based defenses \cite{9463461,christy2025machine,kousar2025deep}, architectural hardening \cite{5948952,castiglione2025securing}, and post-quantum cryptography for IVNs \cite{smith2025post,castiglione2025securing}. Despite this extensive prior work, existing surveys typically focus on individual layers, subsystems, or defensive mechanisms in isolation \cite{jayaratne2025quantitative,8688625}. As vehicles become increasingly software-defined and reliant on APIs, OTA updates, cloud connectivity, and AI/ML-driven control \cite{george2025cyber}, this fragmented treatment complicates end-to-end risk analysis, leaving interactions between access vectors, attack execution, systemic impact, and mitigation strategies implicit. See Section~\ref{sec:2-related-work} for a detailed discussion of prior work and how this survey differs.

This paper addresses this gap by presenting a unified end-to-end taxonomy of
in-vehicle cybersecurity threats. \textbf{By `end-to-end,' we specifically
mean an integrated analysis spanning four stages}: (1) entry points through
which attacks are initiated, (2) attack methodologies employed after access,
(3) systemic impacts that result, and (4) mitigation strategies that address
attacks at each stage. The analysis spans heterogeneous in-vehicle technologies
including CAN, FlexRay, Automotive Ethernet, wireless interfaces (Bluetooth,
 Wi-Fi, cellular), sensors (cameras, LiDAR, radar, GPS), OTA mechanisms, and
cloud services. Prior surveys address the four lifecycle stages selectively or
in isolation, and most are limited to a single protocol or subsystem (Table~\ref{table:related-works}), leaving
the interactions between access vectors, attack execution, and systemic impact
implicit. In contrast, this survey covers both dimensions within a single
framework and enables systematic reasoning across the full attack lifecycle
without composing results from multiple works.
Figure~\ref{fig:taxonomy} shows the %
 end-to-end taxonomy.

\begin{figure}[H]
    \centering
\newcommand{\figsec}[1]{~{\color{black!55}(Section~\ref{#1})}}%
\resizebox{\linewidth}{!}{%
\begin{forest}
  for tree={
    font=\sffamily\footnotesize, %
    line width=0.6pt,
    grow'=0,
    child anchor=west,
    parent anchor=east,
    anchor=west,
    align=left,
    edge={draw, line width=0.6pt},
    edge path={
      \noexpand\path[\forestoption{edge}]
      (!u.parent anchor) -- +(6pt,0) |- (.child anchor) \forestoption{edge label};
    },
    l sep=10pt, %
    s sep=1.5pt, %
  }
[End-to-End Vehicular\\Security Taxonomy, draw, semithick
  [{Introduction\figsec{sec:1-intro}}, edge={gray}]
  [{Related Work\figsec{sec:2-related-work}}, edge={teal!70!black}
    [{\uline{Methodology of the Survey on End-to-End Vehicular Security}\figsec{sec:2.1-methodology}}]
    [{\uline{Attack Modeling and Simulation}\figsec{sec:attack-modeling-simulation}}]
    [{\uline{Survey Structure}\figsec{sec:2.3-survey-structure}}]
  ]
  [{End-to-End Taxonomy: Entry Points,\\Attack Methodologies, Impacts,\\and Mitigations\figsec{sec:3-attacks}}, edge={blue!70!black}
    [{Entry Points\figsec{sec:3.1-entry-points}}
      [{\uline{Entry Points at the Physical Layer}\figsec{sec:3.1.1-physical}}]
      [{\uline{Entry Points at the Data-Link Layer}\figsec{sec:3.1.2-datalink}}]
      [{\uline{Entry Points at the Application Layer}\figsec{sec:3.1.3-LABEL}}]
    ]
    [{Attack Methodologies\figsec{sec:3.2-attack-methodologies}}
      [{\uline{Manipulation}\figsec{sec:3.2.1-manipulation}}]
      [{\uline{Monitoring and Reconnaissance}\figsec{sec:3.2.2-recon}}]
      [{\uline{Disruption}\figsec{sec:3.2.3-disruption}}]
    ]
    [{Attack Impacts\figsec{sec:3.3-attack-impacts}}
      [{\uline{Mechanical and Electrical Anomalies}\figsec{sec:3.3.1-mechanical}}]
      [{\uline{False or Misleading Information}\figsec{sec:3.3.2-false-info}}]
      [{\uline{Information Leakage and Follow-On Attacks}\figsec{sec:3.3.3-leakage}}]
    ]
    [{Attack Detection and Mitigations\figsec{sec:3.4-attack-mitigations}}
      [{\uline{Detection and Mitigations at Attack Entry Points}\figsec{sec:3.4.1-entry-mitigations}}]
      [{\uline{Detection and Mitigations Disrupting Attack Methodologies}\figsec{sec:3.4.2-method-mitigations}}]
      [{\uline{Detection and Mitigations for Impact Containment and Recovery}\figsec{sec:3.4.3-impact-mitigations}}]
      [{\uline{Data-Driven and Semantic Attack Detection}\figsec{sec:3.4.4-data-driven}}]
    ]
  ]
  [{Trends, Research Gaps, and\\Future Directions\figsec{sec:4-research-gaps}}, edge={orange!80!black}
    [{\uline{Entry Points: Expanding and Underexamined Attack Surfaces}\figsec{sec:4.1-gaps-entry}}]
    [{\uline{Attack Methodologies: From Protocol Exploits to AI-Targeted Attacks}\figsec{sec:4.2-gaps-methods}}]
    [{\uline{Impacts: Cascading, Systemic, and Fleet-Level Effects}\figsec{sec:4.3-gaps-impacts}}]
    [{\uline{Mitigations: Detection Emphasis and Emerging Approaches}\figsec{sec:4.4-gaps-mitigations}}]
    [{\uline{Cybersecurity Testbeds and Experimental Evaluation}\figsec{sec:testbeds}}]
  ]
  [{Conclusions\figsec{sec:6-conclusions}}, edge={purple}]
]
\end{forest}}

\caption{%
Structural overview of the paper.}

\label{fig:organization}
\end{figure}
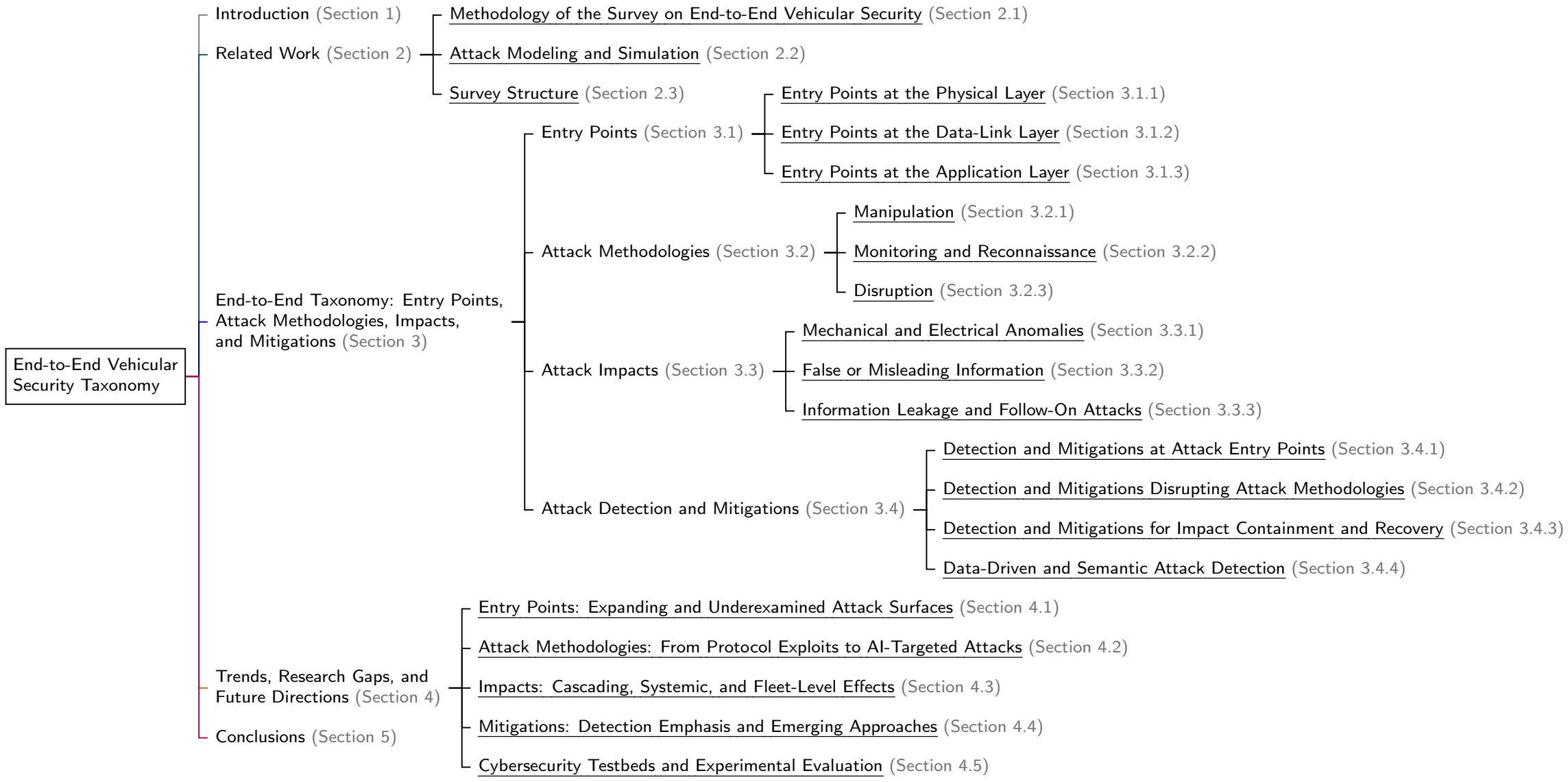

To ground this framework concretely, we use {four}  representative attack chains
as driving examples throughout the survey: the Jeep Cherokee remote compromise
\cite{miller2015remote}, the Tesla Free-Fall exploit \cite{niefreefall2017},
 GPS spoofing against  connected-vehicle safety applications \cite{ying2023gps}{,
and the 2024 Kia connected-vehicle backend/API compromise
\cite{curry2024kia}},
each of which is traced through entry points, methodologies, impacts, and
mitigations in the sections that follow.

{
These attacks are representative rather than exhaustive. The earlier Jeep
Cherokee and Tesla cases remain relevant because vehicle architectures evolve
over long deployment lifecycles, and CAN-based and distributed ECU
architectures will remain in service as newer vehicles adopt Automotive
Ethernet, centralized compute, and domain or zonal controllers. Architectural
evolution therefore introduces new attack surfaces without immediately
eliminating legacy ones. The Kia case complements these earlier examples with
a recent cloud- and API-mediated attack path.
}

 {The remainder of this paper is organized as follows (Figure~\ref{fig:organization}).} First, we provide a literature review of in-vehicle cybersecurity research in Section \ref{sec:2-related-work}. We then analyze attack entry points (Section~\ref{sec:3.1-entry-points}), examine in-vehicle attack methodologies affecting networks, sensors, software, and AI/ML pipelines (Section~\ref{sec:3.2-attack-methodologies}), and assess their impacts on safety, privacy, and operation (Section~\ref{sec:3.3-attack-impacts}). We then review mitigation techniques proposed for both legacy and modern vehicles (Section~\ref{sec:3.4-attack-mitigations}), before identifying open challenges and future research directions relevant to secure, resilient automotive systems (%
 Section~\ref{sec:4-research-gaps}).

\newcolumntype{C}[1]{>{\centering\arraybackslash}p{#1}}
\newcommand{\rothead}[2][black]{\rotatebox{90}{\scriptsize\bfseries\color{#1}\begin{tabular}{@{}l@{\hspace{3pt}}}#2\end{tabular}}}
\begin{table*}[!htbp]
\centering
\caption{
Comparison of Related Work Across Security Dimensions in Vehicular Networks}
\label{table:related-works}
\footnotesize
\setlength{\tabcolsep}{3pt}
\begin{tabularx}{\textwidth}{|C{0.7cm}|C{0.7cm}|C{0.45cm}|C{0.45cm}|C{0.45cm}|C{0.45cm}|C{0.45cm}|C{0.45cm}|>{\raggedright\arraybackslash}X|}
\hline
\textbf{Ref.} & \textbf{Year} & \multicolumn{4}{c|}{\textbf{End-to-End}} & & & \\
\cline{3-6}
 & & \rothead[red]{Entry\\Points} & \rothead[red]{Attack\\Methodologies} & \rothead[red]{Impacts} & \rothead[red]{Mitigations} & \rothead{Comprehensive\\Coverage} & \rothead{Future\\Work} & \textbf{Shortcomings (relative to end-to-end modeling and cross-technology coverage)} \\
\hline

\cite{5940525} & 2011 & \checkmark & \checkmark & $\times$ & \checkmark & $\times$ & $\times$ &
Mitigation and future work limited to entry surfaces; no impact analysis or full end-to-end chain. \\
\hline

\cite{6615528} & 2013 & \checkmark & \checkmark & \checkmark & \checkmark & \checkmark & $\times$ &
Early embedded-network threat survey; predates recent architectures and modern external sensing. \\
\hline

\cite{DBLP:journals/corr/abs-1802-01725} &  2018 & $\times$ & \checkmark & \checkmark & $\times$ & $\times$ & $\times$ &
CAN bus security only; no structured entry-surface taxonomy or system-level mitigation comparison. \\
\hline

\cite{10.1145/3431233} & 2021 & \checkmark & \checkmark & \checkmark & \checkmark & $\times$ & \checkmark &
Comprehensive CAN bus attack and defense survey; does not cover other IVN protocols or subsystems. \\
\hline

\cite{7393668} & 2016 & \checkmark & \checkmark & $\times$ & $\times$ & \checkmark & $\times$ &
IVN architecture and QoS survey; security covered briefly, without taxonomy, impact, or mitigation. \\
\hline

\cite{5948952} & 2011 & $\times$ & $\times$ & $\times$ & $\times$ & \checkmark & \checkmark &
Architecture and standards-oriented survey; does not systematically analyze attack surfaces, threat taxonomies, or mitigation. \\
\hline

\cite{9316663} &  2021 & \checkmark & \checkmark & \checkmark & \checkmark & \checkmark & $\times$ &
ITS/VANET security survey rather than IVN-specific; no dedicated future-work section. \\
\hline

\cite{LE201817} & 2018 & $\times$ & \checkmark & $\times$ & \checkmark & \checkmark & \checkmark &
Automotive application and platform security survey; not IVN-centric; no operational or safety impact. \\
\hline

\cite{8688625} &  2020 & $\times$ & \checkmark & $\times$ & \checkmark & $\times$ & \checkmark &
IVN IDS survey; mitigation-centric, without entry-surface modeling or system-level impact analysis. \\
\hline

\cite{checkoway2011comprehensive} & 2011 & \checkmark & $\times$ & \checkmark & $\times$ & \checkmark & $\times$ &
Remote compromise via external interfaces; no structured taxonomy or mitigation comparison. \\
\hline

\cite{KIM2021102150} & 2021 & $\times$ & \checkmark & \checkmark & \checkmark & \checkmark & \checkmark &
Comprehensive survey of attacks and defenses in autonomous vehicles; not centered on in-vehicle network attacks. \\
\hline

\cite{9619465} & 2022 & $\times$ & \checkmark & $\times$ & \checkmark & \checkmark & $\times$ &
ML-based vehicular security survey; defense-centric, without linking threats, impact, and mitigation. \\
\hline

\cite{9463461} & 2021 & $\times$ & $\times$ & $\times$ & \checkmark & \checkmark & \checkmark &
ML for vehicular networking; security is not the focus; no structured threat or impact modeling. \\

\hline
\cite{limbasiya2022systematic} & 2022 & \checkmark & \checkmark & $\times$ & \checkmark & $\times$ & \checkmark &
CAN, OBD-II, and ECU attacks with detection and prevention; excludes LiDAR, radar, and GPS; no impact analysis.
\\
\hline
\cite{karopoulos2022demystifying} & 2022 & $\times$ & \checkmark & $\times$ & \checkmark & $\times$ & \checkmark &
Meta-survey of IDS taxonomies; no entry-point characterization or safety-consequence modeling. \\
\hline
\cite{info10040148} &  2019 & $\times$ & \checkmark & $\times$ & $\times$ & $\times$ & \checkmark &
Development-oriented attack taxonomy; no evaluation of mitigation strategies or operational impact. \\
\hline
\cite{kifor2024automotive} & 2024 & $\times$ & $\times$ & $\times$ & \checkmark & \checkmark & \checkmark &
ISO/SAE~21434 and UN~R155 compliance survey; attack surfaces and impact are not modeled.
\\

\hline

\cite{thing2016autonomous} &
2016 &
\checkmark &
\checkmark &
$\times$ &
\checkmark &
\checkmark &
$\times$ &
Taxonomy of autonomous-vehicle attacks and defenses; no explicit end-to-end lifecycle chain. \\
\hline

\cite{martinezcruz2021security} &
2021 &
$\times$ &
\checkmark &
$\times$ &
\checkmark &
\checkmark &
\checkmark &
Multi-protocol IVN survey including Automotive Ethernet; does not link entry surfaces to consequences. \\
\hline

\cite{sommer2024combining} &
2024 &
$\times$ &
\checkmark &
$\times$ &
\checkmark &
\checkmark &
$\times$ &
Maps UN~R155 threats to STRIDE, CAPEC, and CWE using real attacks; no end-to-end lifecycle model. \\
\hline

\cite{AESystematicReview2024} &
2024 &
$\times$ &
\checkmark &
\checkmark &
\checkmark &
$\times$ &
\checkmark &
Automotive Ethernet attacks and countermeasures mapped to R155; protocol-specific by design. \\
\hline

This Work & 2026 & \checkmark & \checkmark & \checkmark & \checkmark & \checkmark & \checkmark &
End-to-end survey linking entry points, attack methodologies, impacts, and mitigations across IVNs. \\
\hline

\end{tabularx}
\end{table*}

\section{Related Work}\label{sec:2-related-work}

Existing surveys on vehicular security address important aspects of the problem, yet they differ substantially in scope and structural coverage. Table~\ref{table:related-works} evaluates prior work across {six comparison dimensions}: entry-point characterization, %
 attack-methodology characterization, impact analysis, mitigation comparison, {cross-technology coverage,} and future research discussion. This comparison enables an objective assessment of how comprehensively each survey models the security lifecycle.

Early works such as \cite{5940525} and \cite{6615528} provide foundational overviews of threats and protection mechanisms in embedded automotive networks. These studies characterize entry points and attack %
 methodologies, and in some cases discuss mitigation techniques. However, structured modeling of operational impact is either absent or limited, and explicit linkage across all security dimensions remains incomplete. While \cite{6615528} offers a broad early survey, it predates recent architectural shifts and does not incorporate modern external sensing and connectivity paradigms.

Several %
 works focus narrowly on specific protocols or components. For example, \cite{DBLP:journals/corr/abs-1802-01725} and \cite{10.1145/3431233} concentrate on %
 CAN bus security. Although \cite{10.1145/3431233} provides comprehensive coverage within that protocol and addresses multiple dimensions, its scope remains confined to CAN and does not extend to heterogeneous in-vehicle architectures. Similarly, \cite{8688625} surveys intrusion detection systems for IVNs and compares mitigation strategies, yet does not structurally connect attacks to system-level operational consequences. These protocol- or mechanism-centric surveys provide depth within a limited scope but do not generalize to an end-to-end perspective across subsystems.

Architecture-oriented and networking-focused surveys such as \cite{7393668} and \cite{5948952} emphasize communication technologies, standards, and quality-of-service considerations. Security threats are discussed only briefly or not systematically analyzed. In these works, entry points, attack taxonomies, and mitigation strategies are not integrated within a unified analytical framework. Consequently, the relationship between vulnerabilities and their operational implications remains underdeveloped. 
\cite{kifor2024automotive} addresses this layer through a standards and testing lens,
surveying ISO/SAE~21434 and UN~R155 compliance mechanisms, but does not model attack
surfaces or operational impact within an integrated analytical chain.

Broader ITS and autonomous vehicle surveys, including \cite{9316663} and \cite{KIM2021102150}, provide comprehensive classifications of attacks and defense mechanisms across vehicular and infrastructure domains. However, their primary focus lies on Intelligent Transportation Systems (ITS), Vehicular Ad hoc NETwork (VANET), or autonomous driving contexts rather than %
 IVN security specifically. While they address multiple dimensions, the treatment is distributed across communication, infrastructure, and application layers without an explicit, system-level mapping tailored to IVNs. 
\cite{limbasiya2022systematic} similarly spans Connected and Autonomous Vehicle (CAV) attack types %
and detection and prevention strategies, focusing on CAN, OBD-II, and ECU attacks, yet treats these dimensions in parallel rather
than constructing an explicit %
 entry point$\rightarrow$%
 attack methodology$\rightarrow$%
 impact$\rightarrow$mitigation chain.

Machine learning–centric surveys such as \cite{9619465} and \cite{9463461} evaluate defensive techniques and networking optimizations. These works emphasize mitigation strategies and performance considerations, yet they do not provide structured linkage between attack surfaces, operational impact, and defense mechanisms. Likewise, taxonomy-driven work such as \cite{info10040148} formalizes attack classification for automotive development but does not evaluate mitigation strategies or consequence modeling. \cite{karopoulos2022demystifying} consolidates prior IDS taxonomies into a meta-survey but addresses attack classification and mitigation comparison solely within the IDS lens, leaving entry-point characterization and operational consequence modeling absent.

{
Several established automotive security taxonomies and threat-analysis frameworks organize the threat space at different levels of abstraction. Thing and Wu classify autonomous-vehicle attacks together with corresponding defenses \cite{thing2016autonomous}, while Singh and Kim examine hardware- and software-level security challenges and architectural protection mechanisms for intelligent vehicles \cite{singh2017security}. SARA provides an automotive threat-analysis and risk-assessment framework based on an explicit threat model, attack-method/asset mappings, and attack-tree reasoning \cite{monteuuis2018sara}. The Auto-ISAC Automotive Threat Matrix (ATM) provides an automotive-specific vocabulary of adversary tactics and techniques derived from observed attacks and reproducible automotive security research \cite{autoisac2024atm}. At a more general level, STRIDE provides an established threat classification used within the Security Development Lifecycle \cite{howard2006sdl}.
}

{
Checkoway et al. provide one of the earliest classifications of automotive attack entry points, organizing access by indirect physical, short-range wireless, and long-range wireless channels \cite{checkoway2011comprehensive}. This proximity-based view is retained in this survey as an attacker-proximity attribute alongside the layer-based entry-point classification (Table~\ref{table:entry-points-layered}), rather than as the primary organizing dimension: proximity determines how an entry point is reached, while the physical/data-link/application layer determines where the trust boundary is crossed.
}

{
The stage-based organization used in this survey is also consistent with general-purpose cybersecurity lifecycle models outside the automotive domain. MITRE ATT\&CK includes tactics such as Initial Access, Execution, and Impact \cite{strom2020attack}, while the Lockheed Martin Cyber Kill Chain describes progression through reconnaissance, weaponization, delivery, exploitation, installation, command and control, and actions on objectives \cite{hutchins2011killchain}. These models provide complementary stage-based representations of adversary behavior; our taxonomy adapts this lifecycle perspective to vehicular systems by explicitly linking entry point, attack methodology, operational impact, and mitigation.
}

{
Sommer et al. further connect automotive and general cybersecurity representations by mapping UN~R155 threats to STRIDE, CAPEC, and CWE \cite{sommer2024combining}. The mappings are validated and refined using an Automotive Attack Database containing 361 real-world vehicle attacks and 621 attack steps, thereby connecting abstract automotive threats with concrete attack techniques, vulnerabilities, and mitigations. These models complement rather than replace the lifecycle organization used in this survey.
}

{
The distinction from \cite{info10040148} is particularly important. That work organizes automotive attacks at multiple levels of abstraction so that attack knowledge can support development activities such as threat analysis and security testing. Our framework addresses a different organizational question: rather than refining the classification of an individual attack, it explicitly traces the progression from an entry point through an attack methodology to an operational impact and the corresponding detection or mitigation.
}

{
Protocol-oriented surveys provide another complementary perspective. Martínez-Cruz et al. analyze security issues and countermeasures across multiple in-vehicle communication protocols, including CAN, LIN, FlexRay, MOST, and Automotive Ethernet \cite{martinezcruz2021security}. De Vincenzi et al. focus specifically on Automotive Ethernet, systematically examining attacks, countermeasures, safety implications, and their relationship to UNECE WP.29 R155 \cite{AESystematicReview2024}. These works provide detailed threat-, protocol-, or development-oriented classifications, whereas our taxonomy preserves the relationship among entry point, attack methodology, operational impact, and detection or mitigation across heterogeneous in-vehicle technologies.
}

As summarized in Table~\ref{table:related-works}, the surveyed works address one or several of these security dimensions, but none of the compared surveys systematically integrates all four lifecycle stages across heterogeneous in-vehicle technologies within a single analytical chain. %
In contrast, our work provides a structured end-to-end framework that integrates all four {lifecycle stages} within a unified model.

\begin{figure}[H]
    \centering

    \begin{subfigure}{0.4\linewidth}
        \centering
        \includegraphics[width=\linewidth]{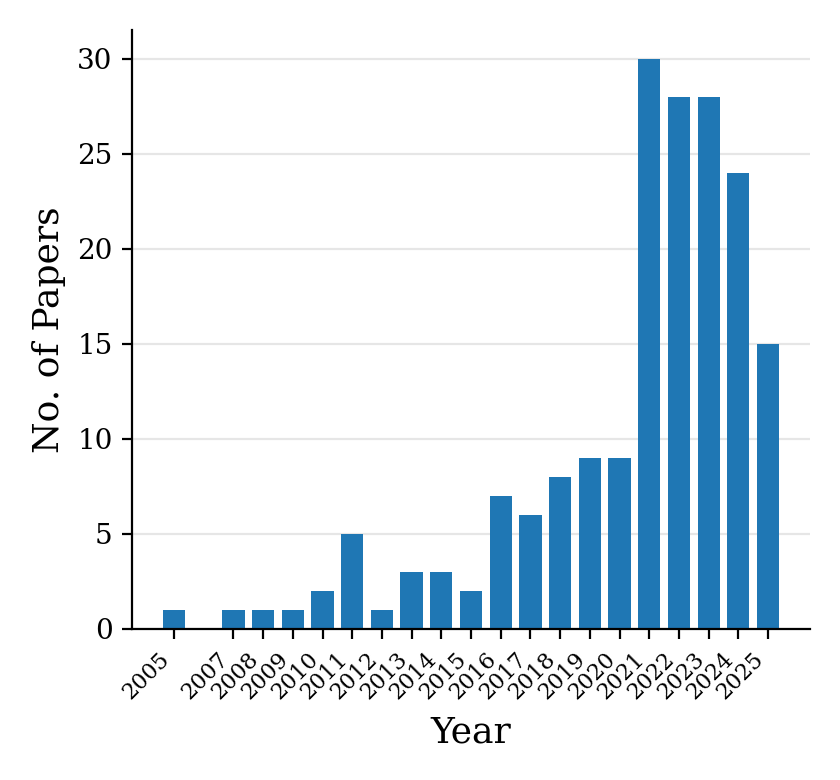}
                \caption{ Surveyed papers by publication year}

    \end{subfigure}
    \hfill
    \begin{subfigure}{0.4\linewidth}
        \centering
        \includegraphics[width=\linewidth]{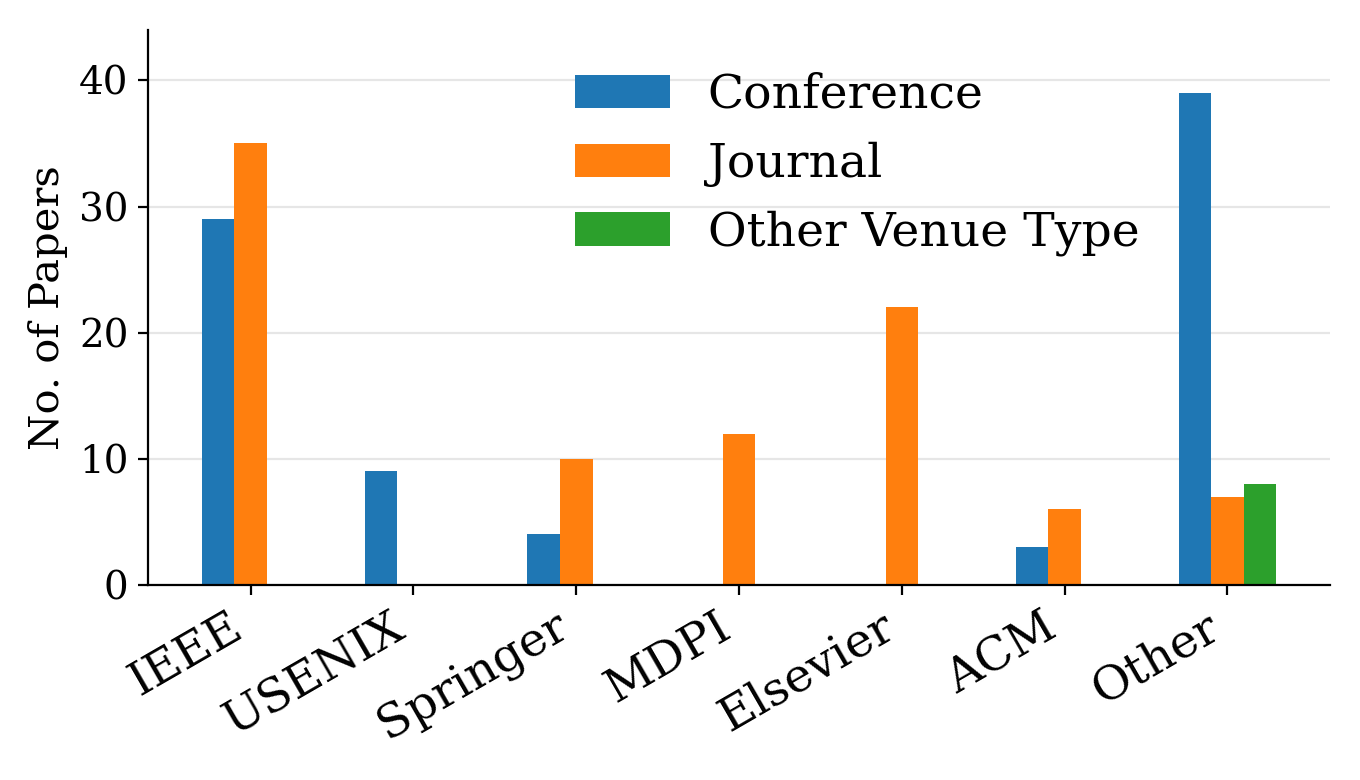}
                \caption{Distribution of collected papers per publication source}

    \end{subfigure}
    \begin{subfigure}{0.5\linewidth}
        \centering
        \includegraphics[width=\linewidth]{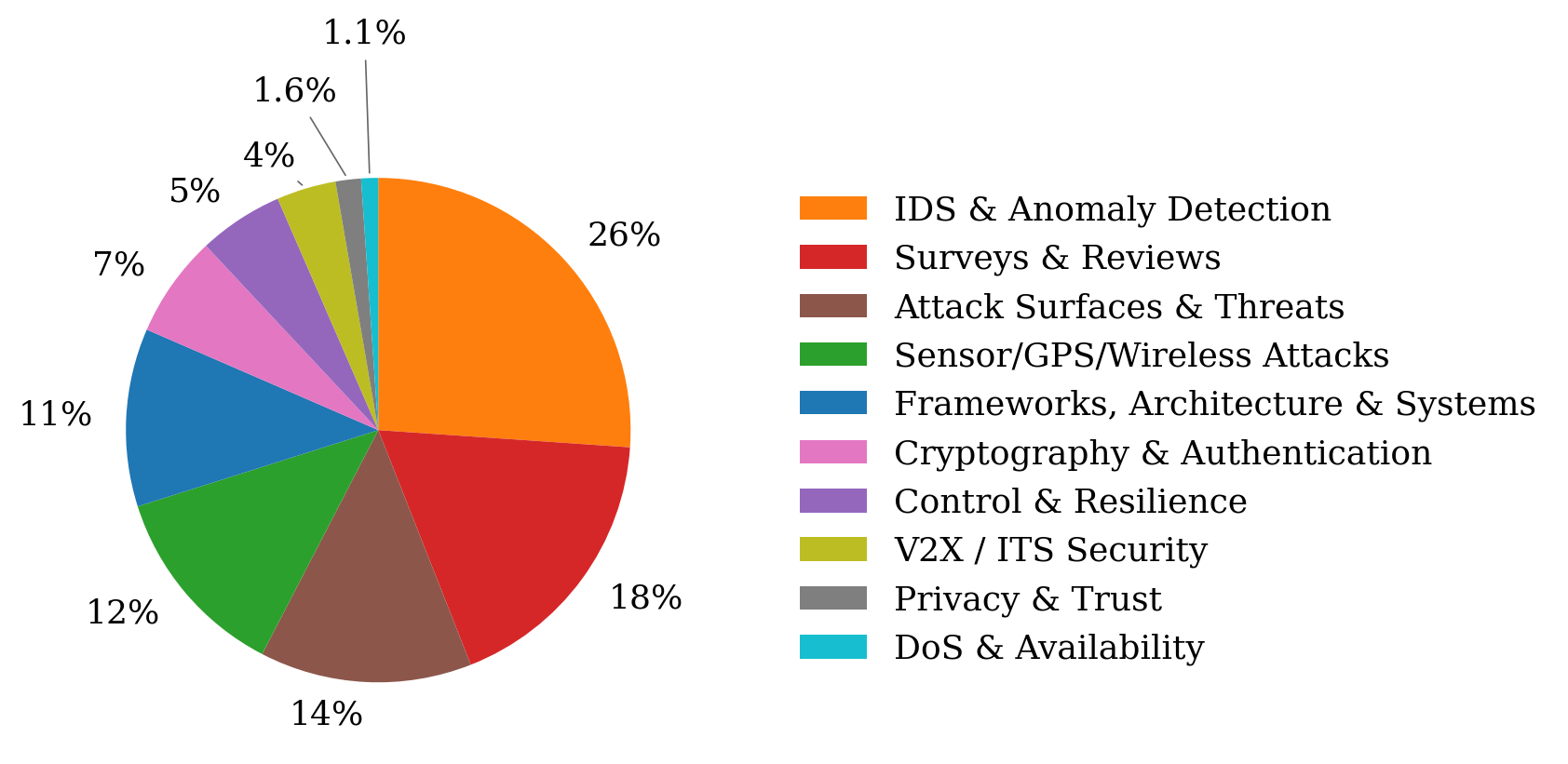}
        \caption{Distribution of research papers by thematic category}
    \end{subfigure}

    \caption{Overview of paper distribution across sources, years, and categories}
    \label{fig:surveyed papers}

\end{figure}

{
The key distinction is therefore not broader coverage of individual attack or defense categories, but the explicit linkage of entry point, attack methodology, operational impact, and mitigation within a single end-to-end analytical chain across heterogeneous in-vehicle technologies.
}

By linking attack surfaces, threat classes, subsystem impact, and corresponding countermeasures across %
 IVNs, this survey formalizes the complete security lifecycle.

To assess coverage across prior surveys, we evaluated each work %
against six dimensions as illustrated in Table~\ref{table:related-works}: (1)~\textbf{Entry Points} whether the survey catalogs external interfaces or physical access paths as attack origins; (2)~\textbf{%
 Attack Methodologies} whether it characterizes the techniques used to exploit those entry points; (3)~\textbf{%
 Impacts} whether it analyzes consequences at the ECU, vehicle, or fleet level; (4)~\textbf{Mitigations} whether it evaluates defensive countermeasures against the attacks it covers; (5)~\textbf{Comprehensive Coverage} whether the survey spans multiple in-vehicle technologies and communication layers rather than a single protocol or subsystem; and (6)~\textbf{Future Work} whether it identifies open problems or research directions. A dimension is marked present~(\checkmark) only if it receives dedicated treatment, defined as a named section, taxonomy, or comparative analysis, not incidental mention.

\subsection{Methodology of the Survey on End-to-End Vehicular Security}\label{sec:2.1-methodology}

In this section, we describe the methodology adopted to conduct the survey on security in vehicular networks. We also present the criteria used to classify and evaluate the existing literature. Figure \ref{fig:surveyed papers} summarizes the surveyed papers.
Our objective was to ensure consistency, traceability, and objective comparison across prior works. The methodology consists of the following steps:

\begin{itemize}
    \item \textbf{Targeted Topic:} The targeted topic was ``Security in Vehicular Networks,'' with emphasis on %
 IVNs, vehicular communication systems, attack modeling, system impact, and mitigation mechanisms.

    \item %
 \textbf{Literature Sources and Manual Search:} %

    {
    We first conducted manual literature searches using Google Scholar, IEEE Xplore, ACM Digital Library, ScienceDirect, SpringerLink, and Wiley Online Library. Searches were iterative and used combinations of terms related to vehicular cybersecurity, in-vehicle networks, attacks, threat modeling, intrusion detection, mitigation, and related topics.
    }

    \item %
 \textbf{Representative Search Queries:} %

    {
    Table~\ref{table:search_queries} lists representative query combinations.
    }

    {
    \begin{table}[!t]
    \centering
    
    \caption{Representative search queries used during the iterative manual literature searches.}
    \label{table:search_queries}
    \small
    \renewcommand{\arraystretch}{1.15}
    \begin{tabularx}{\columnwidth}{|c|X|}
    \hline
    \textbf{\#} & \textbf{Representative Search Query} \\
    \hline
    1 & ``vehicular cybersecurity'' AND attack \\
    \hline
    2 & ``vehicular security'' AND threat \\
    \hline
    3 & ``vehicular security'' AND taxonomy \\
    \hline
    4 & ``in-vehicle network security'' AND attack \\
    \hline
    5 & ``CAN bus attack'' AND security \\
    \hline
    6 & ``CAN security'' AND intrusion detection \\
    \hline
    7 & ``vehicular intrusion detection'' AND mitigation \\
    \hline
    8 & ``vehicular intrusion detection'' AND machine learning \\
    \hline
    9 & ``V2X security'' AND attack \\
    \hline
    10 & ``V2X security'' AND threat \\
    \hline
    11 & ``automotive cybersecurity'' AND mitigation \\
    \hline
    12 & ``in-vehicle network security'' AND defense \\
    \hline
    \end{tabularx}
    \end{table}
    }

    These queries are representative of the combinations used during iterative manual searches and are not intended to reconstruct every historical query verbatim.

    \item \textbf{Citation-Network Expansion:} Papers identified through the manual searches were then used as seed papers in Litmaps. Litmaps was used to expand the literature set by identifying related and connected papers, including citing and cited papers, through citation relationships and recommendations. This expanded the manually identified set and did not replace the manual searches; Litmaps did not make inclusion decisions.

    \item %
 \textbf{Screening and Selection:} %
    {
    Candidate papers identified through the manual searches and Litmaps were reviewed manually by the authors. The abstracts were read and discussed as a group to determine whether each paper was relevant to the scope of the survey, and inclusion decisions were made through this discussion.
    }

    \item \textbf{Selection Criteria:} %

    {
    Papers were retained when they contributed to one or more of the cybersecurity dimensions examined in this survey, including attack surfaces, attack methodologies, operational or safety impacts, mitigation mechanisms, intrusion detection, threat modeling, cybersecurity taxonomies, or relevant vehicular-security architectures and protocols. Studies focused solely on mechanical reliability or non-security networking performance were excluded. Implementation-specific studies were retained when they provided evidence relevant to the cybersecurity dimensions examined in the survey.
    }

    \item \textbf{Collected Literature and Updates:} Following the initial literature discovery, abstract review, and group discussion, 109 papers were initially confirmed for inclusion in the survey. Additional papers were incorporated during subsequent manuscript revisions and literature updates to reflect newer developments and address gaps identified during review. The final manuscript cites %
 202 references, of which 184 are the scholarly papers surveyed here and summarized in Figure~\ref{fig:surveyed papers}; the remaining %
 18 entries are standards, regulations, industry reports, web resources, a book, a technical report, and a white paper. Each of the 184 papers was assigned to a single thematic category (Figure~\ref{fig:surveyed papers}c) according to its primary focus. The surveyed papers appeared through 2025; because of the lag between research, publication, and indexing, the most recent years, particularly 2025, are likely under-represented in Figure~\ref{fig:surveyed papers}a. The main literature collection was conducted from June 28, 2023 to October 28, 2023. Additional papers were incorporated through August 8, 2024, with further literature added during manuscript revision and updating. The retained literature was then used to develop the survey taxonomy, compare prior work, analyze attack entry points, methodologies, impacts, and mitigations, and identify research gaps and future directions.

    \item \textbf{Tracking and Organization:} Google Sheets was used to track candidate papers and inclusion decisions, and Google Drive was used to store and organize the reviewed papers. Table~\ref{table:literature_tools} summarizes the tools used.

    \item \textbf{Evaluation Framework:} Each retained paper was evaluated using a structured rubric consisting of {six} dimensions: (i) treatment of attack surfaces, (ii) structured classification of attack %
 methodologies, (iii) explicit discussion of operational or safety impact, (iv) systematic analysis of mitigation strategies, {(v) coverage across heterogeneous in-vehicle technologies and communication layers, and (vi)} identification of research challenges or future directions.
\end{itemize}

This framework enables an objective comparison of prior surveys and supports the identification of structural gaps in existing literature.

{
\begin{table}[!t]
\centering
\caption{Tools used during literature discovery and screening.}
\label{table:literature_tools}
\small
\renewcommand{\arraystretch}{1.15}

\begin{tabularx}{\columnwidth}{|p{2.0cm}|p{2.5cm}|X|}
\hline
\textbf{Tool} &
\textbf{Purpose} &
\textbf{Use} \\
\hline
Google Scholar &
Manual literature search &
Manual iterative searches and identification of relevant papers. \\
\hline
Publisher databases &
Manual literature search &
IEEE Xplore, ACM Digital Library, ScienceDirect, SpringerLink, and Wiley Online Library. \\
\hline
Litmaps &
Citation-network expansion &
Manually identified papers used as seeds to find related and connected papers through citation relationships and recommendations. \\
\hline
Google Sheets &
Screening/tracking &
Tracking candidate papers and inclusion/exclusion decisions. \\
\hline
Google Drive &
Document management &
Storage and organization of papers reviewed for the survey. \\
\hline
\end{tabularx}
\end{table}
}

\textbf{Research Questions (RQ):}  
For the retained papers, the following research questions were considered:

\begin{itemize}
    \item \textbf{RQ1:} How has research on vehicular security evolved over time?
    \item \textbf{RQ2:} Which aspects of vehicular security (attack surfaces, attack %
 methodologies, impact analysis, mitigation mechanisms) receive the most attention in existing surveys?
    \item \textbf{RQ3:} To what extent do prior surveys provide a structured, system-level treatment rather than isolated analysis of specific components?
\end{itemize}

In response to RQ1, we analyze the temporal distribution of publications to observe trends in vehicular security research. %
 Publication counts stay at five or fewer per year from 2005 through 2015, rise to between 5 and 10 per year during 2016--2020, and increase sharply to 30 in 2021, the peak year, remaining high at 28 in both 2022 and 2023. The lower counts for 2024 (24) and 2025 (15) reflect the collection timeline: the main literature collection ran from June 28, 2023 to October 28, 2023, additional papers were incorporated through August 8, 2024, and later papers were added selectively before submission and during revision. These recent years are therefore under-sampled relative to earlier years, and the drop does not indicate declining research activity.

{
A parallel trend is the increasing use of automotive cybersecurity testbeds and simulation environments. These platforms have evolved from hybrid network simulation \cite{tuohy2016hybrid}, through configurable white-box ECU and real-data CAN testbeds \cite{toyama2018pasta,shi2022cantestbed}, to COTS-ECU hardware platforms \cite{yeo2023vitrobench} and remotely accessible laboratory environments \cite{zachos2024vce} for repeatable attack evaluation and cybersecurity training. We examine this trend and the remaining experimental gaps in Section~\ref{sec:testbeds}.
}

In response to RQ2, we categorize the collected papers according to their primary focus areas and evaluate the balance between attack classification, impact analysis, and mitigation strategies. This analysis highlights that many surveys emphasize either protocol vulnerabilities or defensive mechanisms, while fewer adopt an integrated perspective.

In response to RQ3, we assess whether existing works provide a structured mapping across security dimensions. This evaluation forms the basis for identifying gaps in the literature and motivates the need for a comprehensive, end-to-end survey framework.

\subsubsection*{Taxonomy of the Reviewed Literature}

A high-level analysis of the collected literature allows classification along structured security dimensions. The reviewed works are grouped according to their primary emphasis, including: (i) attack-surface characterization, (ii) attack-method taxonomy, (iii) impact and consequence analysis, (iv) mitigation and defense mechanisms, and (v) taxonomy or framework-oriented studies.

Within each group, subcategories are identified based on protocol focus (e.g., CAN, Automotive Ethernet, V2X), methodological approach (e.g., intrusion detection, cryptographic protection, machine learning), and system scope (e.g., IVN, ITS, autonomous systems). This structured categorization supports objective comparison and highlights gaps addressed by this survey.

\subsection{Attack Modeling and Simulation}
\label{sec:attack-modeling-simulation}
{
Attack modeling provides a structured representation of how an adversary
progresses from initial access to system compromise. Attack trees decompose an
attacker objective into the actions required to achieve it, while attack graphs
represent possible attack paths and dependencies across system components.
Attack-tree methods have been applied to automotive cybersecurity risk
assessment and attack-path analysis \cite{wang2021systematicrisk}. The Meta
Attack Language (MAL) provides a formal framework for defining security-oriented
domain-specific modeling languages and supporting attack-graph analysis
\cite{widel2023mal}. VehicleLang applies MAL to vehicle IT infrastructures and
supports modeling and simulation of automotive attack paths
\cite{katsikeas2022vehiclelang}.

SARA provides an automotive application of attack-tree-based reasoning by combining a threat model, an attack-method/asset mapping, explicit attacker involvement in attack trees, and a risk-assessment procedure \cite{monteuuis2018sara}.

Simulation and testbed environments complement these models by allowing attack
scenarios to be evaluated under controlled conditions. Existing platforms
include CAN-based security testbeds using real vehicle data
\cite{shi2022cantestbed} and VitroBench, which uses commercial automotive ECUs
to support controlled manipulation of in-vehicle communication
\cite{yeo2023vitrobench}. Simulation has also been used to examine how
cyberattacks against connected vehicles can propagate to transportation-level
effects \cite{jayaratne2024simulation}.

These approaches can also provide a mechanism for evaluating the taxonomy
developed in this survey. An attack model can instantiate an entry point,
attack methodology, resulting impact, and corresponding mitigation. Simulation
or testbed execution can then be used to evaluate the modeled progression under
specified system and attack conditions. Explicit specification of the system,
attack path, and experimental conditions also enables the same scenario to be
reproduced across studies.

This survey does not perform simulation-based validation of the proposed
taxonomy. Rather, the taxonomy provides a common structure that can be combined
with attack trees, attack graphs, MAL-based models, and simulation or testbed
environments to construct and evaluate new attack scenarios. Such an approach
can support reproducible evaluation of end-to-end attack paths across different
vehicle architectures and experimental environments.}

\subsection{Survey Structure}\label{sec:2.3-survey-structure}

The remainder of this survey is organized as follows. Section~\ref{sec:3-attacks} presents the end-to-end taxonomy used throughout the paper. Section~\ref{sec:3.1-entry-points} examines attack entry points, Section~\ref{sec:3.2-attack-methodologies} analyzes attack methodologies, Section~\ref{sec:3.3-attack-impacts} examines their operational, safety, availability, and privacy impacts, and Section~\ref{sec:3.4-attack-mitigations} reviews detection and mitigation mechanisms. Section~\ref{sec:4-research-gaps} identifies trends, research gaps, experimental challenges, and future research directions. Finally, Section~\ref{sec:6-conclusions} concludes the survey.
\section{\texorpdfstring{%
 End-to-End Taxonomy: Entry Points, Attack Methodologies, Impacts, and Mitigations}{End-to-End Taxonomy: Entry Points, Attack Methodologies, Impacts, and Mitigations}}\label{sec:3-attacks}

This section outlines the progression of vehicle cyberattacks and their key stages. First, we examine the entry points that allow attackers to gain access. These vulnerabilities serve as the foundation for later actions. Next, we explore the attack methodologies used to exploit the system. This section discusses how attackers move from access to control. We then analyze the impacts of these attacks. This section helps illustrate the real-world consequences of such attacks. Finally, we review the solutions proposed by the community. Each solution addresses specific vulnerabilities and mitigates potential threats. This section helps us understand where most research has been focused and paves the path for the follow-up section, which identifies gaps that require further investigation.

{
Concretely, entry points denote where access is obtained or an untrusted input crosses a trust boundary; attack methodologies denote the adversarial action performed once access is obtained; impacts denote the resulting operational, safety, availability, or privacy consequence; and mitigations denote mechanisms that prevent, detect, contain, or recover from the attack.
}

{
The four-stage taxonomy used in this survey is a lifecycle organization and is complementary to established threat classifications rather than a replacement for them. STRIDE classifies general threat types \cite{howard2006sdl}, while the Automotive Threat Matrix provides automotive-specific adversary tactics and techniques \cite{autoisac2024atm}. Automotive mappings such as \cite{sommer2024combining} further relate vehicle threats to CAPEC, CWE, and observed attack evidence, while \cite{info10040148} provides a development-oriented classification of automotive attacks. These representations can characterize the technique used within the attack-methodology stage of our framework. The additional entry-point, operational-impact, and mitigation stages preserve the surrounding context required to trace how an attack reaches the vehicle, how it is executed, what consequence it produces, and where it can be detected or mitigated.
}

{Four} representative attack chains illustrate the end-to-end framework developed in this survey.
The Jeep Cherokee attack \cite{miller2015remote} entered through a cellular interface, pivoted through the infotainment
system to inject CAN messages, and achieved  remote transmission disablement at highway speed and brake disablement at low speed. The Tesla
Free-Fall attack \cite{niefreefall2017} entered through a %
 Wi-Fi interface, exploited a browser vulnerability to reach
the CAN bus, and enabled persistent ECU compromise. A GPS spoofing attack \cite{ying2023gps} manipulates
GNSS signals to corrupt  localization used by connected-vehicle safety applications such as intersection movement assist without any physical access.
{
The Kia attack entered through connected-vehicle backend APIs and used authorization weaknesses to obtain remote vehicle access \cite{curry2024kia}.
}
We trace each
attack through entry points, attack methodologies, operational impacts, and mitigations in the
sections that follow.

{
Table~\ref{table:representative-attack-chains} summarizes the four examples using the same lifecycle stages applied throughout the survey.
}

\begin{table*}[!t]
\centering
\caption{Representative attack chains mapped to the four-stage taxonomy.}
\label{table:representative-attack-chains}
\small
\renewcommand{\arraystretch}{1.15}

\begin{tabularx}{\textwidth}{|p{2.3cm}|p{2.7cm}|X|X|X|}
\hline

\textbf{Attack} &
\textbf{Entry Point / Surface} &
\textbf{Attack Methodology / Path} &
\textbf{Operational Impact} &
\textbf{Mitigation / Security Lesson} \\
\hline

Jeep Cherokee \cite{miller2015remote} &
Cellular telematics; Internet-facing infotainment/TCU &
Remote service exploit $\rightarrow$ infotainment/TCU pivot $\rightarrow$ CAN injection &
Transmission and brake control; transmission disablement at highway speed, brake disablement at low speed &
OTA patch; no real-time gateway containment at disclosure \\
\hline

Tesla Free-Fall \cite{niefreefall2017} &
 Wi-Fi-exposed browser in the infotainment system &
Browser exploit $\rightarrow$ firmware compromise $\rightarrow$ CAN/ECU access &
Persistent ECU compromise; safety-critical cross-domain impact &
Secure boot and code signing to prevent firmware persistence \\
\hline

GPS spoofing \cite{ying2023gps} &
GNSS receiver; external RF signal &
GNSS spoofing $\rightarrow$ false position/time $\rightarrow$ corrupted localization input &
False or suppressed safety warnings (IMA) &
Trajectory-based anomaly detection \cite{ying2023gps}; GNSS authentication \\
\hline

2024 Kia backend/API compromise \cite{curry2024kia} &
Cloud-facing connected-vehicle backend API &
Authorization flaw $\rightarrow$ account/vehicle access $\rightarrow$ remote commands &
Privacy exposure and unauthorized remote vehicle control &
Backend authorization, ownership validation, and API access control \\
\hline

\end{tabularx}
\end{table*}

{
The modeling and simulation approaches discussed in
Section~\ref{sec:attack-modeling-simulation} provide a mechanism for
instantiating the attack progression represented by this taxonomy. Entry
points define initial access, attack methodologies capture subsequent
adversarial actions, impacts describe the resulting system state, and
mitigations identify mechanisms that %
 {prevent, detect, contain, or recover from the attack}. These
elements can therefore be used to construct attack paths for subsequent
simulation or testbed evaluation.}

{
\begin{table}[!t]
\centering
\caption{Canonical taxonomy and cross-cutting themes used throughout this survey.}
\label{table:canonical-taxonomy}
\small
\renewcommand{\arraystretch}{1.15}
\begin{tabularx}{\columnwidth}{|p{1.8cm}|p{2.6cm}|X|p{0.9cm}|}
\hline
\textbf{Stage} & \textbf{Classification Rule} & \textbf{Categories} & \textbf{\S} \\
\hline
Entry point & Layer at which untrusted input crosses the trust boundary & Physical (ports, sensors/GNSS, EV charging); Data-link (CAN/CAN FD, LIN, FlexRay, Automotive Ethernet); Application (Bluetooth, Wi-Fi, keyless, cellular, V2X, OTA, apps, cloud/API) & 3.1 \\
\hline
Attack methodology & Primary adversarial action performed after access & Manipulation (signal, message, firmware, and control-logic levels); Monitoring and Reconnaissance; Disruption & 3.2 \\
\hline
Impact & Primary consequence produced & Mechanical and Electrical Anomalies; False or Misleading Information; Information Leakage and Follow-On Attacks & 3.3 \\
\hline
Mitigation & Stage addressed, combined with defensive role & At entry points; against methodologies; impact containment and recovery --- each expressed through the roles prevent / detect / contain / recover & 3.4 \\
\hline
\multicolumn{4}{|l|}{\textbf{Cross-cutting themes (not additional taxonomy stages)}} \\
\hline
 Cross-cutting impact properties & Attributes of an impact, not a fifth impact category & Availability loss; safety degradation without explicit failure; cascading cross-domain effects; driver trust erosion; ML/probabilistic failures & 3.3.4 \\
\hline
Research areas & Trends that shape how the four stages are studied and addressed, not a fifth lifecycle stage & Architectural evolution (Automotive Ethernet, zonal/domain E/E architectures); cryptographic foundations (including post-quantum cryptography); regulation and compliance; experimental/testbed evaluation & 4 \\
\hline
\end{tabularx}
\end{table}
}

{
Table~\ref{table:canonical-taxonomy} summarizes the taxonomy and states the classification rule used for each stage. Within the scope of this survey, the four stages provide complete lifecycle coverage: an attack obtains or exploits an entry point, performs an adversarial action, produces an impact, and may be addressed by preventive, detective, containment, or recovery mechanisms. The within-stage categories are organizing classes for the technologies and attacks considered here rather than a claim of a universal cybersecurity ontology; they can be extended as new interfaces, techniques, and architectures emerge. Cross-cutting themes such as architectural evolution, cryptography, regulation, and experimental evaluation are recorded as attributes of an attack chain or as research areas addressed in Section~\ref{sec:4-research-gaps} rather than as additional taxonomy stages.
}

\subsection{Entry Points}\label{sec:3.1-entry-points}

Attack entry points into modern vehicles have expanded with connectivity, ECU consolidation, and feature growth. Adversaries exploit this expansion because many interfaces terminate untrusted inputs and then bridge them into trusted in-vehicle domains. The Jeep Cherokee case exemplifies this progression. In that attack, adversaries reached an Internet-facing infotainment/telematics endpoint over cellular connectivity and then pivoted into the IVN, where legacy bus semantics (e.g., CAN without intrinsic origin authentication) enabled actuator-relevant message injection %
\cite{miller2015remote}. A demarcation failure is therefore exposed: when external protocol endpoints share execution context or routing privilege with safety-relevant buses, remote compromise becomes structurally feasible.
\begin{figure}[t]
    \centering
    \includegraphics[width=0.75\linewidth]{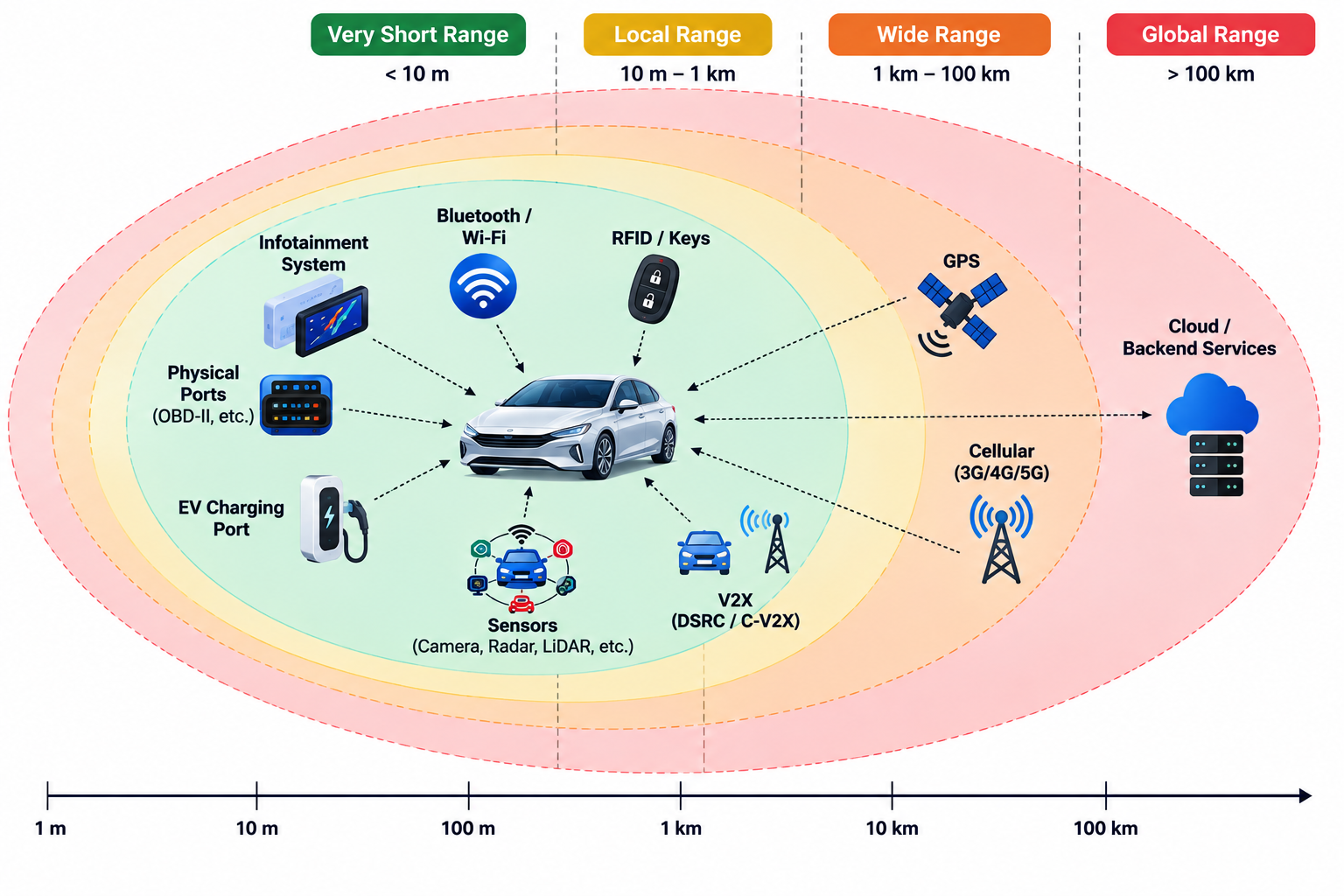}
    \caption{%
Vulnerable ranges for vehicle entry points, including V2X communication.}
    \label{fig:range}
\end{figure}

We describe entry points as trust-boundary crossings in which an untrusted external signal (physical device, RF transmission, sensor stimulus, or cloud-delivered artifact) becomes a trusted internal event (diagnostic session, paired endpoint, fused perception state, or executable update). Figure \ref{fig:range} shows the proximity needed for these attacks. %
 {In practice, an entry point may lead to foothold consolidation in a reachable platform (e.g., infotainment/Telematics Control Unit (TCU)), followed by execution of manipulation, monitoring, or disruption that affects vehicle functions or IVN availability.}

To align with IVN-layer reasoning and to support a compact taxonomy table, we group entry points by (i) physical layer, (ii) data-link layer, and (iii) application layer. For each group, we identify protocols/standards, the gateway component that mediates entry, the trust boundary, and the primary architectural gap. Table \ref{table:entry-points-layered} summarizes these entry points by layer.

\subsubsection{Entry Points at the Physical Layer}\label{sec:3.1.1-physical}
Physical-layer entry points are dominated by (a) wired ports that admit external devices and (b) sensors that convert external stimuli into digital measurements. In both cases, the boundary crossing is immediate: an untrusted artifact or stimulus is accepted by an internal endpoint and then propagated into trusted compute or control paths.

\paragraph{Wired Service and Peripheral Ports (OBD-II, USB, Removable Media)}
\textit{OBD-II:} The OBD-II connector (SAE J1962) provides physical access for maintenance. At the protocol level, diagnostic exchanges commonly use Unified Diagnostic Services (UDS, ISO 14229) over ISO 15765-4 on CAN/CAN FD (ISO 11898). The trust boundary is crossed when an external tester becomes a recognized diagnostic peer, and gateway routing enables access to multiple ECUs. Gaps include permissive diagnostic routing, insufficient authorization of security-critical UDS services (e.g., SecurityAccess and routine control), and the absence of message authentication on CAN frames, which collectively enable injection or reprogramming once physical access is obtained. The operational constraint is proximity; however, prior work anticipates that diagnostic access may become partially wireless, thereby externalizing this boundary and reducing attacker cost \cite{5940525}.

\textit{USB and removable media:} USB ports terminate complex host stacks (USB 2.0/3.x enumeration, mass storage class, and higher-layer parsers) within infotainment/telematics platforms. After enumeration, externally supplied content (media, metadata, application components) enters trusted parsing and service logic. Security gaps concentrate in stack complexity: memory-safety defects, insufficient device-class allowlisting, and vulnerable parsers can yield code execution \cite{checkoway2011comprehensive, 10.1145/3431233}, after which the head unit or TCU can mediate lateral movement toward IVN domains \cite{8370879, 6945510}. There have also been cases where automakers have left unused USB ports, and attackers found them on PC boards \cite{chevybolt_usbhub}.

Removable media update paths expose similar attack surfaces. ISO 9660 CD handling can be abused to deliver crafted payloads into head-unit components that parse disc content \cite{checkoway2011comprehensive}. Unauthorized firmware injection via malicious media becomes feasible when provenance checks, secure boot, and rollback protection are incomplete, thereby translating untrusted media into trusted executable code \cite{6615528}. FM and satellite radio interfaces also form part of the broader connected infotainment and telematics attack surface \cite{LE201817,9316663}.

\paragraph{EV Charging as a Physical-Cyber Interface}
Electric vehicle (EV) charging introduces a physical coupling that may also carry data. Standards such as ISO 15118 (vehicle-to-grid communication) and IEC 61851 signaling can participate in charge negotiation and service orchestration \cite{hu2018review}. The trust boundary is crossed when Electric Vehicle Supply Equipment (EVSE)-side messages and cable-side electronics interact with the vehicle charging ECU, which can then interact with broader IVN functions over internal networks. Prior work reports key-related data skimming and ECU reprogramming through charging-associated pathways \cite{6513548}. Gaps include insufficient mutual authentication and authorization between EV and EVSE, weak compartmentalization of charging subsystems, and unsafe exposure of programming/diagnostic capabilities through charging-related components \cite{sturgess2023cableauth}.

{
Electrification further extends the attack surface beyond the vehicle to charging infrastructure and backend management systems. Protocols such as CHAdeMO and ISO 15118 support communication between the EV and EVSE, while the Open Charge Point Protocol (OCPP) connects charging stations with charging station management systems \cite{chademo2020ocpp,garofalaki2022ocpp}. Vulnerabilities in these interfaces can therefore expose both vehicle-facing and infrastructure-facing communication. Prior work has demonstrated injection, man-in-the-middle, denial-of-service, and code-injection attacks against EV charging communication and OCPP implementations \cite{poudel2023injection,garofalaki2022ocpp}. Compromised charging infrastructure can also affect the distribution grid when attacks alter the operation of multiple charging stations \cite{mokarim2023evcs}.
}

\paragraph{Perception and Sensing Surfaces (LiDAR, Radar, Cameras, GNSS, Tire Pressure Monitoring System (TPMS), Ultrasonic, Microphones)}
\textit{LiDAR:} LiDAR feeds point-cloud measurements into perception stacks, a sensing surface addressed in broader autonomous-vehicle security work \cite{KIM2021102150, 10.1145/3431233}. An adversary can inject or distort optical returns through physical or fusion-level manipulation \cite{Cao2021InvisibleFB, 279980, 10.1145/3460120.3485377}, including object-removal attacks on semantic perception \cite{hau2021object, Zhu2022DemoAL} and spoofing or appearing-object effects relevant to cooperative perception \cite{cao2021automated, zhang2023cooperative, 287252}, so that the pipeline accepts spurious objects, converting untrusted optical energy into trusted internal scene representations.

\textit{Radar:} Radar converts RF reflections into target tracks used by ADAS and therefore constitutes another externally influenced perception component within the broader attack surface of autonomous vehicles \cite{KIM2021102150}.

\textit{Cameras:} Camera pipelines convert optical stimuli into trusted detections, a perception surface treated broadly in autonomous-vehicle security work \cite{KIM2021102150, 10.1145/3431233}. Spoofing and adversarial patterns can induce misclassification and unsafe downstream decisions, including adversarial-pattern and physical-patch spoofing \cite{nassi2021spoofing, sato2021wip} and projected or infrared image spoofing \cite{10.1145/3460120.3484766, sato2023wip}.

\textit{Global Navigation Satellite System (GNSS)/GPS:} GNSS receivers convert civil RF signals into trusted position/time, a surface addressed in broader autonomous- and connected-vehicle security work \cite{KIM2021102150, 10.1145/3431233, 9316663}. Spoofing exploits the absence of intrinsic signal authentication to manipulate reported position \cite{ying2023gps, tang2021fooling}, while jamming denies the signal entirely \cite{abdallah2022not}, enabling external manipulation and disruption without compromising internal endpoints. {The GNSS receiver itself is the physical-layer entry point regardless of how it is attacked; jamming the signal is a Disruption methodology (Section~\ref{sec:3.2-attack-methodologies}), while injecting a spoofed position is a Manipulation methodology, so the entry point does not change with the attack technique used against it.}

\textit{TPMS and ultrasonic:} TPMS telemetry can be read and, in some demonstrations, leveraged for injection pathways \cite{checkoway2011comprehensive, rouf2010security}. TPMS therefore presents a short-range wireless attack surface in which spoofed or adversarial RF messages can influence received tire-pressure telemetry \cite{checkoway2011comprehensive,10.1145/3431233,9316663}. Ultrasonic sensing can be disrupted via interference when signal-processing and fusion checks are weak \cite{KIM2021102150, yan2016can}.

\textit{Microphones:}  Voice-assistant skill hijacking can trigger vehicle commands \cite{Ding2022DemoHC}.

\ifdefined\showcuts
\newcommand{\cutgpsrow}{%
 &
 &
 &
 &
 &
 &
 &
 &
 \\
\hline
}
\else
\newcommand{\cutgpsrow}{}
\fi
\begin{table*}[!htbp]
\centering
\caption{\textcolor{blue}{%
}Entry Points Aligned with Layer-Based Taxonomy}
\label{table:entry-points-layered}
\setlength{\tabcolsep}{3pt}
\adjustbox{max width=\textwidth, max totalheight=0.95\textheight}{%
\begin{tabular}{|>{\raggedright\arraybackslash\hspace{0pt}}p{2.20cm}|>{\raggedright\arraybackslash\hspace{0pt}}p{2.50cm}|>{\raggedright\arraybackslash\hspace{0pt}}p{2.20cm}|>{\raggedright\arraybackslash\hspace{0pt}}p{2.20cm}|>{\raggedright\arraybackslash\hspace{0pt}}p{2.80cm}|>{\raggedright\arraybackslash\hspace{0pt}}p{3.20cm}|>{\raggedright\arraybackslash\hspace{0pt}}p{3.20cm}|>{\raggedright\arraybackslash\hspace{0pt}}p{3.20cm}|>{\raggedright\arraybackslash\hspace{0pt}}p{3.80cm}|}
\hline
Layer & Interface & Access Condition & Attacker Proximity & Target Asset / Privilege & Reported Problems & Representative Contributions & Research Gaps & References \\
\hline

Physical Layer &
OBD-II Port &
Direct physical access to vehicle interior &
Physical &
Tx on CAN; ECU diagnostics access &
Unauthenticated diagnostic access; CAN injection; weak gateway isolation &
IVN compromise via diagnostics; secure gateway evaluation &
No standard secure diagnostics; inconsistent gateways &
\cite{5940525, 6513548, kang2021car, rajapaksha2023improving, SONG2020100198, checkoway2011comprehensive, wang2014vecure} \\
\hline

Physical Layer &
USB Ports &
Physical access; malicious media insertion &
Physical &
Infotainment code execution; IVN pivot &
Media parser exploits; lateral IVN pivot &
Code execution via USB; bridging risks analyzed &
Media validation limited; uneven isolation &
\cite{checkoway2011comprehensive, 10.1145/3431233, 9316663, wolf2007state} \\
\hline

Physical Layer &
Radio/\allowbreak{}CD Player &
Physical proximity; malicious broadcast or media &
Physical / Short-range &
Head-unit compromise; CAN bridge access &
Infotainment decoding vulnerabilities; CAN bridging &
Head-unit compromise demonstrated &
Unpatched legacy units; weak parser verification &
\cite{checkoway2011comprehensive, 6615528, LE201817, 5504804, DBLP:journals/corr/abs-1802-01725} \\
\hline

Physical Layer &
EV Charging Stations &
Physical charging interface; compromised infrastructure &
Physical &
Charging ECU access; firmware/\allowbreak{}channel manipulation &
Charging-channel abuse; malware via cable &
Charger authentication; cable-level security &
Mutual authentication not standardized; immature trust model &
\cite{6513548, sturgess2023cableauth, 8767141, basnet2021exploring} \\
\hline

Physical Layer &
LiDAR, Radar, Cameras, Microphones, TPMS, Ultrasonic Sensors &
Line-of-sight or RF/acoustic proximity &
Short-range &
Perception pipeline manipulation; sensor data injection &
Spoofed readings; false perception &
Laser, RF, and acoustic spoofing demonstrated &
Limited fusion resilience; no robustness benchmarks &
\cite{KIM2021102150, Cao2021InvisibleFB, 10.1145/3431233, el2020cybersecurity, yang2021secure, philipsen2021threats, 10.1145/3460120.3484766, nassi2021spoofing, tang2021fooling, hu2022pass, yan2016can, Ding2022DemoHC, 285475, 8688625, LE201817, kang2020attack} \\
\hline

 {Physical Layer} &
GPS &
Remote signal spoofing &
Remote &
Location/\allowbreak{}time manipulation &
Location spoofing; navigation manipulation; time desynchronization &
Demonstrated GPS spoofing impact; evaluated detection mechanisms &
Robust GNSS authentication not universally deployed; multi-sensor trust calibration limited &
\cite{ying2023gps, tang2021fooling, abdallah2022not} \\
\hline

Data-Link Layer &
CAN/\allowbreak{}CAN FD, LIN, FlexRay, Automotive Ethernet &
Bus attachment (ECU foothold, OBD-II, or compromised gateway) &
Physical / Internal &
Frame injection or spoofing on the IVN bus &
No sender authentication; broadcast arbitration; one foothold suffices &
Frame-level trust assumptions across bus technologies &
Authentication and segmentation not uniformly enforced &
\cite{10.1145/3431233, 7393668} \\
\hline

 {Application Layer} &
Keyless Entry (RFID) &
Short-range RF relay or amplification &
Short-range &
Unlock/\allowbreak{}lock control; authentication bypass &
Relay attacks; key cloning; unauthorized unlock &
RFID reverse engineering; relay attacks demonstrated &
Limited distance bounding; no anti-relay standard &
\cite{checkoway2011comprehensive, DBLP:journals/corr/abs-1802-01725, 9316663, 6513548, el2020cybersecurity, ahmad2020securing, deppenhanced} \\
\hline

 {Application Layer} &
Bluetooth &
Short-range wireless proximity &
Short-range &
Infotainment access; IVN pivot &
Weak pairing authentication; infotainment compromise; IVN pivot &
Analyzed pairing flaws; proposed strengthened authentication &
Unified security policy across OEMs lacking &
 \cite{9833886, KIM2021102150, 7393668, renganathan2022valet} \\
\hline

\cutgpsrow

 {Application Layer} &
Cellular Networks &
Remote Internet-based access &
Remote &
Telematics control; remote IVN access &
Telematics exploitation; remote IVN compromise &
Remote vehicle control demonstrated &
Isolation between telematics and safety domains inconsistent &
\cite{9564848, 6513548, 7393668, checkoway2011comprehensive, miller2015remote} \\
\hline

Application Layer &
V2X (DSRC/\allowbreak{}C-V2X) &
V2X messages from vehicles or roadside units &
Short-range to remote (via RSU) &
V2X unit / TCU; ADAS awareness inputs &
Forged or replayed safety messages; misbehaving authenticated senders &
V2X misbehavior detection surveyed &
Misbehavior detection at scale; IVN trust integration &
\cite{9463461, 9619465, vanderheijden2019misbehavior} \\
\hline

 {Application Layer} &
 Wi-Fi &
Short- to mid-range wireless access &
Short-range &
Infotainment compromise; IVN entry &
Insecure access points; remote code execution &
Validated %
 Wi-Fi-based IVN intrusion; evaluated IDS mechanisms &
Inconsistent secure configuration &
\cite{checkoway2011comprehensive, 10.1145/3431233, info10040148, 9316663, josephlal2019vulnerability, niefreefall2017} \\
\hline

Application Layer &
Automotive Applications &
Local or remote ap\-pli\-ca\-tion-layer interaction &
Remote / Local &
App-level privileges; API access; sandbox escape &
Privilege escalation; insecure APIs; insufficient sandboxing &
App ecosystems analyzed; runtime monitoring &
Little formal verification of third-party apps &
\cite{10.1145/3431233, 8370879, Ding2022DemoHC, lee2019enhanced} \\
\hline

Application Layer &
OTA (Over-the-Air) Updates &
Remote update channel; backend compromise required &
Remote &
Firmware write; OTA partition control &
Tampering; rollback; weak signature checks &
Formally verified OTA frameworks &
No standard secure update lifecycle &
\cite{8688625, seo2023formally, Mahmood2022, Seo2024} \\
\hline

Application Layer &
Cloud / Backend API &
Remote access to backend services &
Remote &
Account and vehicle commands via the backend &
Authorization and ownership-validation flaws &
Demonstrated remote vehicle access through backend APIs &
Backend authorization not systematically evaluated &
\cite{curry2024kia} \\
\hline

\end{tabular}
}
\end{table*}

\subsubsection{Entry Points at the Data-Link Layer}\label{sec:3.1.2-datalink}
Data-link layer entry points arise when adversaries obtain the ability to transmit or observe frames on in-vehicle buses. This category differs from physical and application layers because the bus architecture (especially legacy CAN) tends to remain stable across model years, while exposure pathways evolve. Once bus access is achieved, the absence of intrinsic sender authentication and fine-grained authorization enables injection, spoofing, sniffing, and disruption with comparatively low semantic overhead \cite{5504804}.

\paragraph{In-Vehicle Buses and Diagnostic Transport (CAN/CAN FD, LIN, FlexRay, Automotive Ethernet)}
Legacy CAN/CAN FD (ISO 11898) provides broadcast arbitration without message authentication. Local Interconnect Network (LIN) (ISO 17987) and FlexRay (ISO 17458) exhibit similar trust assumptions at the frame level, although deployment contexts differ. Automotive Ethernet (IEEE 802.3) typically carries IP and middleware such as SOME/IP (Scalable service-Oriented MiddlewarE over IP), DoIP (Diagnostics over Internet Protocol, ISO 13400), and service discovery \cite{10.1145/3431233}. The trust boundary is crossed when an adversary gains any physical or logical attachment to a bus segment (e.g., via an ECU foothold, OBD-II access, or compromised gateway), thereby converting an untrusted node into a frame-originating peer. The security gap is architectural: frame-level authenticity and authorization are not intrinsic to several widely deployed IVN technologies, so a single foothold can enable broad manipulation and disruption unless segmentation and message-level security are enforced \cite{7393668}.

\subsubsection{Entry Points at the Application Layer}\label{sec:3.1.3-LABEL}
Application-layer entry points originate in external networks and software interfaces that terminate high-complexity stacks and then orchestrate translation into IVN actions. The trust boundary is crossed when externally sourced traffic or artifacts are accepted by privileged services on infotainment/TCU platforms, which then act as gateways into internal buses.

\paragraph{Short-Range Wireless Services (Bluetooth, Wi-Fi, Keyless RF Links)}
Bluetooth Classic and Bluetooth Low Energy (BLE) expose profile stacks (e.g., HFP/A2DP/AVRCP, PBAP/MAP, BLE GATT). Trust is granted after pairing and service authorization, which can be undermined by impersonation and negotiation weaknesses (e.g., BIAS+KNOB) or by implementation vulnerabilities %
 \cite{9833886}. The gap is twofold: (i) protocol and policy weaknesses in association and profile authorization \cite{checkoway2011comprehensive, 6513548, 10.1145/3431233} and (ii) memory-safety exposure in complex stacks \cite{info10040148, 9316663}, which can yield code execution and subsequent IVN pivoting via infotainment gateways \cite{8370879, 6945510}.

Wi-Fi integrates IEEE 802.11 connectivity with WPA2/WPA3 and IP services \cite{10.1145/3431233, info10040148, 9316663}. The boundary crossing occurs when an adversary achieves association or reaches exposed IP services, and those services execute within a privileged head-unit/TCU context that can route requests into internal services \cite{6945510}. Gaps include misconfiguration, exposed management endpoints, unpatched stacks, and weak segmentation between Wi-Fi domains and vehicle-control domains \cite{8370879}.

Keyless entry uses short-range RF links (often proprietary LF/UHF designs; Radio-Frequency Identification (RFID)/Near Field Communication (NFC)-like mechanisms appear in some implementations). The trust boundary is crossed when RF exchanges are accepted as authorization events and translated into internal commands. Relay and replay attacks exploit weak distance bounding and weak binding to proximity \cite{checkoway2011comprehensive, 9316663, 6513548}.

\paragraph{Long-Range Telematics and Internet-Facing Services (Cellular)}
Cellular connectivity (2G--5G) enables remote services and backhaul. Vehicle risk concentrates at the endpoint, where Internet-originated traffic reaches the TCU/head unit and then can traverse into IVN domains. The Jeep exploit operationalizes this pathway by leveraging a vulnerable externally reachable service and then crossing the gateway boundary into CAN \cite{miller2015remote}. Gaps include exposed service attack surfaces, weak remote-service authentication/authorization, and implementation vulnerabilities in large network-facing stacks \cite{6513548, LE201817, 10.1145/3431233}, combined with inadequate isolation between telematics and safety-critical domains \cite{9316663, 8370879, 6945510}.

\paragraph{V2X Communications}
V2X interfaces accept external messages that can influence cooperative behaviors. Standards include IEEE 802.11p/IEEE 1609.x (Wireless Access in Vehicular Environments (WAVE)/Dedicated Short-Range Communications (DSRC)), ETSI ITS-G5, and 3GPP C-V2X (PC5 sidelink) \cite{9463461}. The trust boundary is crossed when received messages become trusted inputs to fusion or planning. Gaps include implementation weaknesses in verification stacks, mismanaged credentialing, and denial-of-service susceptibility at RF and protocol-orchestration layers \cite{9619465}. Because authenticated participants may still transmit false information, V2X security also requires misbehavior detection based on message semantics and data consistency in addition to credential-based authentication \cite{vanderheijden2019misbehavior}.

\paragraph{OTA and Software Supply Chain}
OTA delivery uses IP connectivity (cellular/Wi-Fi) and should enforce signed manifests, secure boot, and anti-rollback protections \cite{LE201817}. The trust boundary is crossed when a remote artifact transitions from untrusted network data to trusted executable firmware on ECUs. Attacks target server compromise, key/certificate failures, downgrade pathways, or client validation defects \cite{seo2023formally, Mahmood2022}. OTA also enables rapid remediation after compromise \cite{8688625}. The architectural gap is that Internet-scale reachability becomes coupled to in-vehicle code deployment, which amplifies the consequences of update-trust failure.

\paragraph{Infotainment Apps, Projection, and Voice Services}
Infotainment application ecosystems and projection frameworks (e.g., Android Auto) introduce a software entry point in which third-party packages and runtime content become trusted executable logic within the head unit \cite{10.1145/3431233, 8370879}. Voice and intent interfaces cross the boundary when externally sourced prompts become authorized commands. \cite{Ding2022DemoHC} demonstrates  hijacking of connected-vehicle Alexa skills. Gaps include weak app vetting, overprivileged permissions, insecure IPC, and inadequate origin validation for intents, which can yield footholds that pivot into IVN services via gateway mediation.

\paragraph{Summary and Implications}
The {four} representative chains illustrate the structural asymmetry that defines the entry-point
landscape. The Jeep Cherokee chain \cite{miller2015remote} entered through a cellular telematics interface, the Tesla
Free-Fall attack \cite{niefreefall2017} through a %
 Wi-Fi-exposed browser in the infotainment system, and the GPS
spoofing chain \cite{ying2023gps} through a GNSS receiver.
{
The Kia attack adds a cloud-facing backend API as a fourth entry point \cite{curry2024kia}.
}
These are {four} distinct interfaces, each {crossing a different trust boundary on the path to vehicle functionality}.

Data-link layer weaknesses (e.g., missing intrinsic authentication/authorization on broadcast buses) tend to be stable and difficult to eliminate rapidly because they are embedded in legacy IVN design and backward compatibility constraints. In contrast, physical-layer and application-layer exposure is expanding because manufacturers continue to add ports, sensors, wireless services, cloud features, and update infrastructure, each of which terminates untrusted inputs within high-complexity stacks. As a result, the dominant risk driver is not only what the IVN permits once reached, but also how many pathways now exist to reach it at scale.

As automotive architectures gradually transition from CAN-dominated IVNs toward Automotive Ethernet, the nature of the attack surface is likely to evolve rather than contract. Legacy buses such as CAN expose well-understood data-link–layer weaknesses, including broadcast semantics and limited intrinsic authentication, but their simplicity constrains protocol complexity and interaction patterns. Automotive Ethernet, by contrast, introduces a multi-layer networking substrate that typically carries IP-based services, middleware, and update mechanisms atop IEEE 802.3 links. This shift may enable stronger security mechanisms in principle, such as fine-grained access control and cryptographic authentication at higher layers, yet it also increases software complexity and configuration dependence at gateways and endpoints. As a result, attack vectors may increasingly target network- and application-layer implementations, service orchestration logic, and misconfiguration, rather than low-level frame manipulation alone. From a security perspective, this suggests that while data-link–layer risks may remain relatively stable, physical and application-layer entry points are likely to expand in both number and diversity as Ethernet-based integration progresses, emphasizing the importance of architectural isolation and lifecycle-aware configuration management.

\subsection{Attack Methodologies}\label{sec:3.2-attack-methodologies}

After discussing how attackers gain access to vehicles, we now explore the various types of attacks that malicious actors can deploy. Note that many of these attacks map to multiple methodologies. Our taxonomy offers a generalized, high-level overview of each attack and its primary classification.

{
Manipulation alters signals, messages, software, or system state; monitoring and reconnaissance involve observing or inferring information without directly changing intended operation; and disruption degrades availability, delivery, accessibility, or timeliness. Attacks that span multiple behaviors are grouped below according to their primary adversarial action.
}

\subsubsection{Manipulation}\label{sec:3.2.1-manipulation}
\begin{figure}[t]
    \centering
    \includegraphics[width=0.6\textwidth,trim=180 20 180 20,clip]{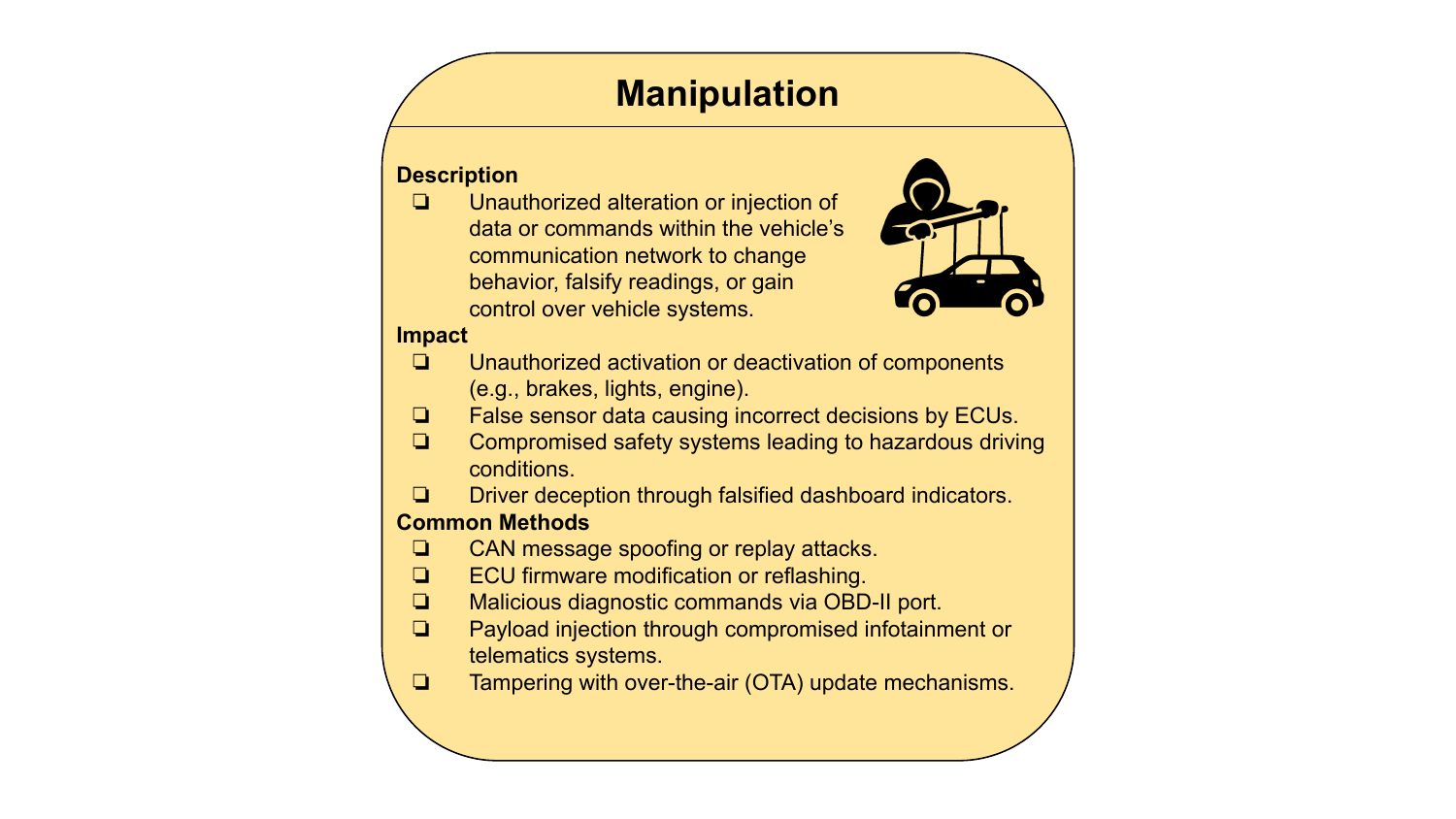}
   \caption{Representative summary of manipulation attacks in connected vehicles, highlighting common ways adversaries alter signals, messages, firmware, or control logic to induce unintended vehicle behavior. The examples shown are illustrative rather than exhaustive; Section~3.2.1 provides the complete discussion of manipulation categories and attack examples.}
    \label{fig:manipulation}
\end{figure}

Manipulation attacks alter signals, messages, code, or control decisions so that the vehicle executes unintended behavior. %
Many attacks span multiple layers, so we group them by the primary object being manipulated: (i) signals and measurements, (ii) messages and frames, (iii) firmware and execution, and (iv) control logic and semantics. Figure~\ref{fig:manipulation} summarizes manipulation attacks, in which adversaries alter data, signals, or control logic to induce unintended vehicle behavior.

\paragraph{%
 Signal- and Measurement-Level Manipulation}
This category targets analog properties, sensor measurements, or derived estimates that upstream components treat as trusted. Attackers can modify these values by direct physical interference, spoofed stimuli, or compromised sensing and acquisition paths. %

{
\textbf{Amplitude shift attacks:} Amplitude-shift attacks modify an in-vehicle signal by adding a positive or negative offset. The underlying temporal pattern of the signal can remain unchanged. Khan et al. model this attack on decoded CAN-derived vehicle signals \cite{9091063}. A compromised ECU changes a feature value while preserving its overall trend. This differs from direct physical-layer manipulation of CAN-wire voltage.
}

{
More advanced attacks vary the perturbation over time to evade detection. Jichici et al.\ define surge, bias, and geometric attacks on in-vehicle signals and show that bias and geometric variants can evade cumulative-sum detection on J1939 heavy-duty vehicle buses \cite{jichici2023control}. Berdich and Groza assess replay, DoS, and fuzzing attacks against ACC and AEB models \cite{berdich2023secure}. Surge attacks push a signal toward a damaging extreme while remaining within an admissible deviation. Bias attacks introduce a small persistent offset. Geometric attacks begin with a small perturbation and progressively increase its magnitude. %
}

{
\textbf{RPM and gear spoofing attacks:} With access to the relevant ECUs or network segments, an attacker injects falsified engine-speed or gear-position messages that other control units accept as legitimate, misleading engine and transmission control \cite{9706416}. Unlike amplitude-shift attacks, which perturb the magnitude of a genuine signal while preserving its trend, these attacks replace the reported state outright.
}

{
\textbf{Sensor spoofing attacks:} Perception and positioning sensors convert external stimuli into trusted measurements, so an attacker who controls the stimulus controls the measurement without compromising any internal software. Demonstrated examples include falsified TPMS pressure frames that exploit unauthenticated sensor broadcasts \cite{rouf2010security}; spoofed GNSS positions that corrupt localization \cite{ying2023gps,tang2021fooling}; timed laser injection that removes real obstacles from LiDAR point clouds \cite{hau2021object,285475}, adds spoofed obstacles \cite{279980,10.1145/3460120.3485377,cao2021automated,zhang2023cooperative,287252}, or distorts point-cloud geometry \cite{279980}, alongside attacks on LiDAR semantic segmentation and camera--LiDAR fusion \cite{Zhu2022DemoAL,279980,Cao2021InvisibleFB}; projected or injected infrared imagery that fabricates obstacles \cite{10.1145/3460120.3484766}; and magnetic interference with ABS wheel-speed sensors \cite{shoukry2013non}. Table~\ref{tab:sensor_attack_requirements} summarizes the equipment and proximity each attack requires.
}

\begin{table}[!htbp]
\centering
\caption{Minimum attacker requirements for sensor-based spoofing attacks.}
\label{tab:sensor_attack_requirements}
\resizebox{\columnwidth}{!}{%
\begin{tabular}{lllll}
\toprule
\textbf{Attack Type} & \textbf{Target Sensor} & \textbf{Equipment} & \textbf{Proximity} & \textbf{Success / Notes} \\
\midrule
TPMS spoofing & Tire pressure sensor & SDR + custom firmware & $<$40\,m & Demonstrated~\cite{rouf2010security} \\
GPS spoofing & GPS receiver &  Simulated attack model & Broadcast &  False IMA warnings (simulation)~\cite{ying2023gps} \\
LiDAR point removal & LiDAR (3D) & Pulsed IR laser + timing circuit & $<$20\,m, LOS &  92.7\% success removing 90\% of target points~\cite{285475} \\
IR image spoofing & Infrared camera &  Off-the-shelf IR light source &  Close range & Fake obstacles demonstrated~\cite{10.1145/3460120.3484766} \\
ABS wheel speed spoofing & Magnetic wheel sensor &  PCB-coil electromagnetic actuator + driver &  In sensor air gap (few mm) & Demonstrated~\cite{shoukry2013non} \\
\bottomrule
\end{tabular}}
\end{table}

\paragraph{%
 Message- and Frame-Level Manipulation}
This category targets the content, timing, origin, and delivery of protocol messages. The attacker’s goal is to inject, replay, impersonate, or otherwise control message flows so that receivers accept incorrect state or commands. %

\textbf{Replay attacks:} A replay attack involves repeating previously valid messages containing outdated data. In IVNs, this can occur when a compromised ECU transmits recorded data instead of current values, leading to repeated operations or pushing the system into unintended states \cite{9091063}.

{
\textbf{Fuzzing attacks:} Fuzzing attacks inject fabricated frames with random or semi-random identifiers and payload values. Random-ID fuzzing introduces identifiers that may not appear in legitimate traffic. Targeted variants instead use valid identifiers and randomize their payloads \cite{verma2024road,shi2022cantestbed}. %
In this context, a fuzzing attack differs from \emph{fuzz testing}. Fuzz testing systematically exercises software or protocols to discover implementation flaws. Fuzzing attacks are also widely used as baseline attack classes in automotive IDS datasets and testbeds \cite{verma2024road,shi2022cantestbed}.
}

\textbf{Message/frame injection attacks:} In CAN, any node can transmit frames and influence arbitration, which enables crafted injection attacks \cite{LE201817,7796898}. Similar injection behavior has been demonstrated on FlexRay \cite{hu2018review} and LIN (e.g., malicious sleep frames) \cite{5940525}. A single crafted CAN frame can also be accepted by one target ECU while other ECUs, including IDS monitors, do not register it \cite{yue2021cancloak}.

\textbf{ECU/node impersonation attacks:} In ECU impersonation, malware on a compromised segment sends spoofed messages claiming to originate from a target ECU. Downstream ECUs may act on falsified information, compromising integrity and safety \cite{272188,10.1145/3431233}.

\textbf{Masquerade attacks:} Masquerade attacks compromise at least two ECUs to manipulate behavior while hiding the compromise. Because many IVN messages are periodic and broadcast, identifiers and intervals are easy to learn, allowing malicious traffic to blend into normal operation for extended periods \cite{197171}.

\textbf{Session hijacking:} Session hijacking gains control over a valid communication session between ECUs or sensors, enabling an attacker to impersonate legitimate components and manipulate critical values (e.g., sensor readings used by the anti-lock braking system (ABS)). %

\textbf{IP spoofing attacks:} IP spoofing forges packet source addresses so that traffic appears to originate from trusted components \cite{9619465}. %

{
\textbf{Injection through trusted projection paths:} A compromised smartphone or malicious application can use a trusted infotainment projection path, such as MirrorLink, to inject messages into the IVN \cite{LE201817,KIM2021102150}; the entry path itself is discussed with application-layer entry points (Section~\ref{sec:3.1.3-LABEL}).
}

{
\textbf{Sybil and illusion attacks:} In cooperative V2X settings, Sybil attacks create multiple fake identities to inject spurious messages or dominate reputation systems \cite{9463461,9619465,9316663}, while illusion attacks manipulate a vehicle's own sensors so that it broadcasts falsified readings that other vehicles accept as trusted \cite{9619465}.
}

\paragraph{%
 Firmware- and Execution-Level Manipulation}
This category targets code and execution context. Attackers aim to introduce or run malicious logic (e.g., via media parsers, third-party apps, compromised update paths, or ECU reprogramming) so that a compromised platform can persist and pivot into other domains.

\textbf{Malware injection:} Malware can be injected by exploiting vulnerabilities in infotainment and communication interfaces (media players, wireless links, diagnostic ports). Crafted content (e.g., malicious media files) can trigger execution and enable follow-on injection into internal networks.  Checkoway et al. \cite{checkoway2011comprehensive} demonstrated multiple such attack paths, emphasizing how increased connectivity expands the feasibility and impact of malware-based attacks.

\textbf{Trojan horse applications:} With general-purpose operating systems and app ecosystems in modern vehicles, malicious or trojanized apps can execute within the vehicle’s security context. App-based footholds can compromise host components and pivot into connected buses \cite{checkoway2011comprehensive}.

\paragraph{%
 Control-Logic and Semantic Manipulation}
This category targets the meaning of actions rather than raw signals or packets. Attackers shape the system’s decisions, safety logic, or higher-level interpretation of inputs.

\textbf{Dirty road patch attacks:} These attacks target lane-keeping and Automatic Lane Centering (ALC) by altering road appearance (e.g., adversarial patches) to mislead camera/deep neural network (DNN)-based lane detection, potentially causing lane departure or reduced driver reaction time \cite{sato2021wip}.

{
\textbf{Camera perception spoofing:} Physical perturbations or projected patterns can cause misclassification or non-detection of road signs and lane markings, including traffic-sign misclassification \cite{nassi2021spoofing,sato2021wip,sato2023wip}, and spoofed GPS input can misdirect region-of-interest selection in vision-based detection \cite{tang2021fooling}.
}

{
\textbf{ACC and CACC manipulation:} Adaptive Cruise Control (ACC) and Cooperative Adaptive Cruise Control (CACC) maintain following distance from local sensing and, for CACC, V2V data. Shifting-stealthy, scaling-stealthy, and time-variant manipulations of these inputs induce erroneous braking or acceleration while remaining protocol-correct \cite{9564677}.
}

\subsubsection{Monitoring and Reconnaissance}\label{sec:3.2.2-recon}
\begin{figure}[t]
\centering
    \includegraphics[width=0.6\textwidth,trim=180 20 180 20,clip]{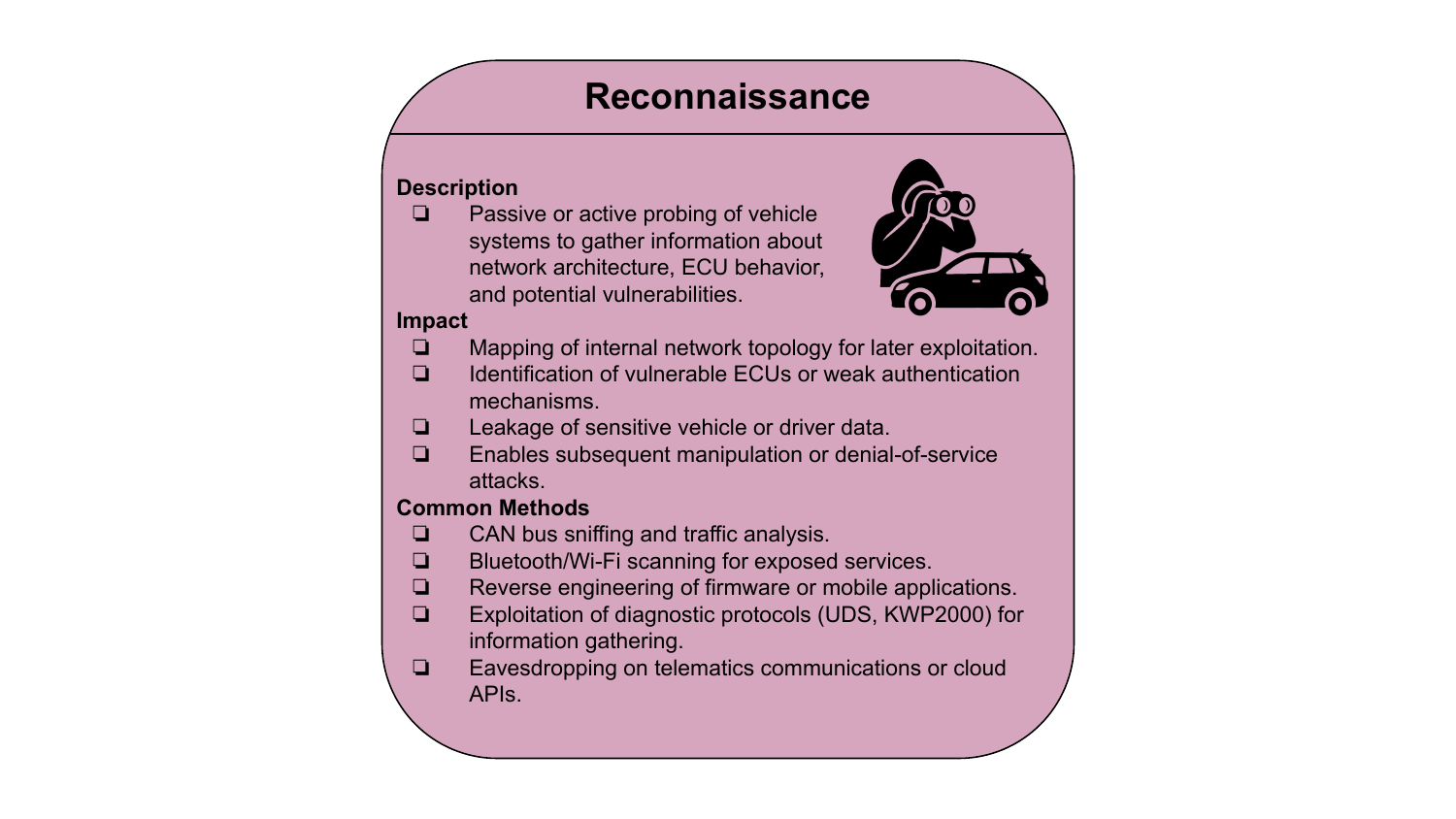}
    \caption{Representative summary of monitoring and reconnaissance attacks in connected vehicles, highlighting common ways adversaries observe traffic, infer system behavior, identify vulnerabilities, or collect sensitive information to support later attacks. The examples shown are illustrative rather than exhaustive; Section~3.2.2 provides the complete discussion of monitoring and reconnaissance attacks.}
    \label{fig:recon}
\end{figure}

Monitoring (reconnaissance) attacks are primarily \emph{passive} or low-interaction attacks that aim to learn internal states, reverse engineer messages, or infer private information without immediately altering vehicle behavior. While often less directly destructive than manipulation or disruption, monitoring attacks are powerful enablers: they provide the knowledge required to mount stealthier injection, impersonation, and targeted disruption attacks. Figure~\ref{fig:recon} summarizes reconnaissance attacks, in which adversaries gather system knowledge to enable subsequent manipulation or disruption.

\paragraph{Internal Monitoring (On-IVN Reconnaissance)}
Internal monitoring assumes the adversary can observe IVN traffic (e.g., via an ECU compromise or a diagnostic foothold) and focuses on learning message semantics, timing, and system behavior.
\textbf{ECU compromise reconnaissance:} Compromising an ECU enables passive reconnaissance. Prior work shows ECUs can be reprogrammed even while a vehicle is in motion \cite{8624397}, implying that an attacker with code execution on an ECU can both modify and observe system behavior. An adversary can monitor IVN traffic, decode frame semantics, and later use this knowledge to mount masquerade or targeted injection attacks.

\textbf{Frame sniffing attacks:} Frame sniffing captures CAN frames to analyze operations, control commands, and sensor data \cite{rajapaksha2023improving,8053478}. This intelligence is commonly used to build injection payloads, learn arbitration timing, and identify high-impact message IDs.

\textbf{TPMS tracking attacks:} TPMS is also vulnerable to tracking. Sensors periodically broadcast identifiers and pressure data. If unencrypted, adversaries can record sensor IDs and track a vehicle over time. Rouf et al. \cite{rouf2010security} found that fixed sensor IDs and lack of cryptography enable tracking; transmissions can be triggered and eavesdropped with low-cost hardware (tens of meters with commodity equipment, with longer ranges possible via directional antennas), raising significant privacy concerns.

\paragraph{External Monitoring (Off-Vehicle Privacy and Inference)}
External monitoring focuses on tracking vehicles or inferring sensitive information using external channels (apps, V2X metadata, side channels) rather than direct IVN compromise.

\textbf{Blue Link personally identifiable information (PII) exfiltration:} The Blue Link application for Hyundai vehicles was shown to transmit sensitive data (e.g., credentials) in plaintext to backend servers, allowing interception and credential capture \cite{KIM2021102150}. With captured credentials, adversaries could unlock vehicles and violate privacy and physical security.

\textbf{Cache side-channel route extraction:} When malware co-resides on the same processor as localization/perception software, cache side channels can infer routes by correlating compute patterns with environment and motion \cite{247674}. This attack class highlights how untrusted apps and shared compute platforms can leak sensitive mobility information.

\textbf{GPS stalking and leading:} GPS stalking/leading exploits GPS velocity traces by tailgating (stalking) or driving ahead (leading)  to reproduce the target’s GPS velocity trace and derive its intra-vehicle key \cite{yang2022gpskey}.

\textbf{Cellular MEMS-based vehicle identification:} Smartphone MEMS sensors (accelerometers/gyroscopes) can reveal vehicle characteristics. Because many apps access these sensors without special permissions, adversaries can infer vehicle type and other traits by analyzing vibration patterns \cite{chenguess}, enabling linkage to demographics or behavioral profiling.

\subsubsection{Disruption}\label{sec:3.2.3-disruption}
\begin{figure}[t]
    \centering
    \includegraphics[width=0.6\textwidth,trim=180 20 180 20,clip]{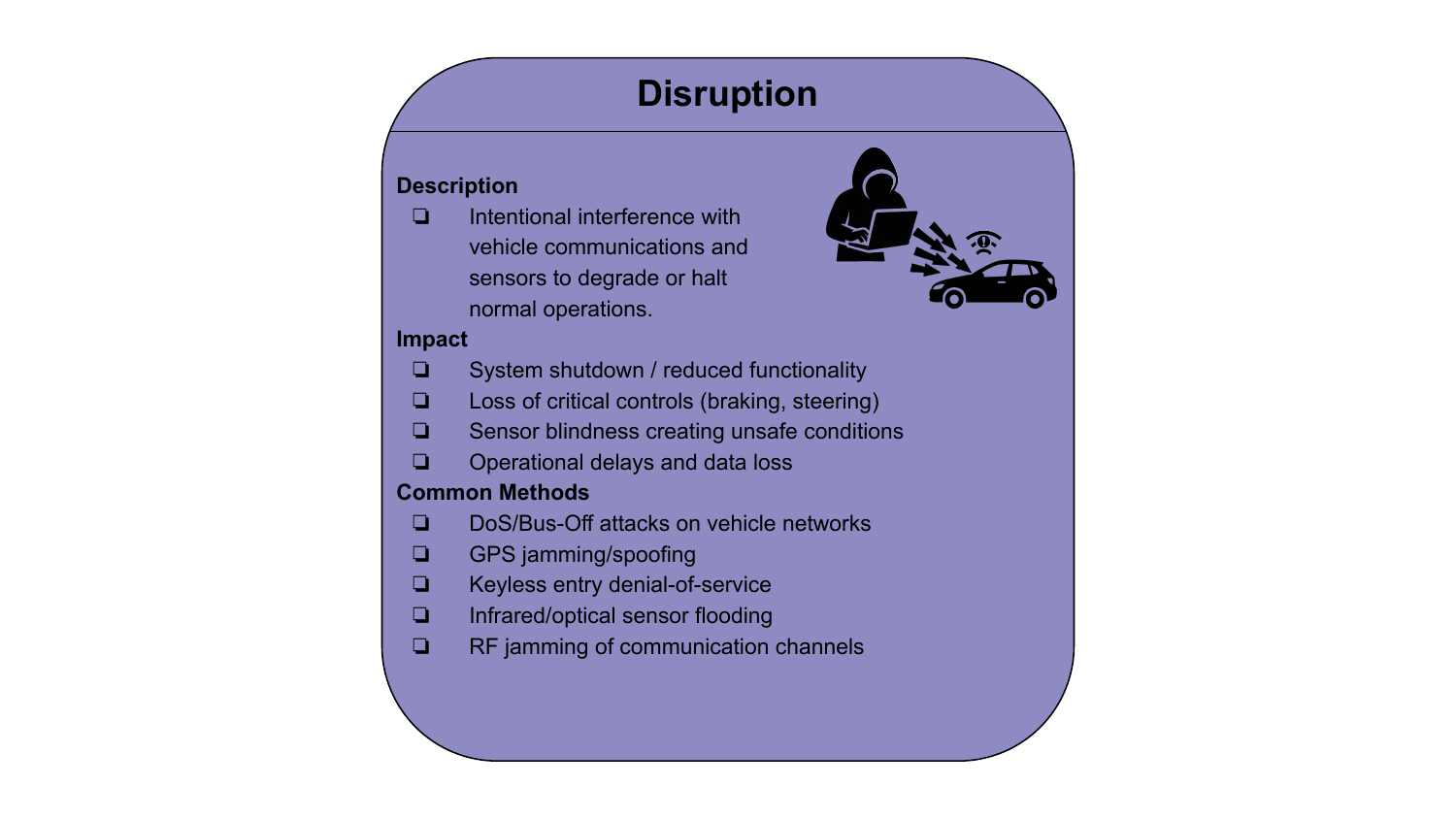}
\caption{Representative summary of disruption attacks in connected vehicles, highlighting common ways adversaries interfere with communication, sensing, or system availability to degrade or prevent normal operation. The examples shown are illustrative rather than exhaustive; Section~3.2.3 provides the complete discussion of disruption attacks.}
    \label{fig:disruption}
\end{figure}

Disruption attacks aim to degrade the availability, reliability, or timeliness of communication and sensing. Unlike many manipulation attacks, disruption may not require deep semantic understanding of message meaning: simply preventing delivery of safety-critical information can be enough to cause unsafe behavior. We separate disruption into internal (on-IVN) and external (signal-/network-facing) forms. Figure~\ref{fig:disruption} summarizes disruption attacks, in which adversaries interfere with communications or sensing to degrade availability and safe operation.

\paragraph{Internal Disruption (On-IVN Availability Attacks)}
Internal disruption targets in-vehicle communication buses and ECU interactions, interfering with the delivery of frames/messages and destabilizing system operation.

\textbf{DoS and Bus-Off attacks:} In CAN-based IVNs, DoS attacks flood the bus with high-priority messages, monopolizing the channel and preventing lower-priority legitimate traffic from being transmitted. Attackers may also overwhelm specific ECUs to render them unresponsive \cite{272188,8642311}. Bus-Off attacks exploit CAN error handling by repeatedly inducing transmission errors (e.g., via bit flips), forcing a node into the Bus-Off state \cite{272188,bloom2021weepingcan,palanca2017stealth}. Disabling critical ECUs can render the vehicle unsafe or uncontrollable \cite{7796898,5504804}.

\textbf{Fabrication and connection exhaustion:} In fabrication, an adversary sends forged messages that overload receivers or create unstable behavior \cite{parno2005challenges}. Connection exhaustion attacks exploit resource limits in higher-layer protocols; for example, in J1939 multi-packet transport, maintaining a connection can prevent other ECUs from establishing their own, causing legitimate packets to be dropped \cite{chatterjee2023exploiting}.

\textbf{Message suspension/drop attacks:} Adversaries intercept and suspend or drop critical messages between nodes, delaying or preventing control actions such as braking or steering~\cite{bloom2021weepingcan, palanca2017stealth}.

\textbf{“Bush Telegraph” and brute-force request flooding:} The “Bush Telegraph” attack exploits indirect or less secure paths across components to propagate disruption from a compromised foothold \cite{9316663}. Brute-force request flooding similarly sends large volumes of valid requests (regardless of priority) to exhaust resources and disrupt normal operation \cite{chatterjee2023exploiting}.

\paragraph{External Disruption (RF/Sensor/Network Availability Attacks)}
External disruption attacks interfere with the external signals and services vehicles rely on, degrading navigation, access control, and cooperative safety systems.

\textbf{GPS jamming:} GPS jamming transmits RF noise on GPS frequencies to prevent accurate position/time solutions. %

\textbf{RF jamming for keyless entry:} RF jammers can prevent vehicles from receiving lock/unlock commands, denying access and creating opportunities for unauthorized entry once jamming ceases \cite{6513548}.

{
Perception attacks that fabricate, remove, or misclassify objects, such as infrared projection, LiDAR point removal, and camera spoofing, alter what the vehicle perceives and are therefore classified as manipulation (Section~\ref{sec:3.2.1-manipulation}). Disruption covers attacks that deny sensing or communication altogether, such as the jamming attacks above.
}

{
\textbf{Grey Hole and Black Hole attacks:} In vehicular ad hoc networks, Grey Hole and Black Hole attacks selectively drop or drop all forwarded packets, disrupting routing and delaying or preventing safety-critical messages \cite{9463461,9619465,9316663}.
}

\subsubsection{Discussion: Methodologies and Implications}

The {four} representative chains trace distinct methodology paths through this taxonomy. The
Jeep Cherokee attack \cite{miller2015remote} combined passive reconnaissance with message-level CAN injection %
 after pivoting through the infotainment system. The Tesla Free-Fall attack
\cite{niefreefall2017} used browser exploitation to escalate to firmware-level compromise%
, converting temporary
wireless access into persistent ECU control. The GPS spoofing chain \cite{ying2023gps} operated at the signal
and measurement level, corrupting navigation inputs without touching the IVN directly.
{
The Kia attack instead exploited application-level authorization and account-management logic in trusted backend services \cite{curry2024kia}.
}

Rather than breaking protocols outright, many attacks take advantage of assumptions embedded in system design, especially the assumption that inputs, once inside the vehicle, are benign.

{
Across these methodology families, manipulation attacks alter what the vehicle accepts as true, monitoring and reconnaissance attacks learn the system without altering it, and disruption attacks deny or delay legitimate operation regardless of message content. Many attacks combine these behaviors, but the primary adversarial action determines their classification in this taxonomy.
}

Monitoring and reconnaissance attacks emphasize that \emph{learning} the system is often a prerequisite to controlling it. Passive sniffing, tracking, and side-channel inference can expose message meaning, driver behavior, and location privacy, and can directly enable stealthier injection and disruption.

Finally, disruption attacks demonstrate that availability and timeliness are first-class safety properties. Flooding, Bus-Off, and message suspension can destabilize IVN operation even without deep semantic knowledge.

Each methodology carries distinct operational, safety, and privacy consequences. In the next section, we analyze the concrete %
 impacts of these attacks, highlighting the functional disruptions, safety hazards, and privacy violations they can cause once these methodologies are successfully deployed.
\subsection{Attack Impacts}\label{sec:3.3-attack-impacts}

In Table~\ref{table:attack-impacts-analytic}, we summarize representative automotive attacks and their consequences. Across the works surveyed in this table, attack impacts consistently fall into three broad categories: (i) attack-induced mechanical and electrical anomalies, (ii) presentation of false or misleading information to the driver, and (iii) information leakage that enables follow-on attacks. This section discusses how each impact class manifests in modern vehicles.

\begin{table*}[!htbp]
\centering
\caption{Analytical Classification of Vehicular Attack Impacts}
\label{table:attack-impacts-analytic}
\setlength{\tabcolsep}{3pt}
\adjustbox{max width=\textwidth, max totalheight=0.95\textheight}{%
\begin{tabular}{|>{\raggedright\arraybackslash}p{2.02cm}|>{\raggedright\arraybackslash}p{2.16cm}|>{\raggedright\arraybackslash}p{1.87cm}|>{\raggedright\arraybackslash}p{1.58cm}|>{\raggedright\arraybackslash}p{2.74cm}|>{\raggedright\arraybackslash}p{2.74cm}|>{\raggedright\arraybackslash}p{2.59cm}|>{\raggedright\arraybackslash}p{3.02cm}|}
\hline
Impact Category & Specific Impact Manifestation & Affected Vehicle Domain & Severity Level & Demonstrated Consequences & Systemic and Cross-Layer Implications & Research Gaps & References \\
\hline

Mechanical and Electrical Anomalies &
Steering Lockout; Brake Lockout; Engine Lockout &
Powertrain / Chassis Control &
Safety-Critical &
Loss of steering/braking authority; engine shutdown during motion &
Immediate crash risk; cascading ECU failures &
Quantitative crash-risk modeling limited; no standardized safety severity metrics &
\cite{miller2015remote, KIM2021102150, info10040148, wolf2007state, mohammadi2022generation, chernikova2019self, 7796898, 8624397, 10.1007/978-3-540-87698-4_21, kang2022detection, kang2020attack, shoukry2013non, 8688625, xu2020robust} \\
\hline

Mechanical and Electrical Anomalies &
Lane Assist Failure; Airbag Disablement &
ADAS / Safety Systems &
Safety-Critical &
Disabled collision avoidance; suppressed airbag deployment &
Failure of fail-safe mechanisms; regulatory liability &
Adversarial-condition safety certification lacking &
\cite{8624397, 5940525, 10.1007/978-3-540-87698-4_21} \\
\hline

Mechanical and Electrical Anomalies &
Head/Tail Lights Disablement; Windshield Wipers Control &
Body Control Module &
Operational / Safety-Critical &
Reduced visibility; signaling failure in traffic &
Accident risk under adverse weather; signaling inconsistency &
Environmental stress testing under attack not evaluated &
\cite{8624397, rajapaksha2023improving} \\
\hline

Mechanical and Electrical Anomalies &
Window Controls; Horn Control; HVAC Control; Cabin Lights Control &
Body / Comfort Systems &
Operational &
Passenger discomfort; nuisance behavior; driver distraction &
Reduced driver trust; potential distraction-based safety degradation &
Human-factor impact under persistent nuisance events underexplored &
\cite{10.1007/978-3-540-87698-4_21, DBLP:journals/corr/abs-1802-01725, wolf2007state, 8624397, info10040148, 5504804} \\
\hline

False or Misleading Information &
Instrument Panel/Dashboard Manipulation; Speedometer Manipulation &
Dashboard / Powertrain Interface &
Safety-Critical &
Incorrect speed reporting; falsified system state feedback &
Driver decision errors; regulatory non-compliance &
Driver-trust modeling under falsified feedback limited &
\cite{8624397, rajapaksha2023improving, 5504804} \\
\hline

False or Misleading Information &
Mileage Gauge Tampering; Removing Speed/RPM Governors &
Powertrain Control &
Economic / Operational &
Fraudulent mileage reporting; unsafe performance modification &
Insurance fraud; unsafe driving conditions &
Tamper-resistant logging and verification standards uneven &
\cite{wolf2007state} \\
\hline

False or Misleading Information &
Infotainment Center Display Manipulation &
Infotainment &
Operational &
Misleading navigation or system information; distraction &
Indirect safety degradation through driver confusion &
Interaction between infotainment deception and ADAS underexplored &
\cite{DBLP:journals/corr/abs-1802-01725} \\
\hline

Information Leakage and Follow-On Attacks &
Compromised In-Vehicle Conversation &
Infotainment / Telematics &
Privacy-Critical &
Unauthorized audio capture; surveillance exposure &
Secondary exploitation; targeted manipulation; social engineering &
End-to-end encrypted in-cabin audio protection not standardized &
\cite{7393668} \\
\hline

Information Leakage and Follow-On Attacks &
Door Lock Tampering; Car Alarm Disablement; Replay-Based Theft &
Body Control / Access Systems &
Security-Critical &
Unauthorized vehicle access; theft execution &
Fleet-level theft campaigns; insurance and economic impact &
Unified anti-theft cryptographic frameworks lacking &
\cite{5940525, 10.1145/3431233, DBLP:journals/corr/abs-1802-01725, wolf2007state, 9316663} \\
\hline

Cross-Cutting Impact %
 Properties &
Cascading Multi-Domain Compromise &
IVN-Wide / Cross-ECU &
Systemic &
Simultaneous control, privacy, and availability degradation &
Propagation across safety and non-safety domains; fleet-scale exploitation potential &
Holistic cross-layer impact modeling and standardized severity taxonomy absent &
\cite{miller2015remote, wolf2007state, 8624397, info10040148} \\
\hline

\end{tabular}
}
\end{table*}

\subsubsection{Mechanical and Electrical Anomalies}\label{sec:3.3.1-mechanical}

\textbf{Steering lockout:}
Steering lockout attacks disable or manipulate steering control, potentially causing loss of vehicle control and severe accidents, particularly at high speeds \cite{miller2015remote, chernikova2019self}. Prior work demonstrates that such attacks can be performed remotely \cite{miller2015remote}, raising concerns for drive-by-wire systems where physical overrides may be limited.

\textbf{Brake lockout:}
Brake lockout attacks %
 {disable or manipulate braking authority by compromising brake-control communication or injecting/replaying messages that affect brake-control modules} \cite{8624397, miller2015remote, shoukry2013non, kang2020attack, kang2022detection, mohammadi2022generation}. The resulting loss of braking capability significantly increases stopping distance and collision risk and may be exploited to force unsafe stops.

\textbf{Engine lockout:}
Engine manipulation enables unauthorized start, shutdown, or forced limp-mode operation %
 \cite{wolf2007state, 8624397}. Such attacks facilitate theft, denial of service, and unsafe loss of propulsion during motion.

\textbf{Lane assist failure:}
 {Attackers can disrupt keep-lane assist functionality through ADAS perception spoofing \cite{nassi2021spoofing}}.

\textbf{Auxiliary actuator disruption:}
Window controls, hazard and exterior lighting, cabin lights, horns, and HVAC functions can be disabled, spoofed, or unintentionally activated through malicious in-vehicle control \cite{5504804,10.1007/978-3-540-87698-4_21}. These demonstrations fall within the broader class of CAN/IVN compromise discussed in early automotive-security work \cite{wolf2007state,5940525} and subsequent CAN/IVN analyses \cite{DBLP:journals/corr/abs-1802-01725,8624397}. While some components are non-critical individually, their disruption can distract drivers, obscure intent signaling, or degrade crash safety.

\subsubsection{False or Misleading Information}\label{sec:3.3.2-false-info}

\textbf{Speedometer and dashboard manipulation:}
Altering CAN payloads carrying instrument cluster data can misrepresent speed, fuel level, or warnings, misleading drivers and increasing accident and violation risk \cite{rajapaksha2023improving, 8624397, 8306904}.

\textbf{Odometer, tachograph, and governor tampering:}
Mileage manipulation and removal of speed or RPM governors enable fraud, regulatory violations, excessive engine wear, and unsafe operation \cite{wolf2007state}.

\textbf{Infotainment and display deception:}
Full or partial compromise of infotainment systems allows attackers to present misleading or distracting content to drivers \cite{DBLP:journals/corr/abs-1802-01725}, degrading situational awareness.

\subsubsection{Information Leakage and Follow-On Attacks}\label{sec:3.3.3-leakage}

\textbf{Eavesdropping and privacy leakage:}
Intercepted in-vehicle communications expose location, sensor data, audio, and video streams, enabling replay, tracking, and more sophisticated attacks \cite{7393668}.

\textbf{Door locks, alarms, and theft enablement:}
Remote unlocking and manipulation of vehicle access-control functions can facilitate unauthorized access \cite{5940525,wolf2007state}, while replay- or relay-style keyless-entry attacks provide another access path \cite{10.1145/3431233,9316663}.

\begin{table}[t]
\centering
\caption{Mapping between attack impacts and system-level implications}
\label{tab:impact-mapping}

\small
\setlength{\tabcolsep}{3pt}

\begin{tabularx}{\textwidth}{l *{5}{>{\centering\arraybackslash}X}}
\hline
Impact Source & %
 Availability Loss & %
 {Safety Degradation Without Explicit Failure} & %
 Cascading Cross-Domain Effects & Driver Trust Erosion & %
 ML / Probabilistic Failures \\
\hline
Mechanical / Electrical Anomalies & $\checkmark$ & $\checkmark$ & $\checkmark$ &  &  \\
\hline
False or Misleading Information &  & $\checkmark$ &  & $\checkmark$ & $\checkmark$ \\
\hline
Information Leakage &  &  & $\checkmark$ &  &  \\
\hline
\end{tabularx}

\end{table}

\subsubsection{Discussion: Cross-Cutting Impact %
 Properties}

The {four} representative attack chains illustrate these cross-cutting impact properties concretely. The
Jeep Cherokee attack \cite{miller2015remote} produced availability loss and safety degradation without explicit failure:
 remote transmission disablement at highway speed and brake disablement at low speed left the vehicle operational in appearance while removing driver
control. The Tesla Free-Fall attack \cite{niefreefall2017} demonstrated cascading behavior, where a %
 Wi-Fi browser
exploit propagated through domain boundaries to reach ECU firmware, converting a connectivity
vulnerability into persistent safety-critical compromise. The GPS spoofing chain
\cite{ying2023gps} exemplifies silent and probabilistic impact:  localization used by connected-vehicle safety applications such as intersection movement assist was corrupted without triggering faults, producing false or suppressed safety warnings while the vehicle continued to operate within nominal bounds.
{
The Kia attack adds a cloud-to-vehicle impact path, combining privacy exposure with unauthorized remote vehicle control \cite{curry2024kia}.
}
Beyond individual components and functions, the impacts described
above reveal several cross-cutting impact properties that shape overall vehicle risk
(see Table \ref{tab:impact-mapping}). 

First, many attacks manifest as \emph{availability loss} at the system level rather than as isolated failures. Engine shutdowns, forced limp modes, brake lockout, or bus flooding can collectively immobilize a vehicle, potentially stranding occupants in unsafe locations. Framing these outcomes as systemic denial of service clarifies that even non-safety-critical components can contribute to safety-relevant unavailability when tightly coupled through the IVN.

Second, a substantial class of impacts involves \emph{safety degradation without explicit failure}. Misleading instrumentation, intermittent actuator behavior, or degraded ADAS functionality may not trigger faults or alerts, yet they increase driver workload and reduce reaction time. These conditions elevate accident likelihood while remaining difficult to diagnose, as the vehicle continues to operate within nominal bounds.

Third, the surveyed impacts frequently exhibit \emph{cascading behavior across domains}. Compromise of infotainment or wireless interfaces can propagate to CAN injection and actuator manipulation; information leakage can enable replay attacks and theft. Such cascades highlight that impact severity is often determined less by the initial entry point than by architectural coupling and insufficient isolation between domains.

Fourth, repeated or subtle manipulation can lead to \emph{erosion of driver trust}. When warnings, displays, or assistance features behave inconsistently, drivers may ignore legitimate alerts or disengage safety systems altogether. This human–vehicle interface breakdown represents an indirect but meaningful safety impact that is not captured by component-level analysis alone.

Finally, as vehicles increasingly integrate learning-enabled perception and control, some impacts become \emph{silent and probabilistic}. Sensor or semantic manipulation may bias model confidence or decision boundaries without producing overt faults, complicating attribution and post-incident analysis. While traditional impacts remain relevant, these properties motivate defenses that consider system-level resilience, cross-layer validation, and safe degradation, rather than focusing solely on preventing individual attacks.

\subsection{Attack {Detection and }Mitigations}\label{sec:3.4-attack-mitigations}

As vehicle architectures become increasingly interconnected, mitigation strategies must address not only isolated vulnerabilities but also the ways attacks propagate across sensing, communication, computation, and control layers. %
As shown in Table~\ref{table:mitigations-analytic}, effective defenses are inherently layered and must be deployed across multiple stages of the attack lifecycle.

{Detection mechanisms complement these defenses by identifying anomalous behavior at attack entry points, during attack execution, and after attacks begin to affect vehicle operation.}

{Mitigation in this survey includes preventive controls, attack detection, impact containment, and recovery; intrusion detection therefore represents one component of the broader defense strategy.}

\ifdefined\showcuts
\newcommand{\cutfhssrow}{%
&
Frequency/RF hopping for keyless integrity &
Physical / Data-Link &
Prevent Manipulation; Preserve Availability &
Less replay/relay; stronger entry &
Hardware, regulatory, and interoperability limits &
Few formal guarantees &
 \\
\cline{2-8}
}
\else
\newcommand{\cutfhssrow}{}
\fi
\begin{table*}[!htbp]
\centering
\caption{Detection and Mitigation Strategies Aligned with the Survey Taxonomy}
\label{table:mitigations-analytic}
\setlength{\tabcolsep}{3pt}
\adjustbox{max width=\textwidth, max totalheight=0.95\textheight}{%
\begin{tabular}{|>{\raggedright\arraybackslash\hspace{0pt}}p{2.38cm}|>{\raggedright\arraybackslash\hspace{0pt}}p{2.55cm}|>{\raggedright\arraybackslash\hspace{0pt}}p{2.12cm}|>{\raggedright\arraybackslash\hspace{0pt}}p{2.04cm}|>{\raggedright\arraybackslash\hspace{0pt}}p{3.06cm}|>{\raggedright\arraybackslash\hspace{0pt}}p{3.06cm}|>{\raggedright\arraybackslash\hspace{0pt}}p{3.06cm}|>{\raggedright\arraybackslash\hspace{0pt}}p{3.57cm}|}
\hline
Detection and Mitigation Category & Defensive Mechanism & Protected Layer(s) & %
 {Defense Objective (Prevent / Detect / Contain / Recover)} & Demonstrated Effectiveness & Deployment Constraints & Research Gaps & References \\
\hline
\multirow{3}{2.38cm}{Detection and Mitigation at Attack Entry Points}&
Secure firmware; sandbox media systems &
Application / Physical &
Prevent Entry; Preserve Integrity &
Less untrusted code execution; stronger integrity &
Legacy ECUs; patch latency &
Uneven secure boot coverage; limited sandbox assurance &
\cite{5504804, iqbal2019towards, 8624397} \\
\cline{2-8}
&
Harden MirrorLink/CAN exposure; authenticate IP traffic &
Application / Data-Link &
Prevent Entry; Prevent Injection &
Fewer projection pivots; less IP spoofing &
Protocol heterogeneity; integration with legacy stacks &
No hardening standard; inconsistent OEM policy &
\cite{LE201817, KIM2021102150} \\
\cline{2-8}

&
Encrypt TPMS, GPS, CAN, and infotainment data &
Data-Link / Application &
Prevent Entry; Preserve Confidentiality &
Less passive disclosure; costlier interception &
Bandwidth/latency overhead; key management &
Scalable key lifecycle; backward compatibility &
\cite{10.1007/978-3-540-87698-4_21, daily2021securing, 287308, KIM2021102150, rouf2010security} \\
\hline

\multirow{4}{2.38cm}{Detection and Mitigations Disrupting Attack Method\-ol\-o\-gies} &
MACs, signatures, session keys; mutual ECU authentication; freshness &
Data-Link / Application &
Detect/\allowbreak{}Prevent Manipulation; Preserve Integrity &
Less spoofing/replay; message provenance &
ECU compute limits; bus timing; credential provisioning &
Lightweight crypto not standardized; mixed legacy fleets &
\cite{parno2005challenges, 9491889, 8270348, 5281251, xie2021security, wang2014vecure, 10049666, 6542519, 197171} \\
\cline{2-8}

&
IDS for timing, voltage, injection, spoofed gear/RPM, masquerading &
Physical / Data-Link / Application &
Detect Manipulation; %
 Detect Monitoring and Reconnaissance &
Detects injection, replay, physical-layer, and spoofed-state anomalies &
Label scarcity; drift; overhead; false positives &
Weak benchmarking; adversarial robustness unclear &
\cite{9091063, 7427089, blevins2021time, shahriar2023cantropy, s23073554, alkhatib2023wip, agbaje2022framework, 6940213, lai2024gradient, fu2024iov, wang2024intrusion, li2024ecf, ibraheem2024anomaly, chandra2024enhanced, altalbe2023enhanced, jichici2023control, singh2023intrusion, zhao2023tree, qiu2023physical, yu2022tce, kandasamy2022detection, andreica2022android, anjum2022vehicle, 9706416, baldini2022detection, 7420322, moriano2022masquerade, wang2021can, lokman2019intrusion} \\
\cline{2-8}

&
GPS anti-spoofing (geofencing); jamming detection &
Physical / Application &
Detect Manipulation; Detect Disruption &
Spoof/jam awareness &
Coverage gaps; environment sensitivity &
No standardized resilient positioning &
\cite{ying2023gps, baldini2022detection, abdallah2022not} \\
\cline{2-8}

\cutfhssrow

&
Sensor fusion and cross-validation; IR filtering; authenticated ACC/CACC data &
Physical / Application &
Detect Manipulation; Preserve Safety &
Detects spoofed perception; resilient cooperative control &
Calibration; compute overhead; sensor availability &
No certified fusion robustness metrics &
\cite{sato2021wip, 10.1145/3460120.3484766, shin2017intelligent, Cao2021InvisibleFB, philipsen2021threats, 9564677, yang2021secure, 9619465} \\
\hline

\multirow{3}{2.38cm}{Detection and Mitigations for Impact Containment and Recovery} &
Redundancy and secure routing for dropped messages &
Data-Link &
Contain Impact; Ensure Recovery &
Tolerates suspension/drop; control continuity &
Bandwidth/latency; routing complexity &
Few safety-grade guarantees under loss &
\cite{LE201817} \\
\cline{2-8}

&
Throttling, queuing, and timeouts against overload/exhaustion &
Application / Data-Link &
Preserve Availability; Contain Impact &
Less effective DoS &
Tuning; may block legitimate bursts &
No assured adaptive policies &
\cite{chatterjee2023exploiting} \\
\cline{2-8}

&
Dynamic IDs and encrypted TPMS broadcasts against tracking &
Physical / Data-Link &
Contain Privacy Impact; Limit Follow-On Attacks &
Less linkability and passive tracking &
Coordination overhead; TPMS receiver compatibility &
No standard identifier rotation &
\cite{rouf2010security} \\
\hline
\end{tabular}
}
\end{table*}

\subsubsection{{Detection and }Mitigations at Attack Entry Points}\label{sec:3.4.1-entry-mitigations}

{
\paragraph{Detection at Attack Entry Points}
Detection at exposed interfaces focuses on identifying anomalous access, communication, or sensor behavior before an attacker establishes a foothold in the IVN. Wireless interfaces can be monitored for abnormal communication or interference patterns, including sustained jamming behavior. Sensor-facing interfaces use plausibility checks and comparison with independent observations to identify manipulated inputs; for example, localization measurements can be checked for trajectory plausibility \cite{ying2023gps}.
}

Entry-point mitigations aim to reduce the likelihood that an attacker can gain initial access through physical interfaces, wireless connectivity, or external data inputs.

\paragraph{Architectural Isolation and Secure Gateways}
Network segmentation and domain separation limit attack propagation by isolating safety-critical ECUs from infotainment, telematics, and third-party application domains \cite{6945510, iqbal2019towards, castiglione2025securing}. Secure gateways enforce message filtering and rate control between domains, ensuring that even compromised non-critical components cannot directly influence control-critical subsystems.

\paragraph{Wireless Interface Hardening}
Wireless entry points such as keyless entry, Bluetooth, cellular, and smartphone integration platforms are protected through protocol-level authentication and frequency agility.  Timestamped rolling codes have been proposed against roll-jam attacks \cite{deppenhanced}, and jammer detection mechanisms identify sustained interference patterns. Protocol hardening and access control mitigate MirrorLink-based CAN injection by restricting which devices can interface with internal networks \cite{LE201817, KIM2021102150}. For vehicular wireless links, physical-layer security schemes using reconfigurable intelligent surfaces have been analyzed to increase secrecy capacity against eavesdroppers \cite{9128438}.

\paragraph{Sensor and External Signal Validation}
 GPS inputs can be checked for plausibility \cite{ying2023gps}, and TPMS messages can be authenticated \cite{rouf2010security}. Cross-referencing localization data with inertial sensors or map constraints further limits spoofing success. These defenses reduce attacker control over navigation, tracking, and safety alerts.

\subsubsection{{Detection and }Mitigations Disrupting Attack Methodologies}\label{sec:3.4.2-method-mitigations}

{
\paragraph{Detection of Attack Methodologies}
Detection after initial access focuses on the behavior produced by an attack. Timing- and traffic-based IDSs identify deviations in message frequency, inter-arrival time, identifiers, and payload behavior \cite{7796898,197171}. Signal-level approaches learn expected signal distributions and detect abnormal temporal or spatial patterns \cite{rajapaksha2023improving}. Runtime monitoring and control-flow integrity target deviations in firmware and execution behavior, while cross-sensor consistency and plausibility checks detect semantic manipulation that preserves valid message syntax \cite{yang2021secure,shin2017intelligent}, and learned predictors detect manipulated ACC/CACC inputs \cite{9564677}.
}

Once access is obtained, attackers exploit weaknesses in signal integrity, message semantics, firmware execution, and control logic. The following mitigations directly target these methodologies.

\paragraph{Signal- and Measurement-Level Defenses}
Signal- and measurement-level attacks can be detected by monitoring deviations from expected signal behavior using statistical or learned temporal models \cite{9091063}. Stealthier manipulations, including surge, bias, and geometric attacks that remain within admissible deviations, require detection based on expected signal dynamics rather than fixed thresholds \cite{jichici2023control}. Attack-resilient state estimation can keep vehicle dynamics stable when sensor measurements are attacked in transit to the controller \cite{rana2017attack}. In perception systems,  software detection of IR-induced artifacts reduces susceptibility to infrared spoofing \cite{10.1145/3460120.3484766}.

AI/ML-based anomaly detection further enhances these defenses by learning expected signal distributions and flagging deviations that remain within physical tolerances but exhibit abnormal temporal or spatial patterns \cite{rajapaksha2023improving}.

\paragraph{Message- and Frame-Level Protections}
Message fabrication, replay, injection, and impersonation attacks are mitigated through cryptographic authentication, timestamps, nonces, and message whitelisting \cite{parno2005challenges, LE201817}. IDS systems monitor traffic rates, identifiers, and payload correlations to detect violations of learned communication baselines \cite{7796898, 197171, 9091063}.

Machine learning–based IDS approaches complement rule-based systems by identifying subtle deviations that preserve protocol correctness but alter operational meaning \cite{7393668}.

\paragraph{Firmware- and Execution-Level Security}
Weak or unenforced access control on ECU diagnostic and reflashing services allows malicious firmware to be loaded; restricting these services and requiring signed firmware updates address this path \cite{5504804}. Message authentication (e.g., message authentication codes and digital signatures) and fingerprinting-based intrusion detection address spoofed or impersonated ECU messages \cite{8624397}. Application sandboxing prevents third-party software from accessing CAN interfaces or safety-critical resources \cite{iqbal2019towards}.

Runtime monitoring and control-flow integrity checks further constrain malicious execution, limiting attacker persistence even after compromise.

\paragraph{Control-Logic and Semantic Validation}
Attacks that preserve syntactic validity while manipulating semantics—such as RPM spoofing, gear spoofing, ACC/CACC manipulation, and lane perception attacks—are mitigated through cross-sensor consistency checks and plausibility validation \cite{yang2021secure, shin2017intelligent}; learned predictors detect manipulated ACC/CACC inputs \cite{9564677}. These defenses ensure that control decisions align with physical constraints and redundant observations.

AI/ML-based perception defenses strengthen this layer by detecting semantic inconsistencies across modalities, monitoring confidence estimates, and identifying out-of-distribution inputs in camera and LiDAR pipelines. Attacks that evade such defenses have been demonstrated \cite{279980, hau2021object, nassi2021spoofing}.

\subsubsection{{Detection and }Mitigations for Impact Containment and Recovery}\label{sec:3.4.3-impact-mitigations}

{
\paragraph{Impact-Level Detection}
At the impact stage, detection shifts from identifying malicious communication to identifying abnormal vehicle or control behavior. Control- and sensor-level monitoring can identify inconsistencies between observed vehicle states, expected physical behavior, and received communication. Such detection provides a trigger for containment, fail-safe operation, or recovery when attacks bypass earlier defenses \cite{yang2021secure,shin2017intelligent}; learned predictors detect manipulated ACC/CACC inputs \cite{9564677}.
}

When preventive defenses fail, impact-focused mitigations limit safety, availability, and privacy consequences.

\paragraph{Availability and Disruption Mitigation}
In SAE J1939 networks, request overload can be mitigated by having ECUs ignore requests that arrive faster than a set rate, and connection exhaustion by terminating transport connections that remain idle beyond a timeout \cite{chatterjee2023exploiting}. For Grey Hole and Black Hole routing attacks, surveyed defenses include machine-learning-based detection \cite{9619465}. These mechanisms preserve critical communication paths under adversarial load.

\paragraph{Fail-Safe Operation and Graceful Degradation}
Fail-safe mechanisms ensure that compromised systems revert to conservative control modes rather than unsafe behavior, particularly for braking, steering, and ACC functions.

\paragraph{Privacy and Information Leakage Mitigation}
Cache side channels can reveal a vehicle's route or location when localization software shares hardware with an attacker's program, so localization software requires strong timing-channel protection \cite{247674}. More broadly, surveyed defenses for autonomous vehicles fall into security architecture, intrusion detection, and anomaly detection \cite{KIM2021102150}.

\subsubsection{Data-Driven and Semantic Attack Detection}\label{sec:3.4.4-data-driven}

{
Data-driven detection learns normal communication, timing, payload, or signal behavior and identifies deviations associated with attacks. Existing approaches range from statistical and machine-learning-based analysis of CAN traffic to models that capture dependencies among vehicle signals \cite{moriano2022masquerade,gao2025signal}.
}

{
Relational approaches extend this analysis by modeling dependencies among signals rather than treating individual observations independently. Signal-correlation methods detect changes in expected relationships among CAN signals \cite{moriano2022masquerade}, while graph-based methods model signal relationships and their temporal behavior for anomaly detection \cite{gao2025signal}. These approaches can identify attacks in which individual messages remain plausible but their relationships with other signals or vehicle states become inconsistent.
}

{
Knowledge-graph-based detection further introduces explicit semantic relationships among vehicle entities, frames, signals, and their dependencies. KG-ID uses an in-vehicle knowledge graph to detect fabrication, masquerade, and fuzzing attacks using frame- and signal-level relationships \cite{sun2025kgid}. Related work combines knowledge-graph-derived semantic information with statistical and deep-learning methods for IVN intrusion detection \cite{alqahtani2025deep}. These approaches extend detection from individual traffic observations toward reasoning over expected relationships among signals and system components.
}

\subsubsection{Discussions}

{Across these stages, detection progresses from identifying anomalous entry behavior, to detecting attack execution, to recognizing abnormal vehicle states. Data-driven and semantic approaches complement these mechanisms by capturing relationships among signals, components, and system behavior.}

While individual mitigations address specific vulnerabilities, no single mechanism provides comprehensive protection. Cryptographic defenses cannot detect semantic manipulation, IDS systems face latency and false-positive trade-offs, and AI-based defenses remain vulnerable to adaptive adversaries. Moreover, security mechanisms must coexist with real-time constraints, functional safety requirements, and certification processes.

Viewed end-to-end, effective mitigation emerges from coordinated deployment across entry points, attack methodologies, and impact containment layers. Open challenges include securing legacy protocols, validating AI robustness under adversarial conditions, and balancing security, safety, and system complexity. Addressing these challenges remains central to the development of resilient and safety-aware in-vehicle architectures. %
Figure~\ref{fig:taxonomy} maps each attack stage to the detection, mitigation, containment, and recovery mechanisms that address it.

\section{Trends, Research Gaps, and Future Directions}\label{sec:4-research-gaps}

The preceding sections of this survey characterize the IVN threat landscape along four
dimensions: entry points, attack methodologies, impacts, and mitigations. Reviewing the
literature as a whole reveals structural patterns in how research attention has been
allocated, where knowledge gaps persist, and how the threat environment is evolving in
response to changes in vehicle architecture, connectivity, and the growing use of
machine learning in vehicle systems. This section summarizes those patterns, identifies
the most significant research deficiencies, and outlines directions that address documented
gaps rather than isolated improvements within already well-studied areas.

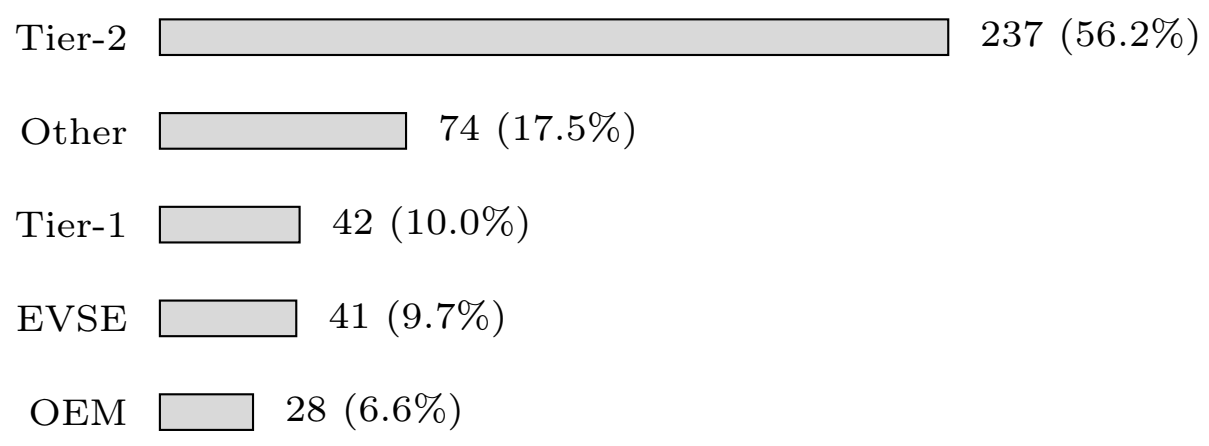
\begin{figure}[!t]
\centering

{\color{black}
\begin{tikzpicture}[x=0.022cm,y=0.62cm,font=\scriptsize]

\def\barh{0.38}

\foreach \name/\val/\pct/\y in {
Tier-2/237/56.2/4,
Other/74/17.5/3,
Tier-1/42/10.0/2,
EVSE/41/9.7/1,
OEM/28/6.6/0}
{
    \node[anchor=east] at (-4,\y+0.19) {\name};

    \fill[black!15]
        (0,\y)
        rectangle
        (\val,\y+\barh);

    \draw[black]
        (0,\y)
        rectangle
        (\val,\y+\barh);

    \node[anchor=west]
        at (\val+4,\y+0.19)
        {\val\ (\pct\%)};
}

\end{tikzpicture}
}

\caption{
Distribution of 422 publicly reported automotive-related CVEs in 2024 by ecosystem source. Percentages are calculated from counts reported by Upstream Security \cite{upstream2025}.
}
\label{fig:upstream_cve_sources}

\end{figure}

A defining characteristic of the period from 2021 to 2026 is the shift of the primary
attack surface from the physical IVN to remote and cloud-connected interfaces. Early
research, including foundational work on CAN bus vulnerabilities,
 was conducted largely under the assumption of physical or short-range access. Reported
incidents have moved in the opposite direction: by 2022,  97\% of attacks were conducted remotely, and telematics and application servers accounted for 35\% of incidents~\cite{upstream2023}. By 2023, telematics and application server attacks represented
43\% of all reported incidents~\cite{upstream2024}, rising further to 66\% in
2024~\cite{upstream2025}. {
Upstream reports 1,147 automotive-related CVEs from 2019 through 2024. Annual disclosures increased from 24 in 2019 to 422 in 2024, including 378 in 2023 \cite{upstream2025}. We further analyzed Upstream's 2024 breakdown by ecosystem source. Tier-2 suppliers account for 237 CVEs (56.2\%), followed by other ecosystem participants with 74 (17.5\%), Tier-1 suppliers with 42 (10.0\%), EVSE with 41 (9.7\%), and OEMs with 28 (6.6\%). Tier-1 and Tier-2 suppliers together account for 66.1\% of the total. Figure~\ref{fig:upstream_cve_sources} summarizes this distribution.
}

{
The distribution shows that disclosed automotive vulnerabilities are not concentrated at OEMs. They extend across the automotive supply chain and into EV charging infrastructure. Upstream attributes the broader CVE growth to increased adoption of connected components, greater vulnerability awareness, and increased research on EV chargers and infotainment systems \cite{upstream2025}.
}

{
These sources map directly onto the entry points in Table~\ref{table:entry-points-layered}: EVSE vulnerabilities correspond to the physical-layer EV charging interface, supplier-provided infotainment and telematics components to application-layer wireless and cellular entry points, and backend and API services to the cloud/API entry point illustrated by the Kia case (Table~\ref{table:representative-attack-chains}).
}

{
Upstream also notes that delays at the National Vulnerability Database affected CVE publication and analysis during 2024. Annual counts should therefore be interpreted as disclosure trends rather than a direct measure of vulnerabilities created during each year \cite{upstream2025}.
}

The body of academic IVN security
research has not fully tracked this shift: CAN-bus IDS publications remain the dominant
research output, while cloud API security, OTA supply chain integrity, and backend
authentication receive comparatively limited treatment~\cite{PMCSurveyDL2023,Bari2023}.

Table~\ref{tab:ivn-trends} summarizes observed trends %
 {across the taxonomy stages and related cross-cutting research areas covered in this survey}, drawing on publication patterns, incident data, and regulatory
milestones from approximately 2021 to 2026.

\begin{table*}[!htbp]
\centering
\caption{Trend Analysis of In-Vehicle Network (IVN) Cybersecurity Research (2021--2026)}
\label{tab:ivn-trends}
\adjustbox{max width=\textwidth, max totalheight=0.88\textheight}{%
\begin{tabular}{|p{3.0cm}|p{3.6cm}|p{3.2cm}|p{2.8cm}|p{3.8cm}|}
\hline
 {\textbf{Research Area}} &
\textbf{Representative Topics} &
\textbf{Estimated Publication Volume (Last 5 Years)} &
\textbf{Trend Direction} &
\textbf{Observed Research Imbalance} \\
\hline

Entry Points &
 {OBD-II/diagnostic access to CAN, infotainment, telematics, cloud APIs, OTA} &
High for CAN and OTA; Low for cloud/API integration &
CAN: Stable; \newline OTA/Cloud/API: Rapidly Increasing &
CAN-focused IDS research dominates; cloud API, OTA supply-chain, and fleet-level attack
surfaces remain systematically understudied~\cite{upstream2024, upstream2026, PMCSurveyDL2023} \\
\hline

Attack Methodologies &
Spoofing, replay, DoS, sensor manipulation, adversarial ML, multi-stage attacks &
High for single-layer attacks; Low for cross-layer and AI-targeted attacks &
Single-layer: Plateau; \newline Cross-layer \& Adversarial ML: Emerging &
Adversarial attacks on ML-based IDS are underexplored; multi-stage and cross-domain
attack chains lack formal models~\cite{Alobaidi2024,Ibrahum2024} \\
\hline

Impacts &
Safety degradation, control loss, privacy leakage, cascading failures, fleet-scale events &
Moderate overall; rapidly increasing for large-scale and fleet-level impacts &
Rapidly Increasing &
Individual-vehicle impact analysis dominates; fleet-scale and cascading failure models
remain scarce despite evidence of ecosystem-level incidents~\cite{upstream2025,upstream2026} \\
\hline

Mitigations &
IDS, anomaly detection, cryptography, segmentation, post-quantum cryptography (PQC), resilience mechanisms &
High for ML-based IDS; Low for PQC and incident response &
IDS: Increasing; \newline Resilience and PQC: Emerging &
Emphasis on detection; containment, coordinated response, and post-quantum migration
strategies are underexplored~\cite{Baccari2024,Nguyen2024,castiglione2025securing} \\
\hline

\multicolumn{5}{|l|}{\textbf{Cross-cutting research areas (see Table~\ref{table:canonical-taxonomy})}} \\
\hline

Architectural Evolution &
Automotive Ethernet, IP-based IVNs, Software-Defined Networking (SDN), zonal/domain electrical/electronic (E/E) architectures &
Moderate; growing since 2022 &
Increasing &
Security implications of Ethernet/Time-Sensitive Networking (TSN) transitions are discussed conceptually; empirical
benchmarking of IDS under real Ethernet IVN conditions remains
limited~\cite{Mozzaquatro2023,AESystematicReview2024} \\
\hline

Cryptographic Foundations &
Lightweight cryptography, authentication protocols, post-quantum cryptography &
Low overall; Very low for post-quantum mechanisms; emerging 2023--2025 &
Emerging &
Long-term cryptographic migration risks are underexplored under real-time IVN constraints;
no automotive-specific PQC standards exist yet~\cite{Chillotti2021,Nguyen2024,castiglione2025securing} \\
\hline

Regulation and Compliance &
ISO/SAE 21434, United Nations Economic Commission for Europe (UNECE) World Forum for Harmonization of Vehicle Regulations (WP.29) R155/R156, lifecycle security governance &
Moderate-to-High since 2021; growing rapidly &
Rapidly Increasing &
Regulatory compliance processes are well documented; independent adversarial validation of
CSMS effectiveness and systemic threat modeling remain
limited~\cite{ISOSAE21434,WP29R155,Zelle2022} \\
\hline

\end{tabular}
}
\end{table*}

{
A second structural change is the transition from distributed, function-specific ECUs connected by low-bandwidth buses toward high-bandwidth Automotive Ethernet backbones and domain- or zonal-controller architectures. This transition changes the security model in several ways. First, high-bandwidth backbones increasingly carry sensor, application, diagnostic, and control traffic over shared infrastructure, increasing the amount and diversity of security-critical data exposed to a common communication fabric. Second, consolidation moves functionality that was previously distributed across multiple ECUs into fewer, more capable controllers. Compromise of such a controller, gateway, or backbone segment can therefore affect multiple functional domains and increase the potential propagation and impact of a successful attack. Third, Ethernet introduces IP-based services, middleware, routing, and configuration mechanisms that shift part of the attack surface from low-level frame manipulation toward service exploitation, misconfiguration, and cross-domain lateral movement \cite{martinezcruz2021security}. At the same time, centralized gateways and zonal controllers provide stronger enforcement points for segmentation, authentication, traffic monitoring, and policy control. Architectural consolidation therefore creates both stronger defensive control points and higher-impact compromise points. The resulting detection and monitoring requirements are discussed further in Section~\ref{sec:architectural-evolution-ids}.
}

\subsection{Entry Points: Expanding and Underexamined Attack Surfaces}\label{sec:4.1-gaps-entry}

Research on IVN entry points has grown substantially, but not uniformly.
Infotainment systems, Bluetooth stacks, and cellular-connected telematics units have been
studied as attack surfaces since the landmark remote exploitation of an unmodified
production vehicle in 2015~\cite{miller2015remote}. OTA update infrastructure has
received increasing attention as its deployment has grown; systematic threat assessment
frameworks for OTA systems have been proposed~\cite{Mahmood2022,Iyieke2024}, and by 2024
manufacturers were resolving 33\% of all vehicle recalls through OTA updates rather than
physical service visits~\cite{Apriorit2025}. Cloud APIs connected to vehicle backend
systems represent a more recently identified and less thoroughly studied class of entry
point. In 2022, automotive API attacks grew by 380\% year-on-year~\cite{upstream2023},
yet systematic security evaluation of automotive API design, authentication mechanisms,
and access control policies remains limited relative to the scale of their deployment.
A recent report shows 92\% of 2024 incidents involved a remote vector and 60\% had
fleet-scale impact~\cite{upstream2025}. Cloud-facing interfaces, including mobile
applications, backend services, and telematics APIs, enable remote vehicle access yet
receive far less scrutiny than %
 IVNs. Systematic threat models and
longitudinal analyses of such APIs are largely missing~\cite{KIM2021102150}. Similarly,
OTA and over-the-top integrations, including third-party infotainment applications,
streaming services, and voice assistants, significantly expand the attack surface without
corresponding security evaluation. These off-board interfaces blur trust boundaries
between cloud, application, and in-vehicle domains, yet remain weakly represented in IVN
security taxonomies~\cite{10.1145/3431233}.

{
Electrification introduces an additional shift in the vehicle attack surface. Charging connects the vehicle to EVSE, charging-network backends, and grid-facing infrastructure, extending trust boundaries beyond conventional in-vehicle and telematics interfaces. As a result, vulnerabilities in EV--EVSE communication or charging-management protocols can propagate beyond an individual vehicle and affect charging infrastructure or grid operation \cite{garofalaki2022ocpp,mokarim2023evcs}.
}

The integration of sensing and AI-driven perception systems in advanced driver assistance
and automated driving functions introduces a class of entry point that is structurally
different from network-facing interfaces. Camera, LiDAR, radar, and GPS inputs feed
directly into ML inference pipelines that influence vehicle control decisions~\cite{Ibrahum2024}.
Physical-world manipulation of these inputs, including adversarial patches and laser illumination, has been demonstrated to produce misclassification in production perception
models without any network-layer access~\cite{Ibrahum2024,Badjie2024}. V2X communication
interfaces, deployed to support cooperative driving functions, extend the wireless attack
surface into the radio domain and introduce new trust dependencies on infrastructure
authenticity~\cite{Baccari2024}. Neither of these interface classes is systematically
addressed in current Threat Analysis and Risk Assessment (TARA) processes, which were
designed primarily around network-facing and physical-access entry points~\cite{ISOSAE21434}.
Beyond software interfaces, non-traditional hardware and external connectivity paths,
including cameras, radar, LiDAR, V2X receivers, and aftermarket OBD-II dongles, function
as indirect gateways into internal networks. The broader literature continues to emphasize
CAN-centric entry points, leaving these alternative surfaces underrepresented~\cite{PMCSurveyDL2023,Bari2023}. This
imbalance suggests that current threat models underestimate how attackers may exploit
heterogeneous and externally exposed components to achieve internal
access~\cite{info10040148,5504804}.

Across all entry point categories, existing studies tend to analyze interfaces in
isolation. Documented incidents show that exploitation frequently chains multiple entry
points: initial access through a web API or mobile application is followed by lateral
movement into the IVN through a gateway or ECU with insufficient access
controls~\cite{upstream2024,upstream2025}. Compound access path modeling that covers
the full chain from external interface to safety-critical IVN component is an identified
gap in the IVN security literature~\cite{miller2015remote}.

\subsection{Attack Methodologies: From Protocol Exploits to AI-Targeted Attacks}\label{sec:4.2-gaps-methods}

The dominant methodology in IVN attack research through the early 2020s was single-stage,
protocol-level exploitation: message injection, denial of service, fuzzing, replay, and
spoofing attacks directed at CAN traffic~\cite{PMCSurveyDL2023}. This body of work is
well-developed and has produced standardized datasets, reproducible experimental setups,
and a clear taxonomy of CAN-layer attack types. The corresponding defensive literature,
comprising ML-based IDS using Long Short-Term Memory (LSTM), Convolutional Neural Network (CNN), Gated Recurrent Unit (GRU), and transformer architectures, has grown
in parallel~\cite{PMCSurveyDL2023,Bari2023,Samir2024,NatureIDS2025}.

Attack methodologies targeting modern vehicles increasingly involve multi-stage,
cross-layer exploit chains, yet defensive models remain largely single-layer and reactive.
Many IVN security approaches assume attacks originate and remain within a single domain
(e.g., CAN message injection), despite real-world incidents demonstrating attack paths
that traverse from infotainment systems or telematics units into internal control networks.
Current intrusion detection frameworks rarely model or correlate these chained stages,
limiting their ability to detect attacks that unfold gradually across
subsystems~\cite{10.1145/3431233}. This gap is particularly pronounced for cross-layer
intrusion detection, where existing defenses tend to focus narrowly on either low-level
network traffic or high-level application behavior, leaving intermediate layers
insufficiently protected. Advanced surveys highlight that such fragmented approaches fail
to address the multidimensional nature of vehicular threats, especially as attackers
exploit interactions between software, network protocols, and hardware trust
assumptions~\cite{8688625,7393668}.

Two methodological directions have received less attention relative to their security
significance. The first is adversarial attacks on ML-based IDS models themselves.
Aloraini et al.~\cite{Alobaidi2024} demonstrated that black-box adversarial inputs
crafted to comply with realistic CAN traffic constraints reduced the F1 score of a
state-of-the-art IVN IDS from 97\% to 79\%, and of a baseline IDS from 95\% to 38\%,
while remaining undetected. The authors note that the IVN security literature has focused
heavily on IDS accuracy but has largely not evaluated IDS robustness under adversarial
conditions~\cite{Alobaidi2024}. As ML-based IDS becomes a standard component of
in-vehicle security architectures, this gap in evaluation methodology has direct
implications for deployment reliability.

The second direction concerns attacks on sensing and perception systems in vehicles
equipped with advanced driver assistance or automated driving functions. Adversarial
perturbations applied to camera images, LiDAR point clouds, or radar returns have been
shown to cause misclassification of road signs, obstacles, and lane markings in deployed
perception models~\cite{Ibrahum2024,Badjie2024}. These attacks differ structurally from
IVN protocol-level exploits: they target the data inputs to ML inference pipelines rather
than network messages, and their effect propagates to the IVN only through the control
commands issued by downstream planning and actuation modules. The two attack classes are
therefore not independent. A successful perception attack produces IVN-level effects, but
they are studied as separate research areas with limited cross-referencing~\cite{Ibrahum2024}.
Integrated threat models that represent the full chain from sensor manipulation to IVN
actuation remain an open problem. AI-targeted attack methodologies more broadly constitute
a distinct and undermodeled threat class:  stealthy falsification of sensor and V2V inputs to ACC/CACC controllers can induce unsafe control actions~\cite{9564677}. Few IVN security mechanisms explicitly address
malicious interference with machine learning pipelines or sensor-driven decision-making,
despite growing reliance on AI in autonomous and assisted driving
systems~\cite{info10040148,9564677}.

A third emerging direction is the use of large language models (LLMs) to automate threat
modeling over IVN topologies. DefenseWeaver~\cite{Yang2025DefenseWeaver} deploys a
multi-agent LLM framework that parses  component-level system configurations and generates function-level
attack trees for specific ECU configurations without relying on static threat libraries.
FEV's TARA Copilot~\cite{FEV2025TARACopilot} applies a similar LLM-based pipeline to
enumerate threat scenarios against UNECE R155 asset and attack surface categories,
including OBD-II diagnostic injection scenarios. Both systems address a documented
scalability problem: manual TARA over complex multi-ECU IVN topologies is labor-intensive
and does not scale to the lifecycle cadence required by ISO/SAE 21434~\cite{Zelle2022}.
The evaluation methodology for LLM-generated threat artifacts, including coverage
completeness and the detection of implausible or infeasible attack paths, has not, to
the best of our knowledge, been established in the literature.

\subsection{Impacts: Cascading, Systemic, and Fleet-Level Effects}\label{sec:4.3-gaps-impacts}

 The impact analyses reviewed in Section~\ref{sec:3.3-attack-impacts} have concentrated on individual-vehicle
scenarios: the consequences of compromised braking, steering, or powertrain control for
the occupants of a single vehicle. %
 Cross-domain propagation and simultaneous multi-vehicle effects receive comparatively little attention in these analyses. For example, an
intrusion originating in infotainment or telematics domains may propagate into powertrain
or ADAS subsystems, yet quantitative models capturing such cross-domain escalation are
rare~\cite{jayaratne2025quantitative}. Industry incident data from the same period shows
a different pattern. The proportion of high- or massive-scale incidents, specifically
those affecting thousands to millions of vehicles or mobility assets, approximately
doubled between 2022 and 2023, reaching nearly 50\% of all reported
incidents~\cite{EETimes2024}. By 2024, 60\% of incidents fell into this category, with
massive-scale events (affecting millions of assets) growing from 5\% to 19\% of the
total~\cite{upstream2025}. In 2024, 409 cybersecurity incidents were recorded across the
automotive sector, up from 295 in 2023~\cite{autoconnectedcar2025}. VicOne estimates the cost of automotive cyberattacks in 2024 at \$22.5\,billion~\cite{VicOne2025}.

The mechanism driving this scale increase is the concentration of vulnerable interfaces
at the backend rather than the vehicle level. Telematics servers, cloud APIs, and OTA
infrastructure serve large populations of vehicles through shared infrastructure; a single
backend compromise can affect every vehicle connected to it. In 2023, 37\% of threat
actor actions targeted multiple Original Equipment Manufacturers (OEMs) simultaneously~\cite{upstream2024}. One ransomware
incident in June 2024 disrupted a software provider serving approximately 15{,}000
automotive dealerships for %
 more than two weeks at an estimated cost of
\$1.02\,billion~\cite{autoconnectedcar2025}. In 2025, ransom-related incidents accounted
for 44\% of all reported automotive incidents,  more than double the 2024 volume~\cite{upstream2026}. Prevailing automotive safety standards emphasize qualitative,
vehicle-centric impact assessments and do not adequately account for fleet-level or
infrastructure-level effects~\cite{jayaratne2025quantitative}. Recent studies demonstrate that cyberattacks can induce
cascading safety incidents, such as telematics-triggered false braking events causing
multi-vehicle collisions, yet such scenarios remain insufficiently addressed by existing
frameworks~\cite{jayaratne2025quantitative}. There is also a lack of cyber-physical consequence modeling
that translates network-level compromises into measurable physical outcomes, including
crashes, injuries, and traffic disruption. Without integrated simulation and evaluation
tools, current assessments risk underestimating the real-world severity of IVN
attacks~\cite{jayaratne2025quantitative}.

These fleet-level impact patterns are not yet systematically represented in academic
impact models. Quantitative frameworks that trace the relationship between an initial
access event, such as a backend API compromise or an OTA distribution server breach,
and the resulting safety or operational consequences across a connected vehicle fleet
remain an open area~\cite{upstream2025,upstream2026}. For vehicles with advanced driving
automation, the impact surface expands further: attacks that manipulate V2X data used
for cooperative perception, or that corrupt shared ML model updates distributed
over-the-air, can in principle affect vehicle behavior across an entire fleet
simultaneously. Neither the impact propagation mechanisms nor the safety consequences of
such attacks have been formally characterized in the literature~\cite{Ibrahum2024,Baccari2024}.

\subsection{Mitigations: Detection Emphasis and Emerging Approaches}\label{sec:4.4-gaps-mitigations}

The four representative attack chains show that mitigation must extend beyond post-compromise detection: it must also include gateway containment and firmware integrity \cite{miller2015remote,niefreefall2017}, sensor resilience \cite{ying2023gps}, and backend and API authorization \cite{curry2024kia}.

Mitigation research in the IVN security literature remains dominated by intrusion detection, particularly deep-learning and graph-based IDS approaches applied to CAN traffic \cite{PMCSurveyDL2023,NatureIDS2025,IVNIDSSurvey2025}. These methods report high accuracy on standard benchmarks, but production deployment remains constrained by the latency, memory, and computation limits of automotive ECUs and by limited validation on production-grade platforms \cite{8688625,5940525}. More recent work moves beyond message-level analysis toward relational and knowledge-graph-based approaches that model dependencies among signals and vehicle entities \cite{moriano2022masquerade,gao2025signal,sun2025kgid,alqahtani2025deep}, though systematic evaluation under production constraints remains limited.

\paragraph{Adversarial robustness of IDS}
Published IDS evaluations predominantly use clean, in-distribution test data \cite{IVNIDSSurvey2025}. Aloraini et al.~\cite{Alobaidi2024} showed that this practice conceals significant vulnerabilities: models that achieve high detection rates on standard benchmarks can be defeated by adversarially crafted inputs consistent with realistic IVN traffic. Adversarial robustness remains uncommon as an explicit design and evaluation requirement in the IVN security literature \cite{Alobaidi2024,Baccari2024}.

\paragraph{Federated learning for distributed IDS}
Raw CAN traffic is commercially sensitive, and cross-OEM data sharing is not established practice. Federated learning has been proposed to enable collaborative IDS training while keeping data local, and demonstrated systems exist \cite{Althunayyan2024HFL}. Cross-platform generalization and multi-protocol coverage remain the primary open problems: models trained on one vehicle platform generalize poorly to others, and no evaluation to date covers both CAN and Automotive Ethernet traffic simultaneously \cite{IVNIDSSurvey2025}.

\label{sec:architectural-evolution-ids}
\paragraph{Architectural evolution and IDS adaptation}

The transition from CAN-centric IVN architectures to Ethernet, TSN, and zonal backbones has proceeded faster than the corresponding adaptation of IDS research. Automotive Ethernet aggregates higher-volume, heterogeneous traffic, including IP services, diagnostics, sensor streams, and control functions, over shared infrastructure \cite{martinezcruz2021security,AESystematicReview2024}, while centralized gateways and zonal controllers increase both enforcement opportunities (e.g., SecOC, MACsec, IPsec, TLS \cite{lauser2024security}) and the potential blast radius of a compromise. IDS must therefore move beyond CAN-only, single-bus monitoring toward architectures capable of detecting cross-domain lateral movement \cite{Haeckel2025}.

Future detection architectures must be multi-protocol and topology-aware, correlating behavior across CAN/CAN FD, Automotive Ethernet, gateways, and zonal controllers while preserving real-time performance and distinguishing malicious behavior from legitimate high-volume traffic. Empirical IDS evaluation under real, production-like Ethernet IVN conditions remains largely undocumented \cite{Mozzaquatro2023}.

\paragraph{Post-quantum cryptography}
Long vehicle operational lifetimes (a vehicle manufactured in 2025 may remain in service beyond 2040) motivate quantum-resistant cryptography for IVN security \cite{Chillotti2021}, but adoption is constrained by the bandwidth and payload limits of protocols such as CAN, where large keys must be fragmented across messages \cite{smith2025post}. NIST finalized initial post-quantum standards in August 2024: ML-KEM (formerly CRYSTALS-Kyber), ML-DSA (formerly CRYSTALS-Dilithium), and SLH-DSA (formerly SPHINCS\texttt{+}). Automotive evaluation remains early-stage, for example a segmented CAN architecture using CRYSTALS-Kyber for session-key establishment \cite{castiglione2025securing}, and automotive-specific migration and compliance guidance remains immature: no dedicated PQC requirements yet exist for automotive use, including within ISO/SAE 21434, and migration paths for HSMs not designed for post-quantum key sizes remain unaddressed \cite{Apriorit2025PQC}.

\paragraph{Resilience and recovery}
The defenses reviewed in Section~\ref{sec:3.4-attack-mitigations} concentrate largely on detection, with limited emphasis on coordinated response, containment, reconfiguration, redundancy, and safe recovery. Comprehensive, vehicle-wide resilience architectures remain comparatively underdeveloped, a gap analogous to the lack of defense-in-depth models in traditional IT systems \cite{5948952}.

\paragraph{Regulatory and standardization landscape}

The regulatory framework for automotive cybersecurity has developed substantially since 2021. ISO/SAE 21434 defines engineering requirements for cybersecurity risk management across the vehicle lifecycle \cite{ISOSAE21434}, and UNECE R155 (cybersecurity management) and R156 (software-update management) are now mandatory for new vehicles \cite{WP29R155}. Together, these instruments have driven OEMs and Tier-1 suppliers to establish formal Cybersecurity Management Systems and increasingly automated TARA processes, including LLM-based TARA pipelines that address the scalability limits of manual TARA over complex IVN topologies, as conditions of market access \cite{Zelle2022,Yang2025DefenseWeaver,FEV2025TARACopilot,Zhu2024,Seo2024}.

Current standards primarily address software and network cybersecurity processes: ISO/SAE 21434 and UNECE R155 do not specify requirements for the integrity of ML training data, the authenticity of OTA model updates, or the security of V2X inputs used in automated-driving functions, and ISO 21448 (SOTIF) was not designed to cover adversarial manipulation of perception inputs \cite{Zhu2024,Ibrahum2024}. Validating the completeness and correctness of AI-generated TARA artifacts, including adversarial evaluation against known attack scenarios, is a related emerging open problem.

\subsection{{Cybersecurity Testbeds and Experimental Evaluation}}
\label{sec:testbeds}

{
Automotive cybersecurity testbeds provide a controlled environment for reproducing attacks and evaluating defenses without the safety and access constraints of experiments on production vehicles. Their design has evolved substantially. Tuohy et al. combine real and simulated traffic in a hybrid environment, enabling repeatable network-level experimentation \cite{tuohy2016hybrid}. PASTA instead uses configurable white-box ECUs and simulated vehicle functions to provide an open and adaptable platform for security evaluation and training \cite{toyama2018pasta}. Shi et al. use real vehicle traces to reproduce CAN behavior and evaluate six attack scenarios, providing greater correspondence with observed vehicle traffic while remaining primarily CAN-focused \cite{shi2022cantestbed}. VitroBench increases hardware fidelity through COTS ECUs and supports sniffing, injection, isolation, fuzzing, and attack reproduction across CAN, CAN FD, and LIN environments \cite{yeo2023vitrobench}. VCE/CLaaS extends testbed access through cloud-connected laboratory infrastructure, allowing students and researchers to execute automotive cybersecurity experiments remotely \cite{zachos2024vce}. At a broader level, simulation and co-simulation frameworks have also been used to study how a vehicle-level cyberattack propagates into transportation-network effects \cite{jayaratne2024simulation}.
}

\begingroup

\begin{table*}[!htbp]
\centering
\caption{Comparison of representative automotive cybersecurity testbeds and simulation environments.}
\label{table:automotive-testbeds}
\small
\setlength{\tabcolsep}{4pt}
\begin{tabularx}{\textwidth}{>{\raggedright\arraybackslash}p{2.2cm} >{\raggedright\arraybackslash}p{2.6cm} >{\raggedright\arraybackslash}X >{\raggedright\arraybackslash}X}
\hline
\textbf{Platform} &
\textbf{Environment} &
\textbf{Primary Capability} &
\textbf{Evaluation / Limitation} \\
\hline

Tuohy et al. \cite{tuohy2016hybrid}
&
Hybrid simulation
&
Combines real and simulated network traffic for controlled IVN experiments.
&
Repeatable and flexible, but changes the structure and operating environment of the original IVN.
\\

PASTA \cite{toyama2018pasta}
&
Configurable white-box ECUs
&
Supports modification of ECU firmware, IVN configuration, and security mechanisms without physical vehicle actuators.
&
Open and adaptable for research and training, but does not reproduce the full complexity and proprietary behavior of production vehicles.
\\

Shi et al. \cite{shi2022cantestbed}
&
Real-data CAN testbed
&
Replays real vehicle traces and supports six controlled CAN attack scenarios.
&
Preserves realistic time-series behavior, but evaluation is primarily limited to CAN and the supplied vehicle traces.
\\

VitroBench \cite{yeo2023vitrobench}
&
COTS ECU testbed
&
Supports sniffing, packet injection, ECU isolation, fuzzing, and attack reproduction using production ECUs.
&
Provides higher hardware fidelity and reproducibility, but remains dependent on the available ECU and network configuration.
\\

VCE/CLaaS \cite{zachos2024vce}
&
Remote hardware testbed
&
Provides cloud-based access to automotive cybersecurity experiments and programmable OBD-II vehicle simulation.
&
Improves accessibility and training support, but experiments remain bounded by the laboratory hardware exposed through the remote platform.
\\

Jayaratne et al. \cite{jayaratne2024simulation}
&
Co-simulation
&
Connects vehicle cyberattack scenarios with transportation-network simulation to evaluate systemic consequences.
&
Captures effects beyond an individual vehicle, but abstracts ECU- and IVN-level implementation details.
\\

\hline
\end{tabularx}
\end{table*}
\endgroup

{
These platforms expose a tradeoff among realism, reproducibility, accessibility, and experimental scale. Software and co-simulation environments provide repeatable and scalable experiments but abstract hardware-specific behavior. Hardware-assisted and COTS-ECU platforms improve fidelity but increase cost, platform dependence, and configuration complexity. Remote-access environments improve accessibility and training but remain constrained by the capabilities of the underlying physical testbed. Existing reviews similarly show that automotive cybersecurity testbeds remain heterogeneous in architecture, objectives, and evaluation methodology \cite{luo2022cybersecuritytesting}. The remaining gap is therefore not the absence of testbeds, but the lack of standardized environments in which the same end-to-end attack can be reproduced across heterogeneous architectures. Common attack scenarios, datasets or traces, metrics, and experiment descriptions are needed to enable comparison across CAN/CAN FD, Automotive Ethernet, wireless interfaces, sensing systems, EV charging interfaces, and backend services.
}

\subsection*{Outlook and Research Directions}

Considered together, the trends documented in this section indicate that the IVN
security problem has expanded beyond the protocol-level characterization that defined
its early research phase. The attack surface now includes remote and cloud-connected
interfaces, AI-driven perception and decision-making pipelines, and shared backend
infrastructure serving large vehicle populations. The scale of documented incidents
reflects this expansion: the automotive industry recorded 409 cybersecurity incidents in
2024, up 39\% from 2023, with a cumulative total of  1,877 documented cases
since 2010~\cite{autoconnectedcar2025}. %

 {The research gaps identified across the taxonomy and related cross-cutting areas translate
into a set of concrete directions for future work.} These are grounded in documented deficiencies
rather than extrapolation.

The most significant structural gap in the mitigation dimension is the absence of IDS
architectures that span multiple IVN protocols and integrate network-layer signals with
sensing and perception data. Current IDS proposals operate within a single protocol
layer, predominantly CAN, and do not cover attacks that originate in sensor inputs or
ML inference pipelines and propagate to the IVN through vehicle control
logic~\cite{IVNIDSSurvey2025,Alobaidi2024}. Developing detection frameworks that
operate across this full chain, and that are evaluated for robustness against adversarial
evasion, addresses the two most directly documented gaps in the IDS literature.

In the entry point dimension, compound access path modeling covering the chain from
external interface to safety-critical IVN component is not yet established. Systematic
security evaluation of cloud API authentication, OTA supply chain integrity, and V2X
data authenticity in automotive deployments has each been identified as an area where
analysis does not match incident data on attack
frequency~\cite{upstream2024,upstream2025,Mahmood2022}.

In the mitigation dimension, federated IDS training has been demonstrated in principle
but not at a scale or across a platform diversity that would support cross-OEM deployment.
The open problems of cross-platform generalization and multi-protocol coverage, documented
in a 2025 IVN IDS survey~\cite{IVNIDSSurvey2025}, define the minimum requirements for
FL-based IDS to provide fleet-level security coverage.

 {As a cross-cutting mitigation issue}, the mismatch between the operational lifetimes of deployed
vehicles and the timeline of post-quantum cryptographic standardization is a documented
and quantifiable risk~\cite{Chillotti2021,Apriorit2025PQC}. Automotive-specific PQC
migration frameworks that address ECU hardware constraints, AUTOSAR integration, and
real-time latency requirements have not yet been produced by the research community,
despite the availability of NIST-standardized algorithms since 2024.

 {As a cross-cutting regulatory issue}, extending coverage to the security of ML model update
pipelines and V2X data integrity for automated driving functions addresses a gap that
has been explicitly identified in safety-security co-analysis
research~\cite{Zhu2024,Ibrahum2024}. The existing ISO/SAE 21434 and UNECE R155
frameworks do not specify requirements for these areas, and the corresponding threat
scenarios are not covered by current TARA practice.

Addressing these gaps requires both technical contributions, including new detection
architectures, protocol-spanning evaluation datasets, and PQC integration studies, as
well as methodological contributions, including adversarial evaluation standards for
ML-based IDS and validation frameworks for AI-generated TARA artifacts. The vehicle
population to which these contributions must eventually apply is large and growing: an
estimated 95\% of new vehicles will be connected by 2030~\cite{Yang2025DefenseWeaver},
and the security properties of their IVN components will increasingly determine the
safety of automated driving functions.

{
Table~\ref{table:research_directions} consolidates the principal gaps identified above and the corresponding directions for future research.
}

\begin{table*}[!htbp]
\centering
\caption{Summary of research directions identified in this survey.}
\label{table:research_directions}
\small
\renewcommand{\arraystretch}{1.15}

\begin{tabularx}{\textwidth}{|p{3.0cm}|X|X|}
\hline

\textbf{Research Area} &
\textbf{Current Gap} &
\textbf{Research Direction} \\
\hline

Entry points and trust boundaries &
Cloud APIs, OTA supply chains, V2X interfaces, and compound access paths remain less studied than CAN-centric entry points \cite{upstream2024,upstream2025,Mahmood2022}. &
Model complete external-to-IVN attack paths. Evaluate authorization, ownership validation, update trust, and V2X authenticity. \\
\hline

Attack methodologies &
Multi-stage and cross-layer attacks remain weakly represented. Adversarial attacks on ML defenses and perception pipelines are also under-evaluated \cite{Alobaidi2024,Ibrahum2024}. &
Correlate attacks across application, network, sensing, and control layers. Evaluate defenses under adaptive and adversarial conditions. \\
\hline

Operational impacts &
Fleet-scale, infrastructure-level, and cyber-physical consequences are not quantified consistently \cite{jayaratne2025quantitative,upstream2025}. &
Connect initial compromise to vehicle safety, traffic disruption, fleet effects, and infrastructure impact. \\
\hline

Detection and mitigation &
IDS evaluation remains largely CAN-centric. Cross-protocol, cross-platform, and cross-OEM generalization is limited \cite{IVNIDSSurvey2025,Althunayyan2024HFL}. &
Develop protocol-spanning and cross-domain detection. Evaluate generalization, adversarial robustness, and real-time deployment. \\
\hline

Experimental evaluation &
Testbeds remain heterogeneous, limiting reproduction of the same end-to-end attack across architectures \cite{luo2022cybersecuritytesting}. &
Standardize attack scenarios, datasets or traces, metrics, and experiment descriptions across heterogeneous vehicle systems. \\
\hline

Cross-cutting: architecture, cryptography, and assurance &
Ethernet/zonal architectures, PQC migration, ML update security, and V2X integrity introduce requirements not fully covered by current assurance practice \cite{AESystematicReview2024,Apriorit2025PQC,Zhu2024,Ibrahum2024}. &
Evaluate layered Ethernet and zonal protections. Develop automotive PQC migration paths. Extend TARA and assurance to ML updates and V2X data integrity. \\
\hline

\end{tabularx}
\end{table*}

\FloatBarrier

\section{Conclusions}\label{sec:6-conclusions}

As vehicles transition into highly connected, software-centric platforms, the cybersecurity assumptions underlying %
 IVNs no longer hold. Increased connectivity through telematics, cloud services, and OTA updates has expanded viable entry points, enabling attackers to reach internal systems through indirect, low-complexity access paths~\cite{upstream2023,upstream2024}.

While intrusion detection and CAN-focused defenses are well studied, comparatively less work addresses cross-layer attacks, fleet-level cascading impacts, real-time response, and post-quantum cryptography for IVNs. Real-time safety requirements conflict with advanced detection overhead, OEM fragmentation prevents unified standards, and long vehicle lifecycles (10-15 years) lock in legacy protocols despite accumulated vulnerabilities.

Securing next-generation vehicles therefore will require system-level designs that account for interaction effects, real-time constraints, and evolving architectures. This taxonomy provides researchers a structured approach to cross-layer threat modeling, OEMs a framework for identifying defense gaps, and policymakers guidance on evolving standards beyond functional safety to address cyber-physical interactions. The framework and bibliography presented here aim to guide future research toward resilient, deployable security architectures for increasingly connected and autonomous vehicles.

\section{Declaration of generative AI and AI-assisted technologies in the manuscript preparation process}

During the preparation of this work the authors used Claude and OpenAI models for grammar checks and adversarial reviews. After using this tool/service, the authors reviewed and edited the content as needed and take full responsibility for the content of the published article.

\bibliographystyle{elsarticle-num}
\bibliography{references-new, references-major-new, references}

\begin{thebibliography}{100}
\expandafter\ifx\csname url\endcsname\relax
  \def\url#1{\texttt{#1}}\fi
\expandafter\ifx\csname urlprefix\endcsname\relax\def\urlprefix{URL }\fi
\expandafter\ifx\csname href\endcsname\relax
  \def\href#1#2{#2} \def\path#1{#1}\fi

\bibitem{rathore2022vehicle}
R.~S. Rathore, C.~Hewage, O.~Kaiwartya, J.~Lloret, In-vehicle communication
  cyber security: challenges and solutions, Sensors 22~(17) (2022) 6679.

\bibitem{canino2025cybersecurity}
N.~Canino, P.~Dini, S.~Mazzetti, D.~Rossi, S.~Saponara, E.~Soldaini,
  Cybersecurity of automotive wired networking systems: Evolution, challenges,
  and countermeasures, Electronics 14~(3) (2025) 471.

\bibitem{george2025cyber}
D.~George, S.~Pavithra, J.~Das, Cyber-resilient autonomous vehicles: Securing
  networks and enhancing decision-making with next-gen security measures,
  Results in Engineering 28 (2025) 107179.

\bibitem{santa2025automotive}
V.~S. Barletta, D.~Caivano, C.~Catalano, M.~De~Vincentiis, M.~Scalera,
  Automotive intelligence for supporting vehicle-soc analysts, Computer
  Standards \& Interfaces 96 (2025) 104088.

\bibitem{kousar2025deep}
A.~Kousar, S.~Ahmed, Z.~A. Khan, A deep learning approach for real-time
  intrusion mitigation in automotive controller area networks, World Electric
  Vehicle Journal 16~(9) (2025) 492.

\bibitem{christy2025machine}
C.~Christy, A.~Nirmala, A.~M.~O. Teena, A.~I. Amali, Machine learning based
  multi-stage intrusion detection system and feature selection ensemble
  security in cloud assisted vehicular ad hoc networks, Scientific Reports
  15~(1) (2025) 27058.

\bibitem{jayaratne2025quantitative}
D.~N.~D. Jayaratne, Q.~Lu, A.~Rakib, M.~A. Ramli, R.~M. Mepparambath, S.~A.
  Shaikh, H.~N. Nguyen, A quantitative methodology for systemic impact
  assessment of cyber threats in connected vehicles, Computers \& Security 160
  (2026) 104729.

\bibitem{checkoway2011comprehensive}
S.~Checkoway, D.~McCoy, B.~Kantor, D.~Anderson, H.~Shacham, S.~Savage,
  K.~Koscher, A.~Czeskis, F.~Roesner, T.~Kohno, Comprehensive experimental
  analyses of automotive attack surfaces, in: 20th USENIX security symposium
  (USENIX Security 11), 2011.

\bibitem{wright2011hacking}
A.~Wright, Hacking cars, Communications of the ACM 54~(11) (2011) 18--19.

\bibitem{6615528}
I.~Studnia, V.~Nicomette, E.~Alata, Y.~Deswarte, M.~Kaâniche, Y.~Laarouchi,
  Survey on security threats and protection mechanisms in embedded automotive
  networks, in: 2013 43rd Annual IEEE/IFIP Conference on Dependable Systems and
  Networks Workshop (DSN-W), 2013, pp. 1--12.

\bibitem{10.1145/3431233}
E.~Aliwa, O.~Rana, C.~Perera, P.~Burnap, Cyberattacks and countermeasures for
  in-vehicle networks, ACM Comput. Surv. 54~(1) (2021) Article 21.
\newblock \href {https://doi.org/10.1145/3431233} {\path{doi:10.1145/3431233}}.

\bibitem{info10040148}
F.~Sommer, J.~Dürrwang, R.~Kriesten, Survey and classification of automotive
  security attacks, Information 10~(4) (2019).
\newblock \href {https://doi.org/10.3390/info10040148}
  {\path{doi:10.3390/info10040148}}.

\bibitem{DBLP:journals/corr/abs-1802-01725}
O.~Avatefipour, H.~Malik, State-of-the-art survey on in-vehicle network
  communication (can-bus) security and vulnerabilities, CoRR abs/1802.01725
  (2018).
\newblock \href {http://arxiv.org/abs/1802.01725} {\path{arXiv:1802.01725}}.

\bibitem{8688625}
W.~Wu, R.~Li, G.~Xie, J.~An, Y.~Bai, J.~Zhou, K.~Li, A survey of intrusion
  detection for in-vehicle networks, IEEE Transactions on Intelligent
  Transportation Systems 21~(3) (2020) 919--933.

\bibitem{7393668}
W.~Zeng, M.~A.~S. Khalid, S.~Chowdhury, In-vehicle networks outlook:
  Achievements and challenges, IEEE Communications Surveys \& Tutorials 18~(3)
  (2016) 1552--1571.

\bibitem{LE201817}
V.~H. Le, J.~{den Hartog}, N.~Zannone, Security and privacy for innovative
  automotive applications: A survey, Computer Communications 132 (2018) 17--41.

\bibitem{KIM2021102150}
K.~Kim, J.~S. Kim, S.~Jeong, J.-H. Park, H.~K. Kim, Cybersecurity for
  autonomous vehicles: Review of attacks and defense, Computers \& Security 103
  (2021) 102150.

\bibitem{9463461}
F.~Tang, B.~Mao, N.~Kato, G.~Gui, Comprehensive survey on machine learning in
  vehicular network: Technology, applications and challenges, IEEE
  Communications Surveys \& Tutorials 23~(3) (2021) 2027--2057.

\bibitem{5948952}
G.~Karagiannis, O.~Altintas, E.~Ekici, G.~Heijenk, B.~Jarupan, K.~Lin, T.~Weil,
  Vehicular networking: A survey and tutorial on requirements, architectures,
  challenges, standards and solutions, IEEE Communications Surveys \& Tutorials
  13~(4) (2011) 584--616.

\bibitem{castiglione2025securing}
A.~Castiglione, T.~Elia, Securing in-vehicle communications through
  post-quantum cryptography and network segmentation, Computers and Electrical
  Engineering 126 (2025) 110488.

\bibitem{smith2025post}
S.~Smith, K.~Owens, K.~Kozan, Post quantum cryptography on intravehicular
  networks, Tech. Rep. 2025-01-0451, SAE Technical Paper (2025).
\newblock \href {https://doi.org/10.4271/2025-01-0451}
  {\path{doi:10.4271/2025-01-0451}}.

\bibitem{miller2015remote}
C.~Miller, C.~Valasek,
  \href{https://www.ioactive.com/pdfs/IOActive_Remote_Car_Hacking.pdf}{Remote
  exploitation of an unaltered passenger vehicle}, Black Hat USA 2015 white
  paper, IOActive (2015).
\newline\urlprefix\url{https://www.ioactive.com/pdfs/IOActive_Remote_Car_Hacking.pdf}

\bibitem{niefreefall2017}
S.~Nie, L.~Liu, Y.~Du, Free-fall: Hacking tesla from wireless to can bus, Black
  Hat USA 2017 Briefings white paper (2017).

\bibitem{ying2023gps}
J.~Ying, Y.~Feng, Q.~A. Chen, Z.~Mao, Gps spoofing attack detection on
  intersection movement assist using one-class classification, in: Symposium on
  Vehicle Security and Privacy (VehicleSec), 2023.

\bibitem{curry2024kia}
N.~Rivera, S.~Curry, J.~Rhinehart, I.~Carroll, K.~Lugo, Hacking kia: Remotely
  controlling cars with just a license plate,
  \url{https://samcurry.net/hacking-kia}, 20 Sep 2024.

\bibitem{5940525}
P.~Kleberger, T.~Olovsson, E.~Jonsson, Security aspects of the in-vehicle
  network in the connected car, in: 2011 IEEE Intelligent Vehicles Symposium
  (IV), 2011, pp. 528--533.

\bibitem{9316663}
A.~Lamssaggad, N.~Benamar, A.~S. Hafid, M.~Msahli, A survey on the current
  security landscape of intelligent transportation systems, IEEE Access 9
  (2021) 9180--9208.

\bibitem{9619465}
A.~Talpur, M.~Gurusamy, Machine learning for security in vehicular networks: A
  comprehensive survey, IEEE Communications Surveys \& Tutorials 24~(1) (2022)
  346--379.

\bibitem{limbasiya2022systematic}
T.~Limbasiya, K.~Z. Teng, S.~Chattopadhyay, J.~Zhou, A systematic survey of
  attack detection and prevention in connected and autonomous vehicles,
  Vehicular Communications 37 (2022) 100515.
\newblock \href {https://doi.org/10.1016/j.vehcom.2022.100515}
  {\path{doi:10.1016/j.vehcom.2022.100515}}.

\bibitem{karopoulos2022demystifying}
G.~Karopoulos, G.~Kambourakis, E.~Chatzoglou, J.~L. Hern{\'a}ndez-Ramos,
  V.~Kouliaridis, Demystifying in-vehicle intrusion detection systems: A survey
  of surveys and a meta-taxonomy, Electronics 11~(7) (2022) 1072.
\newblock \href {https://doi.org/10.3390/electronics11071072}
  {\path{doi:10.3390/electronics11071072}}.

\bibitem{kifor2024automotive}
C.~V. Kifor, A.~Popescu, Automotive cybersecurity: A survey on frameworks,
  standards, and testing and monitoring technologies, Sensors 24~(18) (2024)
  6139.
\newblock \href {https://doi.org/10.3390/s24186139}
  {\path{doi:10.3390/s24186139}}.

\bibitem{thing2016autonomous}
V.~L.~L. Thing, J.~Wu, Autonomous vehicle security: A taxonomy of attacks and
  defences, in: 2016 IEEE International Conference on Internet of Things
  (iThings) and IEEE Green Computing and Communications (GreenCom) and IEEE
  Cyber, Physical and Social Computing (CPSCom) and IEEE Smart Data
  (SmartData), IEEE, 2016, pp. 164--170.
\newblock \href
  {https://doi.org/10.1109/iThings-GreenCom-CPSCom-SmartData.2016.52}
  {\path{doi:10.1109/iThings-GreenCom-CPSCom-SmartData.2016.52}}.

\bibitem{martinezcruz2021security}
A.~Mart{\'i}nez-Cruz, K.~A. Ram{\'i}rez-Guti{\'e}rrez, C.~Feregrino-Uribe,
  A.~Morales-Reyes, Security on in-vehicle communication protocols: Issues,
  challenges, and future research directions, Computer Communications 180
  (2021) 1--20.
\newblock \href {https://doi.org/10.1016/j.comcom.2021.08.027}
  {\path{doi:10.1016/j.comcom.2021.08.027}}.

\bibitem{sommer2024combining}
F.~Sommer, M.~Gierl, R.~Kriesten, F.~Kargl, E.~Sax, Combining cyber security
  intelligence to refine automotive cyber threats, ACM Transactions on Privacy
  and Security 27~(2) (2024) 1--34.
\newblock \href {https://doi.org/10.1145/3644075} {\path{doi:10.1145/3644075}}.

\bibitem{AESystematicReview2024}
M.~De~Vincenzi, G.~Costantino, I.~Matteucci, F.~Fenzl, C.~Plappert, R.~Rieke,
  D.~Zelle, A systematic review on security attacks and countermeasures in
  automotive {Ethernet}, ACM Computing Surveys 56~(6) (2024).
\newblock \href {https://doi.org/10.1145/3637059} {\path{doi:10.1145/3637059}}.

\bibitem{singh2017security}
M.~Singh, S.~Kim, Security analysis of intelligent vehicles: Challenges and
  scope, in: 2017 International SoC Design Conference (ISOCC), 2017, pp.
  13--14.
\newblock \href {https://doi.org/10.1109/ISOCC.2017.8368805}
  {\path{doi:10.1109/ISOCC.2017.8368805}}.

\bibitem{monteuuis2018sara}
J.-P. Monteuuis, A.~Boudguiga, J.~Zhang, H.~Labiod, A.~Servel, P.~Urien,
  {SARA}: Security automotive risk analysis method, in: Proceedings of the 4th
  ACM Workshop on Cyber-Physical System Security, 2018, pp. 3--14.
\newblock \href {https://doi.org/10.1145/3198458.3198465}
  {\path{doi:10.1145/3198458.3198465}}.

\bibitem{autoisac2024atm}
{Automotive Information Sharing and Analysis Center (Auto-ISAC)}, Automotive
  threat matrix ({ATM}), \url{https://atm.automotiveisac.com/}, accessed:
  September 17, 2026 (2024).

\bibitem{howard2006sdl}
M.~Howard, S.~Lipner, The Security Development Lifecycle: {SDL}, a Process for
  Developing Demonstrably More Secure Software, Microsoft Press, Redmond, WA,
  USA, 2006.

\bibitem{strom2020attack}
B.~E. Strom, A.~Applebaum, D.~P. Miller, K.~C. Nickels, A.~G. Pennington, C.~B.
  Thomas, {MITRE ATT\&CK}: Design and philosophy, Tech. Rep. MP180360R1, The
  MITRE Corporation, McLean, VA (2020).

\bibitem{hutchins2011killchain}
E.~M. Hutchins, M.~J. Cloppert, R.~M. Amin, Intelligence-driven computer
  network defense informed by analysis of adversary campaigns and intrusion
  kill chains, in: Proceedings of the 6th International Conference on
  Information Warfare and Security, 2011, pp. 113--125.

\bibitem{tuohy2016hybrid}
S.~Tuohy, M.~Glavin, E.~Jones, C.~Hughes, L.~Kilmartin, Hybrid testbed for
  simulating in-vehicle automotive networks, Simulation Modelling Practice and
  Theory 66 (2016) 193--211.
\newblock \href {https://doi.org/10.1016/j.simpat.2016.01.001}
  {\path{doi:10.1016/j.simpat.2016.01.001}}.

\bibitem{toyama2018pasta}
T.~Toyama, T.~Yoshida, H.~Oguma, T.~Matsumoto, {PASTA}: Portable automotive
  security testbed with adaptability, in: Black Hat Europe 2018, London, UK,
  2018, pp. 1--10.

\bibitem{shi2022cantestbed}
D.~Shi, L.~Kou, C.~Huo, T.~Wu, \href{https://doi.org/10.1155/2022/7176194}{A
  {CAN} bus security testbed framework for automotive cyber-physical systems},
  Wireless Communications and Mobile Computing 2022 (2022) 7176194.
\newblock \href {https://doi.org/10.1155/2022/7176194}
  {\path{doi:10.1155/2022/7176194}}.
\newline\urlprefix\url{https://doi.org/10.1155/2022/7176194}

\bibitem{yeo2023vitrobench}
A.~K.~T. Yeo, M.~E. Garbelini, S.~Chattopadhyay, J.~Zhou,
  \href{https://doi.org/10.1016/j.vehcom.2023.100649}{{VitroBench}:
  Manipulating in-vehicle networks and {COTS} {ECUs} on your bench: A
  comprehensive test platform for automotive cybersecurity research}, Vehicular
  Communications 43 (2023) 100649.
\newblock \href {https://doi.org/10.1016/j.vehcom.2023.100649}
  {\path{doi:10.1016/j.vehcom.2023.100649}}.
\newline\urlprefix\url{https://doi.org/10.1016/j.vehcom.2023.100649}

\bibitem{zachos2024vce}
M.~Zachos, P.~Satam, R.~Naama, Vehicle cyber engineering ({VCE}) testbed with
  {CLaaS} (cyber-security labs as a service), SAE Technical Papers (2024).
\newblock \href {https://doi.org/10.4271/2024-01-2796}
  {\path{doi:10.4271/2024-01-2796}}.

\bibitem{wang2021systematicrisk}
Y.~Wang, Y.~Wang, H.~Qin, H.~Ji, Y.~Zhang, J.~Wang,
  \href{https://doi.org/10.1007/s42154-021-00140-6}{A systematic risk
  assessment framework of automotive cybersecurity}, Automotive Innovation
  4~(3) (2021) 253--261.
\newblock \href {https://doi.org/10.1007/s42154-021-00140-6}
  {\path{doi:10.1007/s42154-021-00140-6}}.
\newline\urlprefix\url{https://doi.org/10.1007/s42154-021-00140-6}

\bibitem{widel2023mal}
W.~Widel, S.~Hacks, M.~Ekstedt, P.~Johnson, R.~Lagerstr{\"o}m,
  \href{https://doi.org/10.1016/j.cose.2023.103284}{The meta attack language --
  a formal description}, Computers \& Security 130 (2023) 103284.
\newblock \href {https://doi.org/10.1016/j.cose.2023.103284}
  {\path{doi:10.1016/j.cose.2023.103284}}.
\newline\urlprefix\url{https://doi.org/10.1016/j.cose.2023.103284}

\bibitem{katsikeas2022vehiclelang}
S.~Katsikeas, P.~Johnsson, S.~Hacks, R.~Lagerstr{\"o}m,
  \href{https://doi.org/10.1016/j.cose.2022.102705}{{VehicleLang}: A
  probabilistic modeling and simulation language for modern vehicle {IT}
  infrastructures}, Computers \& Security 117 (2022) 102705.
\newblock \href {https://doi.org/10.1016/j.cose.2022.102705}
  {\path{doi:10.1016/j.cose.2022.102705}}.
\newline\urlprefix\url{https://doi.org/10.1016/j.cose.2022.102705}

\bibitem{jayaratne2024simulation}
D.~N.~D. Jayaratne, S.~H. Kamtam, S.~A. Shaikh, M.~A. Ramli, Q.~Lu, R.~M.
  Mepparambath, H.~N. Nguyen, A.~Rakib,
  \href{https://doi.org/10.1016/j.simpat.2024.103005}{A simulation framework
  for automotive cybersecurity risk assessment}, Simulation Modelling Practice
  and Theory 136 (2024) 103005.
\newblock \href {https://doi.org/10.1016/j.simpat.2024.103005}
  {\path{doi:10.1016/j.simpat.2024.103005}}.
\newline\urlprefix\url{https://doi.org/10.1016/j.simpat.2024.103005}

\bibitem{8370879}
X.~Li, Y.~Yu, G.~Sun, K.~Chen, Connected vehicles' security from the
  perspective of the in-vehicle network, IEEE Network 32~(3) (2018) 58--63.

\bibitem{6945510}
S.~Seifert, R.~Obermaisser, Secure automotive gateway — secure communication
  for future cars, in: 2014 12th IEEE International Conference on Industrial
  Informatics (INDIN), 2014, pp. 213--220.

\bibitem{chevybolt_usbhub}
{Chevy Bolt Forum}, Center console removal and hidden usb hub,
  \url{https://www.chevybolt.org/threads/center-console-removal-and-hidden-usb-hub.22233/},
  accessed: Feb. 1, 2026 (2020).

\bibitem{hu2018review}
Q.~Hu, F.~Luo, Review of secure communication approaches for in-vehicle
  network, International Journal of Automotive Technology 19 (2018) 879--894.

\bibitem{6513548}
F.~Sagstetter, M.~Lukasiewycz, S.~Steinhorst, M.~Wolf, A.~Bouard, W.~R. Harris,
  S.~Jha, T.~Peyrin, A.~Poschmann, S.~Chakraborty, Security challenges in
  automotive hardware/software architecture design, in: 2013 Design, Automation
  \& Test in Europe Conference \& Exhibition (DATE), 2013, pp. 458--463.

\bibitem{sturgess2023cableauth}
J.~Sturgess, S.~K{\"o}hler, S.~Birnbach, I.~Martinovic, Cableauth: a biometric
  second factor authentication scheme for electric vehicle charging, in:
  Symposium on Vehicle Security and Privacy (VehicleSec), Internet Society,
  2023.
\newblock \href {https://doi.org/10.14722/vehiclesec.2023.23041}
  {\path{doi:10.14722/vehiclesec.2023.23041}}.

\bibitem{chademo2020ocpp}
{Open Charge Alliance}, {CHAdeMO Association},
  \href{https://www.chademo.com/wp2016/wp-content/uploads/2020/12/Using%20OCPP%20with%20CHAdeMO%20Final.pdf}{Using
  {OCPP} with {CHAdeMO}}, White paper (Dec. 2020).
\newline\urlprefix\url{https://www.chademo.com/wp2016/wp-content/uploads/2020/12/Using%20OCPP%20with%20CHAdeMO%20Final.pdf}

\bibitem{garofalaki2022ocpp}
Z.~Garofalaki, D.~Kosmanos, S.~Moschoyiannis, D.~Kallergis, C.~Douligeris,
  Electric vehicle charging: A survey on the security issues and challenges of
  the {Open Charge Point Protocol (OCPP)}, IEEE Communications Surveys \&
  Tutorials 24~(3) (2022) 1504--1533.
\newblock \href {https://doi.org/10.1109/COMST.2022.3184448}
  {\path{doi:10.1109/COMST.2022.3184448}}.

\bibitem{poudel2023injection}
S.~Poudel, J.~E. Baugh, A.~Takiddin, M.~Ismail, S.~S. Refaat, Injection attacks
  and detection strategy in front-end vehicle-to-grid communication, in: 2023
  IEEE International Conference on Communications, Control, and Computing
  Technologies for Smart Grids (SmartGridComm), Glasgow, United Kingdom, 2023,
  pp. 1--6.
\newblock \href {https://doi.org/10.1109/SmartGridComm57358.2023.10333927}
  {\path{doi:10.1109/SmartGridComm57358.2023.10333927}}.

\bibitem{mokarim2023evcs}
A.~Mokarim, G.~B. Gaggero, M.~Marchese, Evaluation of the impact of
  cyber-attacks against electric vehicle charging stations in a low voltage
  distribution grid, in: 2023 IEEE International Conference on Communications,
  Control, and Computing Technologies for Smart Grids (SmartGridComm), 2023,
  pp. 1--7.
\newblock \href {https://doi.org/10.1109/SmartGridComm57358.2023.10333896}
  {\path{doi:10.1109/SmartGridComm57358.2023.10333896}}.

\bibitem{Cao2021InvisibleFB}
Y.~Cao, N.~Wang, C.~Xiao, D.~Yang, J.~Fang, R.~Yang, Q.~A. Chen, M.~Liu, B.~Li,
  Invisible for both camera and lidar: Security of multi-sensor fusion based
  perception in autonomous driving under physical-world attacks, 2021 IEEE
  Symposium on Security and Privacy (SP) (2021) 176--194.

\bibitem{279980}
R.~S. Hallyburton, Y.~Liu, Y.~Cao, Z.~M. Mao, M.~Pajic, Security analysis of
  {Camera-LiDAR} fusion against {Black-Box} attacks on autonomous vehicles, in:
  31st USENIX Security Symposium (USENIX Security 22), USENIX Association,
  Boston, MA, 2022, pp. 1903--1920.

\bibitem{10.1145/3460120.3485377}
Y.~Zhu, C.~Miao, T.~Zheng, F.~Hajiaghajani, L.~Su, C.~Qiao, Can we use
  arbitrary objects to attack lidar perception in autonomous driving?, in:
  Proceedings of the 2021 ACM SIGSAC Conference on Computer and Communications
  Security, CCS '21, Association for Computing Machinery, New York, NY, USA,
  2021, p. 1945–1960.

\bibitem{hau2021object}
Z.~Hau, K.~T. Co, S.~Demetriou, E.~C. Lupu, Object removal attacks on
  lidar-based 3d object detectors, in: Third International Workshop on
  Automotive and Autonomous Vehicle Security (AutoSec), 2021, arXiv:2102.03722.

\bibitem{Zhu2022DemoAL}
Y.~Zhu, C.~Miao, F.~Hajiaghajani, M.~Huai, L.~Su, C.~Qiao, Demo: Attacking
  lidar semantic segmentation in autonomous driving, Fourth International
  Workshop on Automotive and Autonomous Vehicle Security (AutoSec) (2022).

\bibitem{cao2021automated}
Y.~Cao, J.~Ma, K.~Fu, S.~Rampazzi, Z.~M. Mao, Demo: Automated tracking system
  for {LiDAR} spoofing attacks on moving targets, in: Third International
  Workshop on Automotive and Autonomous Vehicle Security (AutoSec), 2021.
\newblock \href {https://doi.org/10.14722/autosec.2021.23019}
  {\path{doi:10.14722/autosec.2021.23019}}.

\bibitem{zhang2023cooperative}
H.~Zhang, Z.~Li, S.~Cheng, A.~Clark, Cooperative perception for safe control of
  autonomous vehicles under lidar spoofing attacks, in: Symposium on Vehicle
  Security and Privacy (VehicleSec), 2023.
\newblock \href {https://doi.org/10.14722/vehiclesec.2023.23066}
  {\path{doi:10.14722/vehiclesec.2023.23066}}.

\bibitem{287252}
Q.~Xiao, X.~Pan, Y.~Lu, M.~Zhang, J.~Dai, M.~Yang, Exorcising wraith:
  Protecting {LiDAR-based} object detector in automated driving system from
  appearing attacks, in: 32nd USENIX Security Symposium (USENIX Security 23),
  USENIX Association, Anaheim, CA, 2023, pp. 2939--2956.

\bibitem{nassi2021spoofing}
B.~Nassi, D.~Nassi, R.~Ben-Netanel, Y.~Elovici, Spoofing mobileye 630's video
  camera using a projector, in: Third International Workshop on Automotive and
  Autonomous Vehicle Security (AutoSec), 2021.
\newblock \href {https://doi.org/10.14722/autosec.2021.23007}
  {\path{doi:10.14722/autosec.2021.23007}}.

\bibitem{sato2021wip}
T.~Sato, J.~Shen, N.~Wang, Y.~J. Jia, X.~Lin, Q.~A. Chen, Wip: Deployability
  improvement, stealthiness user study, and safety impact assessment on real
  vehicle for dirty road patch attack, in: Third International Workshop on
  Automotive and Autonomous Vehicle Security (AutoSec), 2021.
\newblock \href {https://doi.org/10.14722/autosec.2021.23027}
  {\path{doi:10.14722/autosec.2021.23027}}.

\bibitem{10.1145/3460120.3484766}
W.~Wang, Y.~Yao, X.~Liu, X.~Li, P.~Hao, T.~Zhu, I can see the light: Attacks on
  autonomous vehicles using invisible lights, in: Proceedings of the 2021 ACM
  SIGSAC Conference on Computer and Communications Security, CCS '21,
  Association for Computing Machinery, New York, NY, USA, 2021, p. 1930–1944.

\bibitem{sato2023wip}
T.~Sato, S.~H. Bhupathiraju, M.~Clifford, T.~Sugawara, Q.~A. Chen, S.~Rampazzi,
  Wip: Infrared laser reflection attack against traffic sign recognition
  systems, in: Symposium on Vehicle Security and Privacy (VehicleSec), 2023.

\bibitem{tang2021fooling}
K.~Tang, J.~Shen, Q.~A. Chen, Fooling perception via location: a case of
  region-of-interest attacks on traffic light detection in autonomous driving,
  in: Workshop on Automotive and Autonomous Vehicle Security (AutoSec), 2021.

\bibitem{abdallah2022not}
A.~A. Abdallah, Z.~M. Kassas, C.~Lee, Demo: I am not afraid of the {GPS}
  jammer: Exploiting cellular signals for accurate ground vehicle navigation in
  a {GPS}-denied environment, in: Workshop on Automotive and Autonomous Vehicle
  Security (AutoSec), 2022.
\newblock \href {https://doi.org/10.14722/autosec.2022.23049}
  {\path{doi:10.14722/autosec.2022.23049}}.

\bibitem{rouf2010security}
I.~Rouf, R.~Miller, H.~Mustafa, T.~Taylor, S.~Oh, W.~Xu, M.~Gruteser,
  W.~Trappe, I.~Seskar, Security and privacy vulnerabilities of $\{$In-Car$\}$
  wireless networks: A tire pressure monitoring system case study, in: 19th
  USENIX Security Symposium (USENIX Security 10), 2010.

\bibitem{yan2016can}
C.~Yan, W.~Xu, J.~Liu, Can you trust autonomous vehicles: Contactless attacks
  against sensors of self-driving vehicle, DEF CON 24, Las Vegas (2016).

\bibitem{Ding2022DemoHC}
W.~Ding, L.~Cheng, X.~Mi, Z.~Zhao, H.~Hu, Demo \#10: Hijacking connected
  vehicle alexa skills, in: Fourth International Workshop on Automotive and
  Autonomous Vehicle Security (AutoSec), 2022.

\bibitem{kang2021car}
H.~Kang, B.~I. Kwak, Y.~H. Lee, H.~Lee, H.~Lee, H.~K. Kim, Car hacking and
  defense competition on in-vehicle network, in: Third International Workshop
  on Automotive and Autonomous Vehicle Security (AutoSec), 2021.

\bibitem{rajapaksha2023improving}
S.~Rajapaksha, H.~Kalutarage, M.~O. Al-Kadri, A.~Petrovski, G.~Madzudzo,
  Improving in-vehicle networks intrusion detection using on-device transfer
  learning, in: Symposium on Vehicle Security and Privacy (VehicleSec), 2023.

\bibitem{SONG2020100198}
H.~M. Song, J.~Woo, H.~K. Kim, In-vehicle network intrusion detection using
  deep convolutional neural network, Vehicular Communications 21 (2020) 100198.

\bibitem{wang2014vecure}
Q.~Wang, S.~Sawhney, Vecure: A practical security framework to protect the can
  bus of vehicles, in: 2014 International Conference on the Internet of Things
  (IOT), IEEE, 2014, pp. 13--18.

\bibitem{wolf2007state}
M.~Wolf, A.~Weimerskirch, T.~Wollinger, State of the art: Embedding security in
  vehicles, EURASIP Journal on Embedded Systems 2007 (2007) 1--16.

\bibitem{5504804}
K.~Koscher, A.~Czeskis, F.~Roesner, S.~Patel, T.~Kohno, S.~Checkoway, D.~McCoy,
  B.~Kantor, D.~Anderson, H.~Shacham, S.~Savage, Experimental security analysis
  of a modern automobile, in: 2010 IEEE Symposium on Security and Privacy,
  2010, pp. 447--462.

\bibitem{8767141}
R.~Gottumukkala, R.~Merchant, A.~Tauzin, K.~Leon, A.~Roche, P.~Darby,
  Cyber-physical system security of vehicle charging stations, in: 2019 IEEE
  Green Technologies Conference(GreenTech), 2019, pp. 1--5.
\newblock \href {https://doi.org/10.1109/GreenTech.2019.8767141}
  {\path{doi:10.1109/GreenTech.2019.8767141}}.

\bibitem{basnet2021exploring}
M.~Basnet, M.~H. Ali, Exploring cybersecurity issues in 5g enabled electric
  vehicle charging station with deep learning, IET Generation, Transmission \&
  Distribution 15~(24) (2021) 3435--3449.

\bibitem{el2020cybersecurity}
Z.~El-Rewini, K.~Sadatsharan, N.~Sugunaraj, D.~F. Selvaraj, S.~J. Plathottam,
  P.~Ranganathan, Cybersecurity attacks in vehicular sensors, IEEE Sensors
  Journal 20~(22) (2020) 13752--13767.

\bibitem{yang2021secure}
T.~Yang, C.~Lv, A secure sensor fusion framework for connected and automated
  vehicles under sensor attacks, IEEE Internet of Things Journal 9~(22) (2022)
  22357--22365.

\bibitem{philipsen2021threats}
S.~G. Philipsen, B.~Andersen, B.~Singh, Threats and attacks to modern vehicles,
  in: 2021 IEEE International Conference on Internet of Things and Intelligence
  Systems (IoTaIS), IEEE, 2021, pp. 22--27.

\bibitem{hu2022pass}
Z.~Hu, J.~Shen, S.~Guo, X.~Zhang, Z.~Zhong, Q.~A. Chen, K.~Li, Pass: A
  system-driven evaluation platform for autonomous driving safety and security,
  in: Workshop on Automotive and Autonomous Vehicle Security (AutoSec), 2022.

\bibitem{285475}
Y.~Cao, S.~H. Bhupathiraju, P.~Naghavi, T.~Sugawara, Z.~M. Mao, S.~Rampazzi,
  You can{\textquoteright}t see me: Physical removal attacks on {LiDAR-based}
  autonomous vehicles driving frameworks, in: 32nd USENIX Security Symposium
  (USENIX Security 23), USENIX Association, Anaheim, CA, 2023, pp. 2993--3010.

\bibitem{kang2020attack}
L.~Kang, H.~Shen, Attack detection and mitigation for sensor and can bus
  attacks in vehicle anti-lock braking systems, in: 2020 29th International
  Conference on Computer Communications and Networks (ICCCN), IEEE, 2020, pp.
  1--9.

\bibitem{ahmad2020securing}
U.~Ahmad, H.~Song, A.~Bilal, M.~Alazab, A.~Jolfaei, Securing smart vehicles
  from relay attacks using machine learning, The Journal of Supercomputing 76
  (2020) 2665--2682.

\bibitem{deppenhanced}
Z.~Depp, H.~B. Tulay, C.~E. Koksal, Enhanced vehicular roll-jam attack using a
  known noise source, in: Symposium on Vehicle Security and Privacy
  (VehicleSec), 2023.
\newblock \href {https://doi.org/10.14722/vehiclesec.2023.23037}
  {\path{doi:10.14722/vehiclesec.2023.23037}}.

\bibitem{9833886}
D.~Antonioli, M.~Payer, On the insecurity of vehicles against protocol-level
  bluetooth threats, in: 2022 IEEE Security and Privacy Workshops (SPW), 2022,
  pp. 353--362.

\bibitem{renganathan2022valet}
V.~Renganathan, E.~Yurtsever, Q.~Ahmed, A.~Yener, Valet attack on privacy: a
  cybersecurity threat in automotive bluetooth infotainment systems,
  Cybersecurity 5~(1) (2022) 30.

\bibitem{9564848}
K.~Zhang, A.~Olmsted, Examining autonomous vehicle operating systems
  vulnerabilities using a cyber-physical approach, in: 2021 IEEE International
  Intelligent Transportation Systems Conference (ITSC), 2021, pp. 976--981.

\bibitem{vanderheijden2019misbehavior}
R.~W. van~der Heijden, S.~Dietzel, T.~Leinm{\"u}ller, F.~Kargl, Survey on
  misbehavior detection in cooperative intelligent transportation systems, IEEE
  Communications Surveys \& Tutorials 21~(1) (2019) 779--811.
\newblock \href {https://doi.org/10.1109/COMST.2018.2873088}
  {\path{doi:10.1109/COMST.2018.2873088}}.

\bibitem{josephlal2019vulnerability}
E.~F.~M. Josephlal, S.~Adepu, Vulnerability analysis of an automotive
  infotainment system's wifi capability, in: 2019 IEEE 19th International
  Symposium on High Assurance Systems Engineering (HASE), IEEE, 2019, pp.
  241--246.

\bibitem{lee2019enhanced}
Y.~Lee, S.~Woo, J.~Lee, Y.~Song, H.~Moon, D.~H. Lee, Enhanced android
  app-repackaging attack on in-vehicle network, Wireless Communications and
  Mobile Computing 2019~(1) (2019) 5650245.

\bibitem{seo2023formally}
J.~Seo, J.~Kwak, S.~Kim, Formally verified software update management system in
  automotive, in: Symposium on Vehicle Security and Privacy (VehicleSec), 2023.
\newblock \href {https://doi.org/10.14722/vehiclesec.2023.23087}
  {\path{doi:10.14722/vehiclesec.2023.23087}}.

\bibitem{Mahmood2022}
S.~Mahmood, H.~N. Nguyen, S.~A. Shaikh, Systematic threat assessment and
  security testing of automotive over-the-air ({OTA}) updates, Vehicular
  Communications 35 (2022) 100468.
\newblock \href {https://doi.org/10.1016/j.vehcom.2022.100468}
  {\path{doi:10.1016/j.vehcom.2022.100468}}.

\bibitem{Seo2024}
J.~Seo, J.~Kwak, S.~Kim, Development for high-assurance software update
  management system complying with {UN R156}, IEEE Access 12 (2024)
  135811--135830.

\bibitem{9091063}
Z.~Khan, M.~Chowdhury, M.~Islam, C.-Y. Huang, M.~Rahman, Long short-term memory
  neural network-based attack detection model for in-vehicle network security,
  IEEE Sensors Letters 4~(6) (2020) 1--4.

\bibitem{jichici2023control}
C.~Jichici, A.~Berdich, A.~Musuroi, B.~Groza, Control system level intrusion
  detection on j1939 heavy-duty vehicle buses, IEEE Transactions on Industrial
  Informatics 20~(2) (2024) 2029--2041.

\bibitem{berdich2023secure}
A.~Berdich, B.~Groza, Cyberattacks on adaptive cruise controls and emergency
  braking systems: Adversary models, impact assessment, and countermeasures,
  IEEE Transactions on Reliability 73~(2) (2024) 1216--1230.
\newblock \href {https://doi.org/10.1109/TR.2024.3373810}
  {\path{doi:10.1109/TR.2024.3373810}}.

\bibitem{9706416}
K.~Agrawal, T.~Alladi, A.~Agrawal, V.~Chamola, A.~Benslimane, Novelads: A novel
  anomaly detection system for intra-vehicular networks, IEEE Transactions on
  Intelligent Transportation Systems 23~(11) (2022) 22596--22606.

\bibitem{shoukry2013non}
Y.~Shoukry, P.~Martin, P.~Tabuada, M.~Srivastava, Non-invasive spoofing attacks
  for anti-lock braking systems, in: Cryptographic Hardware and Embedded
  Systems-CHES 2013: 15th International Workshop, Santa Barbara, CA, USA,
  August 20-23, 2013. Proceedings 15, Springer, 2013, pp. 55--72.

\bibitem{verma2024road}
M.~E. Verma, R.~A. Bridges, M.~D. Iannacone, S.~C. Hollifield, P.~Moriano,
  S.~C. Hespeler, B.~Kay, F.~L. Combs, A comprehensive guide to {CAN} {IDS}
  data and introduction of the {ROAD} dataset, PLOS ONE 19~(1) (2024) e0296879.
\newblock \href {https://doi.org/10.1371/journal.pone.0296879}
  {\path{doi:10.1371/journal.pone.0296879}}.

\bibitem{7796898}
A.~Taylor, S.~Leblanc, N.~Japkowicz, Anomaly detection in automobile control
  network data with long short-term memory networks, in: 2016 IEEE
  International Conference on Data Science and Advanced Analytics (DSAA), 2016,
  pp. 130--139.

\bibitem{yue2021cancloak}
L.~Yue, Z.~Li, T.~Yin, C.~Zhang, Cancloak: Deceiving two ecus with one frame,
  in: Third International Workshop on Automotive and Autonomous Vehicle
  Security (AutoSec), 2021.

\bibitem{272188}
B.~Groza, L.~Popa, P.-S. Murvay, Y.~Elovici, A.~Shabtai, {CANARY} - a reactive
  defense mechanism for controller area networks based on active {RelaYs}, in:
  30th USENIX Security Symposium (USENIX Security 21), USENIX Association,
  2021, pp. 4259--4276.

\bibitem{197171}
K.-T. Cho, K.~G. Shin, Fingerprinting electronic control units for vehicle
  intrusion detection, in: 25th USENIX Security Symposium (USENIX Security 16),
  USENIX Association, Austin, TX, 2016, pp. 911--927.

\bibitem{9564677}
S.-L. Wang, S.-Y. Wu, C.-C. Lin, S.~Boddupalli, P.-J. Chang, C.-W. Lin, C.-S.
  Shih, S.~Ray, Deep-learning-based intrusion detection for autonomous
  vehicle-following systems, in: 2021 IEEE International Intelligent
  Transportation Systems Conference (ITSC), 2021, pp. 865--872.

\bibitem{8624397}
K.~B. Kelarestaghi, M.~Foruhandeh, K.~Heaslip, R.~Gerdes, Intelligent
  transportation system security: Impact-oriented risk assessment of in-vehicle
  networks, IEEE Intelligent Transportation Systems Magazine 13~(2) (2021)
  91--104.

\bibitem{8053478}
J.~Liu, S.~Zhang, W.~Sun, Y.~Shi, In-vehicle network attacks and
  countermeasures: Challenges and future directions, IEEE Network 31~(5) (2017)
  50--58.

\bibitem{247674}
M.~Luo, A.~C. Myers, G.~E. Suh, Stealthy tracking of autonomous vehicles with
  cache side channels, in: 29th USENIX Security Symposium (USENIX Security 20),
  USENIX Association, 2020, pp. 859--876.

\bibitem{yang2022gpskey}
E.~Yang, S.~Fang, Gpskey: Gps-based secret key establishment for intra-vehicle
  environment, in: Workshop on Automotive and Autonomous Vehicle Security
  (AutoSec), 2022.

\bibitem{chenguess}
D.~Chen, M.~D. Pes{\'e}, K.~G. Shin, Guess which car type i am driving:
  Information leak via driving apps, in: Proceedings Inaugural International
  Symposium on Vehicle Security \& Privacy, Internet Society San Diego, CA,
  USA, 2023.

\bibitem{8642311}
O.~Y. Al-Jarrah, C.~Maple, M.~Dianati, D.~Oxtoby, A.~Mouzakitis, Intrusion
  detection systems for intra-vehicle networks: A review, IEEE Access 7 (2019)
  21266--21289.

\bibitem{bloom2021weepingcan}
G.~Bloom, Weepingcan: A stealthy can bus-off attack, in: Workshop on Automotive
  and Autonomous Vehicle Security (AutoSec), 2021.

\bibitem{palanca2017stealth}
A.~Palanca, E.~Evenchick, F.~Maggi, S.~Zanero, A stealth, selective, link-layer
  denial-of-service attack against automotive networks, in: Detection of
  Intrusions and Malware, and Vulnerability Assessment: 14th International
  Conference, DIMVA 2017, Bonn, Germany, July 6-7, 2017, Proceedings 14,
  Springer, 2017, pp. 185--206.

\bibitem{parno2005challenges}
B.~Parno, A.~Perrig, Challenges in securing vehicular networks, in: Workshop on
  hot topics in networks (HotNets-IV), Maryland, USA, 2005, pp. 1--6.

\bibitem{chatterjee2023exploiting}
R.~Chatterjee, S.~Mukherjee, J.~Daily, Exploiting transport protocol
  vulnerabilities in sae j1939 networks, in: Symposium on Vehicle Security and
  Privacy (VehicleSec), 2023.

\bibitem{mohammadi2022generation}
A.~Mohammadi, H.~Malik, M.~Abbaszadeh, Generation of can-based wheel lockup
  attacks on the dynamics of vehicle traction, in: Workshop on Automotive and
  Autonomous Vehicle Security (AutoSec), 2022.

\bibitem{chernikova2019self}
A.~Chernikova, A.~Oprea, C.~Nita-Rotaru, B.~Kim, Are self-driving cars secure?
  evasion attacks against deep neural networks for steering angle prediction,
  in: 2019 IEEE Security and Privacy Workshops (SPW), IEEE, 2019, pp. 132--137.

\bibitem{10.1007/978-3-540-87698-4_21}
T.~Hoppe, S.~Kiltz, J.~Dittmann,
  \href{https://doi.org/10.1007/978-3-540-87698-4_21}{Security threats to
  automotive can networks --- practical examples and selected short-term
  countermeasures}, in: Proceedings of the 27th International Conference on
  Computer Safety, Reliability, and Security, SAFECOMP '08, Springer-Verlag,
  Berlin, Heidelberg, 2008, p. 235–248.
\newblock \href {https://doi.org/10.1007/978-3-540-87698-4_21}
  {\path{doi:10.1007/978-3-540-87698-4_21}}.
\newline\urlprefix\url{https://doi.org/10.1007/978-3-540-87698-4_21}

\bibitem{kang2022detection}
L.~Kang, H.~Shen, Detection and mitigation of sensor and can bus attacks in
  vehicle anti-lock braking systems, ACM Transactions on Cyber-Physical Systems
  (TCPS) 6~(1) (2022) 1--24.

\bibitem{xu2020robust}
X.~Xu, X.~Li, P.~Dong, Y.~Liu, H.~Zhang, Robust reset speed synchronization
  control for an integrated motor-transmission powertrain system of a connected
  vehicle under a replay attack, IEEE Transactions on Vehicular Technology
  70~(6) (2021) 5524--5536.

\bibitem{8306904}
W.~Choi, K.~Joo, H.~J. Jo, M.~C. Park, D.~H. Lee, Voltageids: Low-level
  communication characteristics for automotive intrusion detection system, IEEE
  Transactions on Information Forensics and Security 13~(8) (2018) 2114--2129.

\bibitem{iqbal2019towards}
S.~Iqbal, A.~Haque, M.~Zulkernine, Towards a security architecture for
  protecting connected vehicles from malware, in: 2019 IEEE 89th Vehicular
  Technology Conference (VTC2019-Spring), IEEE, 2019, pp. 1--5.

\bibitem{daily2021securing}
J.~Daily, D.~Nnaji, B.~Ettlinger, Securing can traffic on j1939 networks, in:
  Third International Workshop on Automotive and Autonomous Vehicle Security
  (AutoSec), 2021.

\bibitem{287308}
K.~Serag, R.~Bhatia, A.~Faqih, M.~O. Ozmen, V.~Kumar, Z.~B. Celik, D.~Xu,
  {ZBCAN}: A {Zero-Byte} {CAN} defense system, in: 32nd USENIX Security
  Symposium (USENIX Security 23), USENIX Association, Anaheim, CA, 2023, pp.
  6893--6910.

\bibitem{9491889}
D.~Wang, S.~Ganesan, Automotive network security, in: 2021 IEEE International
  Conference on Electro Information Technology (EIT), 2021, pp. 193--196.
\newblock \href {https://doi.org/10.1109/EIT51626.2021.9491889}
  {\path{doi:10.1109/EIT51626.2021.9491889}}.

\bibitem{8270348}
B.~Groza, P.-S. Murvay, Security solutions for the controller area network:
  Bringing authentication to in-vehicle networks, IEEE Vehicular Technology
  Magazine 13~(1) (2018) 40--47.
\newblock \href {https://doi.org/10.1109/MVT.2017.2736344}
  {\path{doi:10.1109/MVT.2017.2736344}}.

\bibitem{5281251}
H.~Zhu, R.~Lu, X.~Shen, X.~Lin, Security in service-oriented vehicular
  networks, IEEE Wireless Communications 16~(4) (2009) 16--22.
\newblock \href {https://doi.org/10.1109/MWC.2009.5281251}
  {\path{doi:10.1109/MWC.2009.5281251}}.

\bibitem{xie2021security}
G.~Xie, L.~T. Yang, Y.~Liu, H.~Luo, X.~Peng, R.~Li, Security enhancement for
  real-time independent in-vehicle can-fd messages in vehicular networks, IEEE
  Transactions on Vehicular Technology 70~(6) (2021) 5244--5253.

\bibitem{10049666}
W.~Wu, J.~Dai, H.~Huang, Q.~Zhao, G.~Zeng, R.~Li, A digital watermark method
  for in-vehicle network security enhancement, IEEE Transactions on Vehicular
  Technology 72~(7) (2023) 8398--8408.
\newblock \href {https://doi.org/10.1109/TVT.2023.3247180}
  {\path{doi:10.1109/TVT.2023.3247180}}.

\bibitem{6542519}
C.-W. Lin, A.~Sangiovanni-Vincentelli, Cyber-security for the controller area
  network (can) communication protocol, in: 2012 International Conference on
  Cyber Security, 2012, pp. 1--7.
\newblock \href {https://doi.org/10.1109/CyberSecurity.2012.7}
  {\path{doi:10.1109/CyberSecurity.2012.7}}.

\bibitem{7427089}
H.~M. Song, H.~R. Kim, H.~K. Kim, Intrusion detection system based on the
  analysis of time intervals of can messages for in-vehicle network, in: 2016
  International Conference on Information Networking (ICOIN), 2016, pp. 63--68.
\newblock \href {https://doi.org/10.1109/ICOIN.2016.7427089}
  {\path{doi:10.1109/ICOIN.2016.7427089}}.

\bibitem{blevins2021time}
D.~H. Blevins, P.~Moriano, R.~A. Bridges, M.~E. Verma, M.~D. Iannacone, S.~C.
  Hollifield, Time-based can intrusion detection benchmark, arXiv preprint
  arXiv:2101.05781 (2021).

\bibitem{shahriar2023cantropy}
M.~H. Shahriar, W.~Lou, Y.~T. Hou, Cantropy: Time series feature
  extraction-based intrusion detection systems for controller area networks,
  in: Symposium on Vehicle Security and Privacy (VehicleSec), 2023.

\bibitem{s23073554}
J.~Khan, D.-W. Lim, Y.-S. Kim, Intrusion detection system can-bus in-vehicle
  networks based on the statistical characteristics of attacks, Sensors 23~(7)
  (2023).
\newblock \href {https://doi.org/10.3390/s23073554}
  {\path{doi:10.3390/s23073554}}.

\bibitem{alkhatib2023wip}
N.~Alkhatib, L.~Achaji, M.~Mushtaq, H.~Ghauch, J.-L. Danger, Wip: Amica:
  Attention-based multi-identifier model for asynchronous intrusion detection
  on controller area networks, in: Symposium on Vehicle Security and Privacy
  (VehicleSec), 2023.

\bibitem{agbaje2022framework}
P.~Agbaje, A.~Anjum, A.~Mitra, G.~Bloom, H.~Olufowobi, A framework for
  consistent and repeatable controller area network ids evaluation, in: Fourth
  International Workshop on Automotive and Autonomous Vehicle Security
  (AutoSec), 2022.

\bibitem{6940213}
H.~Sedjelmaci, S.~M. Senouci, M.~A. Abu-Rgheff, An efficient and lightweight
  intrusion detection mechanism for service-oriented vehicular networks, IEEE
  Internet of Things Journal 1~(6) (2014) 570--577.
\newblock \href {https://doi.org/10.1109/JIOT.2014.2366120}
  {\path{doi:10.1109/JIOT.2014.2366120}}.

\bibitem{lai2024gradient}
Y.~Lai, J.~Wei, Y.~Chen, Gradient correlation based detection of adversarial
  attacks on vehicular networks, Computer Networks 255 (2024) 110868.

\bibitem{fu2024iov}
M.~Fu, P.~Wang, M.~Liu, Z.~Zhang, X.~Zhou, Iov-bert-ids: Hybrid network
  intrusion detection system in iov using large language models, IEEE
  Transactions on Vehicular Technology 74~(2) (2025) 1909--1921.

\bibitem{wang2024intrusion}
S.~Wang, Y.~Wang, B.~Zheng, J.~Cheng, Y.~Su, Y.~Dai, Intrusion detection system
  for vehicular networks based on mobilenetv3, IEEE Access (2024).

\bibitem{li2024ecf}
S.~Li, Y.~Cao, H.~J. Hadi, F.~Hao, F.~B. Hussain, L.~Chen, Ecf-ids: an enhanced
  cuckoo filter-based intrusion detection system for in-vehicle network, IEEE
  Transactions on Network and Service Management (2024).

\bibitem{ibraheem2024anomaly}
A.~Ibraheem, Z.~Sheng, G.~Parisis, Anomaly detection and classification for
  sdn-enabled in-vehicle network using network tomography-based deep learning,
  in: 2024 IEEE Wireless Communications and Networking Conference (WCNC), IEEE,
  2024, pp. 1--6.

\bibitem{chandra2024enhanced}
L.~S. Chandra, D.~K. Singh, An enhanced intrusion detection system model using
  machine learning algorithm, in: 2024 International Conference on Emerging
  Systems and Intelligent Computing (ESIC), IEEE, 2024, pp. 106--111.

\bibitem{altalbe2023enhanced}
A.~Altalbe, Enhanced intrusion detection in in-vehicle networks using advanced
  feature fusion and stacking-enriched learning, IEEE Access 12 (2024)
  2045--2056.

\bibitem{singh2023intrusion}
N.~Singh, R.~Agarwal, Intrusion detection system for smart vehicles using
  machine learning algorithms, in: 2023 international conference on
  communication, security and artificial intelligence (ICCSAI), IEEE, 2023, pp.
  749--753.

\bibitem{zhao2023tree}
Y.~Zhao, Z.~Bi, G.~Xu, C.~Wang, G.~Xu, Tree-ids: An incremental intrusion
  detection system for connected vehicles, in: 2023 IEEE 48th Conference on
  Local Computer Networks (LCN), IEEE, 2023, pp. 1--4.

\bibitem{qiu2023physical}
S.~Qiu, J.~Fei, H.~Yang, Y.~Xiao, X.~Zhang, A physical fingerprint-based
  intrusion detection and localization in fieldbus network, in: 2023 4th
  International Seminar on Artificial Intelligence, Networking and Information
  Technology (AINIT), IEEE, 2023, pp. 294--302.

\bibitem{yu2022tce}
Z.~Yu, Y.~Liu, G.~Xie, R.~Li, S.~Liu, L.~T. Yang, Tce-ids: Time interval
  conditional entropy-based intrusion detection system for automotive
  controller area networks, IEEE Transactions on Industrial Informatics 19~(2)
  (2023) 1185--1195.

\bibitem{kandasamy2022detection}
J.~Kandasamy, S.~Arunagirinathan, P.~Sivaraj, M.~Pameela, G.~T. Subham,
  R.~Nagarajan, Detection of cyber attack in electric vehicles using alstm
  based machine learning, in: 2022 4th International Conference on Inventive
  Research in Computing Applications (ICIRCA), IEEE, 2022, pp. 596--600.

\bibitem{andreica2022android}
T.~Andreica, C.-D. Curiac, C.~Jichici, B.~Groza, Android head units vs.
  in-vehicle ecus: performance assessment for deploying in-vehicle intrusion
  detection systems for the can bus, IEEE Access 10 (2022) 95161--95178.

\bibitem{anjum2022vehicle}
A.~Anjum, P.~Agbaje, S.~Hounsinou, H.~Olufowobi, In-vehicle network anomaly
  detection using extreme gradient boosting machine, in: 2022 11th
  Mediterranean Conference on Embedded Computing (MECO), IEEE, 2022, pp. 1--6.

\bibitem{baldini2022detection}
G.~Baldini, Detection of cybersecurity spoofing attacks in vehicular networks
  with recurrence quantification analysis, Computer Communications 191 (2022)
  486--499.

\bibitem{7420322}
A.~Taylor, N.~Japkowicz, S.~Leblanc, Frequency-based anomaly detection for the
  automotive can bus, in: 2015 World Congress on Industrial Control Systems
  Security (WCICSS), 2015, pp. 45--49.
\newblock \href {https://doi.org/10.1109/WCICSS.2015.7420322}
  {\path{doi:10.1109/WCICSS.2015.7420322}}.

\bibitem{moriano2022masquerade}
P.~Moriano, R.~A. Bridges, M.~D. Iannacone, Detecting {CAN} masquerade attacks
  with signal clustering similarity, in: Fourth International Workshop on
  Automotive and Autonomous Vehicle Security (AutoSec), 2022.
\newblock \href {https://doi.org/10.14722/autosec.2022.23028}
  {\path{doi:10.14722/autosec.2022.23028}}.

\bibitem{wang2021can}
J.~Wang, X.~Mo, A can bus anomaly detection based on flxgboost algorithm, in:
  2021 IEEE 23rd Int Conf on High Performance Computing \& Communications; 7th
  Int Conf on Data Science \& Systems; 19th Int Conf on Smart City; 7th Int
  Conf on Dependability in Sensor, Cloud \& Big Data Systems \& Application
  (HPCC/DSS/SmartCity/DependSys), IEEE, 2021, pp. 1558--1564.

\bibitem{lokman2019intrusion}
S.-F. Lokman, A.~T. Othman, M.-H. Abu-Bakar, Intrusion detection system for
  automotive controller area network (can) bus system: a review, EURASIP
  Journal on Wireless Communications and Networking 2019~(1) (2019) 1--17.

\bibitem{shin2017intelligent}
J.~Shin, Y.~Baek, Y.~Eun, S.~H. Son, Intelligent sensor attack detection and
  identification for automotive cyber-physical systems, in: 2017 IEEE Symposium
  series on computational intelligence (SSCI), IEEE, 2017, pp. 1--8.

\bibitem{9128438}
A.~U. Makarfi, K.~M. Rabie, O.~Kaiwartya, X.~Li, R.~Kharel, Physical layer
  security in vehicular networks with reconfigurable intelligent surfaces, in:
  2020 IEEE 91st Vehicular Technology Conference (VTC2020-Spring), 2020, pp.
  1--6.
\newblock \href {https://doi.org/10.1109/VTC2020-Spring48590.2020.9128438}
  {\path{doi:10.1109/VTC2020-Spring48590.2020.9128438}}.

\bibitem{rana2017attack}
M.~M. Rana, Attack resilient wireless sensor networks for smart electric
  vehicles, IEEE sensors letters 1~(2) (2017) 1--4.

\bibitem{gao2025signal}
F.~Gao, J.~Liu, C.~Li, Z.~Gao, R.~Zhao, Signal-relationship-aware explainable
  intrusion detection in controller area networks using graph transformers,
  Knowledge-Based Systems 328 (2025) 114237.
\newblock \href {https://doi.org/10.1016/j.knosys.2025.114237}
  {\path{doi:10.1016/j.knosys.2025.114237}}.

\bibitem{sun2025kgid}
H.~Sun, J.~Wang, J.~Weng, W.~Tan, {KG-ID}: Knowledge graph-based intrusion
  detection on in-vehicle network, IEEE Transactions on Intelligent
  Transportation Systems 26~(4) (2025) 4988--5000.
\newblock \href {https://doi.org/10.1109/TITS.2025.3530155}
  {\path{doi:10.1109/TITS.2025.3530155}}.

\bibitem{alqahtani2025deep}
H.~Alqahtani, G.~Kumar, Deep learning-based intrusion detection system for
  in-vehicle networks with knowledge graph and statistical methods,
  International Journal of Machine Learning and Cybernetics 16~(5--6) (2025)
  3539--3555.
\newblock \href {https://doi.org/10.1007/s13042-024-02465-0}
  {\path{doi:10.1007/s13042-024-02465-0}}.

\bibitem{upstream2023}
{Upstream Security}, 2023 global automotive cybersecurity report,
  \url{https://upstream.auto/reports/2023report/} (2023).

\bibitem{upstream2024}
{Upstream Security}, 2024 global automotive cybersecurity report,
  \url{https://upstream.auto/reports/global-automotive-cybersecurity-report-2024/}
  (2024).

\bibitem{upstream2025}
{Upstream Security}, 2025 automotive \& smart mobility global cybersecurity
  report: Bridging the cybersecurity gap in automotive \& smart mobility,
  \url{https://upstream.auto/reports/global-automotive-cybersecurity-report-2025/},
  accessed: 2026-09-17 (2025).

\bibitem{PMCSurveyDL2023}
F.~Luo, J.~Wang, X.~Zhang, Y.~Jiang, Z.~Li, C.~Luo, In-vehicle network
  intrusion detection systems: a systematic survey of deep learning-based
  approaches, PeerJ Computer Science 9 (2023) e1648.

\bibitem{Bari2023}
B.~S. Bari, K.~Yelamarthi, S.~Ghafoor, Intrusion detection in vehicle
  {Controller Area Network} ({CAN}) bus using machine learning: A comparative
  performance study, Sensors 23~(7) (2023) 3610.
\newblock \href {https://doi.org/10.3390/s23073610}
  {\path{doi:10.3390/s23073610}}.

\bibitem{upstream2026}
{Upstream Security}, 2026 global automotive and smart mobility cybersecurity
  report,
  \url{https://upstream.auto/reports/global-automotive-cybersecurity-report/}
  (2026).

\bibitem{Alobaidi2024}
F.~Aloraini, A.~Javed, O.~Rana, Adversarial attacks on intrusion detection
  systems in in-vehicle networks of connected and autonomous vehicles, Sensors
  24~(12) (2024) 3848.
\newblock \href {https://doi.org/10.3390/s24123848}
  {\path{doi:10.3390/s24123848}}.

\bibitem{Ibrahum2024}
A.~D.~M. Ibrahum, et~al., Deep learning adversarial attacks and defenses in
  autonomous vehicles: A systematic literature review from a safety
  perspective, Artificial Intelligence Review (2024).
\newblock \href {https://doi.org/10.1007/s10462-024-11014-8}
  {\path{doi:10.1007/s10462-024-11014-8}}.

\bibitem{Baccari2024}
S.~Baccari, M.~Hadded, H.~Ghazzai, H.~Touati, M.~Elhadef, Anomaly detection in
  connected and autonomous vehicles: A survey, analysis, and research
  challenges, IEEE Access 12 (2024) 19250--19276.
\newblock \href {https://doi.org/10.1109/ACCESS.2024.3361829}
  {\path{doi:10.1109/ACCESS.2024.3361829}}.

\bibitem{Nguyen2024}
H.~P. Nguyen, Y.~Chen, Lightweight, post-quantum secure cryptography based on
  {Ascon}: Hardware implementation in automotive applications, Electronics
  13~(22) (2024) 4550.
\newblock \href {https://doi.org/10.3390/electronics13224550}
  {\path{doi:10.3390/electronics13224550}}.

\bibitem{Mozzaquatro2023}
W.~Toghuj, N.~Turab, Automotive {Ethernet} architecture and security:
  Challenges and technologies, International Journal of Electrical and Computer
  Engineering 13~(5) (2023) 5209--5221.

\bibitem{Chillotti2021}
T.~Fritzmann, J.~Vith, D.~Fl{\'o}rez, J.~Sep{\'u}lveda, Post-quantum
  cryptography for automotive systems, Microprocessors and Microsystems 87
  (2021) 104379.
\newblock \href {https://doi.org/10.1016/j.micpro.2021.104379}
  {\path{doi:10.1016/j.micpro.2021.104379}}.

\bibitem{ISOSAE21434}
{ISO/SAE 21434:2021} --- road vehicles: Cybersecurity engineering,
  \url{https://www.iso.org/standard/70918.html} (2021).

\bibitem{WP29R155}
{UNECE World Forum for Harmonization of Vehicle Regulations}, {UN Regulation
  No. 155} --- cybersecurity and cybersecurity management system,
  \url{https://unece.org/transport/documents/2021/03/standards/un-regulation-no-155-cybersecurity-and-cybersecurity-management}
  (2021).

\bibitem{Zelle2022}
D.~Zelle, C.~Plappert, R.~Rieke, D.~Scheuermann, C.~Krau{\ss}, {ThreatSurf}: A
  method for automated threat surface assessment in automotive cybersecurity
  engineering, Microprocessors and Microsystems 90 (2022) 104461.
\newblock \href {https://doi.org/10.1016/j.micpro.2022.104461}
  {\path{doi:10.1016/j.micpro.2022.104461}}.

\bibitem{Iyieke2024}
V.~Iyieke, H.~Jadidbonab, A.~Rakib, J.~Bryans, D.~Dhaliwal, O.~Kosmas, An
  adaptable security-by-design approach for ensuring a secure over the air
  ({OTA}) update in modern vehicles, Computers \& Security 150 (2025) 104268.
\newblock \href {https://doi.org/10.1016/j.cose.2024.104268}
  {\path{doi:10.1016/j.cose.2024.104268}}.

\bibitem{Apriorit2025}
{Apriorit}, Cybersecurity risks of automotive {OTA} updates,
  \url{https://www.apriorit.com/dev-blog/cybersecurity-risks-of-ota-automotive}
  (2025).

\bibitem{Badjie2024}
B.~Badjie, J.~Cec{\'i}lio, A.~Casimiro, Adversarial attacks and countermeasures
  on image classification-based deep learning models in autonomous driving
  systems: A systematic review, ACM Computing Surveys 57~(1) (2024) 1--52.

\bibitem{Samir2024}
S.~S. Ben~Hassane, R.~Martin, H.~Touati, M.~Hadded, H.~Ghazzai, Machine
  learning-based intrusion detection for securing in-vehicle {CAN} bus
  communication, SN Computer Science 5~(8) (2024) 1082.
\newblock \href {https://doi.org/10.1007/s42979-024-03465-1}
  {\path{doi:10.1007/s42979-024-03465-1}}.

\bibitem{NatureIDS2025}
R.~Rai, J.~Grover, P.~Sharma, A.~Pareek, Securing the {CAN} bus using deep
  learning for intrusion detection in vehicles, Scientific Reports 15~(1)
  (2025) 13820.
\newblock \href {https://doi.org/10.1038/s41598-025-98433-x}
  {\path{doi:10.1038/s41598-025-98433-x}}.

\bibitem{Yang2025DefenseWeaver}
Y.~Yang, et~al., Automating function-level {TARA} for automotive full-lifecycle
  security (2025).
\newblock \href {http://arxiv.org/abs/2504.18083} {\path{arXiv:2504.18083}}.

\bibitem{FEV2025TARACopilot}
{FEV Group}, Automation of threat analysis and risk assessment ({TARA}) using
  artificial intelligence,
  \url{https://fev.io/automation-of-threat-analysis-and-risk-assessment-tara-using-artificial-intelligence/}
  (2026).

\bibitem{EETimes2024}
{EE Times}, Automotive cybersecurity: A review of 2023,
  \url{https://www.eetimes.com/automotive-cybersecurity-a-review-of-2023/}
  (2024).

\bibitem{autoconnectedcar2025}
{Auto Connected Car News}, Cyber attacks on autos rises more than tripled in
  2024,
  \url{https://www.autoconnectedcar.com/2025/02/cyber-attacks-on-autos-rises-more-than-tripled-in-2024/}
  (2025).

\bibitem{VicOne2025}
{VicOne}, Shifting gears for 2025: The next generation of automotive
  cybersecurity challenges,
  \url{https://vicone.com/blog/shifting-gears-for-2025-the-next-generation-of-automotive-cybersecurity-challenges}
  (2025).

\bibitem{IVNIDSSurvey2025}
M.~Althunayyan, A.~Javed, O.~Rana, A survey of learning-based intrusion
  detection systems for in-vehicle network (2025).
\newblock \href {http://arxiv.org/abs/2505.11551} {\path{arXiv:2505.11551}}.

\bibitem{Althunayyan2024HFL}
M.~Althunayyan, A.~Javed, O.~Rana, A robust multi-stage intrusion detection
  system for in-vehicle network security using hierarchical federated learning,
  Vehicular Communications 49 (2024) 100837.
\newblock \href {https://doi.org/10.1016/j.vehcom.2024.100837}
  {\path{doi:10.1016/j.vehcom.2024.100837}}.

\bibitem{lauser2024security}
T.~Lauser, D.~Zelle, D.~Kern, C.~Krau{\ss}, L.~V{\"o}lker, Security protocols
  for ethernet-based in-vehicle communication, in: 2024 IEEE Vehicular
  Networking Conference (VNC), IEEE, 2024, pp. 148--155.
\newblock \href {https://doi.org/10.1109/VNC61989.2024.10575984}
  {\path{doi:10.1109/VNC61989.2024.10575984}}.

\bibitem{Haeckel2025}
T.~H{\"a}ckel, P.~Meyer, F.~Korf, T.~C. Schmidt, Securing future in-vehicle
  networks: Monitoring and control for {Ethernet} backbones, in: Engineering
  Safe and Trustworthy Cyber Physical Systems: Essays Dedicated to Werner Damm
  on the Occasion of His 71st Birthday, Springer, 2025, pp. 163--182.
\newblock \href {https://doi.org/10.1007/978-3-031-97537-0_10}
  {\path{doi:10.1007/978-3-031-97537-0_10}}.

\bibitem{Apriorit2025PQC}
{Apriorit}, Post-quantum cryptography in automotive,
  \url{https://www.apriorit.com/dev-blog/post-quantum-cryptography-in-automotive}
  (2025).

\bibitem{Zhu2024}
Y.~Li, W.~Liu, Q.~Liu, X.~Zheng, K.~Sun, C.~Huang, Complying with {ISO 26262}
  and {ISO/SAE 21434}: A safety and security co-analysis method for intelligent
  connected vehicles, Sensors 24~(6) (2024) 1848.
\newblock \href {https://doi.org/10.3390/s24061848}
  {\path{doi:10.3390/s24061848}}.

\bibitem{luo2022cybersecuritytesting}
F.~Luo, X.~Zhang, Z.~Yang, Y.~Jiang, J.~Wang, M.~Wu, W.~Feng, Cybersecurity
  testing for automotive domain: A survey, Sensors 22~(23) (2022) 9211.
\newblock \href {https://doi.org/10.3390/s22239211}
  {\path{doi:10.3390/s22239211}}.

\end{thebibliography}

\end{document}